\documentclass[journal ,12]{IEEEtran}

\usepackage{amsmath,epsfig,amssymb,verbatim,amsopn,cite,subfigure,multirow}
\usepackage{balance}
\usepackage{footnote}
\usepackage{algorithm,algorithmic}
\usepackage[usenames,dvipsnames]{color}
\usepackage[all]{xy}  %%% used to make block diagram
\usepackage{url}
\usepackage{xurl}
\usepackage{adjustbox}
\usepackage{enumitem}
\usepackage{cite,algorithm,algorithmic,amsmath,amssymb,amsthm,empheq,mhsetup}
\usepackage{array}
\usepackage{makecell}
\usepackage{subfigure,amsfonts,balance}
\usepackage{epstopdf}
\usepackage{setspace}
\usepackage{xcolor}
\usepackage{multirow}
\usepackage{array}
\usepackage{tikz}

\usepackage{pifont}% http://ctan.org/pkg/pifont
\usepackage{array}
\newcolumntype{P}[1]{>{\centering\arraybackslash}p{#1}}
\newcolumntype{M}[1]{>{\centering\arraybackslash}m{#1}}
\newtheorem{Remark}{Remark}

  {\proof}{\proofend}
\newtheorem{proposition}{Proposition}

\newcounter{mytempeqcounter}

\newcommand{\qa}{{\bf a}}
\newcommand{\qb}{{\bf b}}

\newcommand{\qe}{{\bf e}}

\newcommand{\qg}{{\bf g}}

\newcommand{\qr}{{\bf r}}

\newcommand{\qv}{{\bf v}}
\newcommand{\qw}{{\bf w}}

\newcommand{\qy}{{\bf y}}

\newcommand{\qG}{{\bf G}}

\newcommand{\qI}{{\bf I}}

\newcommand{\qZ}{{\bf Z}}

\newcommand{\ettall}{\emph{et al.}}

\newcommand{\Sn}{\sigma_n^2}

\newcommand{\dl}{\mathtt{dl}}
\newcommand{\ul}{\mathtt{ul}}

\newcommand{\FD}{\mathsf{FD}}
\newcommand{\HD}{\mathsf{HD}}
\newcommand{\SI}{\mathtt{SI}}

\newcommand{\FDCF}{\mathtt{FDCF}}
\newcommand{\FDCEL}{\mathtt{FDCel}}
\newcommand{\HDCEL}{\mathtt{HDCel}}
\newcommand{\NAFD}{\mathtt{NAFD}}
\newcommand{\DSHDceldl}{\mathtt{DS^{HDCell}_\dl}}
\newcommand{\DSHDcelul}{\mathtt{DS^{HDCell}_\ul}}
\newcommand{\DSFDceldl}{\mathtt{DS^{FDCell}_\dl}}
\newcommand{\DSFDcelul}{\mathtt{DS^{FDCell}_\ul}}
\newcommand{\DSFDCFdl}{\mathtt{DS^{FDCF}_\dl}}
\newcommand{\DSFDCFul}{\mathtt{DS^{FDCF}_\ul}}
\newcommand{\DSNAFDdl}{\mathtt{DS^{NAFD}_\dl}}
\newcommand{\DSNAFDul}{\mathtt{DS^{NAFD}_\ul}}

\newcommand{\PtotFDU}{\mathtt{P_{tot,\ul}^\FDCEL}}
\newcommand{\PtotFDD}{\mathtt{P_{tot,\dl}^\FDCEL}}
\newcommand{\PtotHDU}{\mathtt{P_{tot,\ul}^\HDCEL}}
\newcommand{\PtotHDD}{\mathtt{P_{tot,\dl}^\HDCEL}}

\newcommand{\PtotFD}{\mathtt{P_{tot}^\FDCEL}}
\newcommand{\PtotHD}{\mathtt{P_{tot}^\HDCEL}}
\newcommand{\PtotCFFC}{\mathtt{P_{tot}^\FDCF}}
\newcommand{\PtotNAFD}{\mathtt{P_{tot}^\NAFD}}
\newcommand{\MAPDL}{\mathcal{M}_{\dl}}
\newcommand{\MAPUL}{\mathcal{M}_{\ul}}

\newcommand{\Ntx}{N_{\mathtt{tx}}^\BS}
\newcommand{\Nrx}{N_{\mathtt{rx}}^\BS}
\newcommand{\NtxAP}{N_{\mathtt{tx}}^\AP}
\newcommand{\NrxAP}{N_{\mathtt{rx}}^\AP}

\newcommand{\hgmkpd}{\hat{\qg}_{mk'}^{\dl}}
\newcommand{\tgmkd}{\tilde{\qg}_{mk}^{\dl}}

\newcommand{\gmkd}{\qg_{mk}^{\dl}}

\newcommand{\hgmkd}{\hat{\qg}_{mk}^{\dl}}

\newcommand{\hgmlu}{\hat{\qg}_{m\ell}^{\ul}}

\newcommand{\gmlu}{\qg_{m\ell}^{\ul}}
\newcommand{\gamdmk}{\gamma_{mk}^{\dl}}
\newcommand{\gamdmkp}{\gamma_{mk'}^{\dl}}
\newcommand{\gamuml}{\gamma_{m\ell}^{\ul}}

\newcommand{\betamkd}{\beta_{mk}^{\dl}}

\newcommand{\betakldu}{\beta_{k\ell}^{\mathtt{du}}}
\newcommand{\betamlu}{\beta_{m\ell}^{\ul}}

\newcommand{\alphml}{\alpha_{m\ell}}

\newcommand{\KU}{\mathcal{K}_{\ul}}
\newcommand{\KD}{\mathcal{K}_{\dl}}

\newcommand{\MR}{\mathtt{MR}}

\newcommand{\ZF}{\mathtt{ZF}}

\newcommand{\SE}{\mathtt{SE}}
\newcommand{\EE}{\mathtt{EE}}

\newcommand{\Pbfixdm}{\mathtt{P}_{\AP,m}^{\mathtt{fh,fixed}}}
\newcommand{\Pbtm}{\mathtt{P}_{\AP,m}^{\mathtt{fh,td}}}
\newcommand{\Etot}{\mathtt{E_{tot}}}
\newcommand{\Ptot}{\mathtt{P_{tot}}}
\newcommand{\SEtot}{\mathtt{SE_{tot}}}
\newcommand{ \PSICBS}{\mathtt{P_{BS}^{SIC}}}
\newcommand{ \PSICAP}{\mathtt{P_{AP}^{SIC}}}
\newcommand{ \PBSstat}{\mathtt{P_{BS}^{sta}}}
\newcommand{ \PAPstat}{\mathtt{P_{AP}^{sta}}}

\newcommand{ \UEk}{\mathtt{UE}_{k,\ul}}
\newcommand{ \BS}{\mathtt{BS}}
\newcommand{ \AP}{\mathtt{AP}}
\newcommand{ \data}{\mathtt{dat}}
\newcommand{ \PBSk}{\mathtt{P}_{\mathtt{BS}}^{\data}}
\newcommand{ \PAPm}{\mathtt{P}_{\mathtt{AP},m}^{\data}}

\newcommand{ \PAPbhm}{\mathtt{P}_{\mathtt{AP},m}^{\mathtt{fh}}}

\newcommand{ \circuit}{\mathtt{cir}}
\newcommand{ \Pulcir}{\mathtt{P}_{\ul,k}^{\circuit}}
\newcommand{ \PAPcir}{\mathtt{P}_{\AP,m}^{\circuit}}
\newcommand{ \PBScir}{\mathtt{P}_{\BS}^{\circuit}}

\newcommand{ \PAPbhNAFD}{\mathtt{P}_{\mathtt{fh}}^{\NAFD}}
\newcommand{ \PULUE}{\mathtt{P}_{\ul,k}^{\data}}
\newcommand{ \PULUEDyn}{\mathtt{P}_{\ul,k}^\mathtt{dyn}}
\newcommand{ \PULUESta}{\mathtt{P}_{\ul,k}^\mathtt{sta}}
\newcommand{ \PBSsdyndl}{\mathtt{P_{BS,dl}^{dyn}}}
\newcommand{ \PBSsdynul}{\mathtt{P_{BS,ul}^{dyn}}}
\newcommand{ \PAPdyndl}{\mathtt{P_{AP,dl}^{dyn}}}
\newcommand{ \PAPdynul}{\mathtt{P_{AP,ul}^{dyn}}}
\newcommand{\CCI}{\mathtt{CCI}}
\newcommand{\CCIu}{\mathtt{CCI}_{\ul}}
\newcommand{\CCId}{\mathtt{CCI}_{\dl}}
\newcommand{\BBI}{\mathtt{IBSI}}
\newcommand{\NOISE}{\Sn}
\newcommand{\UTDI}{\mathtt{UDI}}

\newcommand{\CCIuCF}{\mathtt{CCI}_{\ul}^{\FDCF}}
\newcommand{\CCIdCF}{\mathtt{CCI}_{\dl}^{\FDCF}}
\newcommand{\IAPICF}{\mathtt{IAPI}^{\FDCF}}
\newcommand{\IAPINAFD}{\mathtt{IAPI}^{\mathtt{NAFD}}}
\newcommand{\UDIdCF}{\mathtt{UDI}^{\FDCF}}
\newcommand{\UDICel}{\mathtt{UDI}^{\mathtt{Cel}}}
\newcommand{\UDIdNAFD}{\mathtt{UDI}^{\mathtt{NAFD}}}
\newcommand{\CCIuNA}{\mathtt{CCI}_{\ul}^{\mathtt{NAFD}}}
\newcommand{\CCIdNA}{\mathtt{CCI}_{\dl}^{\mathtt{NAFD}}}
\newcommand{\Intrad }{\CCI_{\mathtt{IntraCell,\dl}}}
\newcommand{\Interd }{\CCI_{\mathtt{InterCell,\dl}}}
\newcommand{\Intrau }{\CCI_{\mathtt{IntraCell,\ul}}}
\newcommand{\Interu }{\CCI_{\mathtt{InterCell,\ul}}}

\newcommand{\hgmk}{\hat{\qg}_{mk_u}^{\ul}}
\newcommand{\SIBS}{\mathtt{SI_{BS}}}
\newcommand{\SIAP}{\mathtt{SI_{AP}}}

\newcommand{\SEQoS}{\mathcal{S}_\dl^o}

\DeclareMathOperator{\MM}{\mathcal{M}}

\DeclareMathOperator{\K}{\mathcal{K}}

\newcommand{\Ex}{\mathbb{E}}

\newcommand{\Sm}{\mathcal{S}_m^{\dl}}
\newcommand{\Wm}{\mathcal{W}_m^{\dl}}
\newcommand{\Smu}{\mathcal{S}_m^{\ul}}
\newcommand{\Wmu}{\mathcal{W}_m^{\ul}}
\newcommand{\Zk}{\mathcal{Z}_k^{\dl}}
\newcommand{\Tk}{\mathcal{T}_k^{\dl}}
\newcommand{\Zqdl}{\mathcal{Z}_q^{\dl}}
\newcommand{\Tqdl}{\mathcal{T}_q^{\dl}}
\newcommand{\Zlup}{\mathcal{Z}_{\ell}^{\ul}}
\newcommand{\Tlup}{\mathcal{T}_{\ell}^{\ul}}
\newcommand{\wmkdl}{\qv_{mk}^{\dl}}
\newcommand{\wmkdlzf}{\qv_{mk}^{\dl,\ZF}}
\newcommand{\wmkdlmr}{\qv_{mk}^{\dl,\MR}}
\newcommand{\wmlulzf}{\qv_{m\ell}^{\ul,\ZF}}
\newcommand{\wmlulmr}{\qv_{m\ell}^{\ul,\MR}}
\newcommand{\wikdl}{\qv_{ik}^{\dl}}
\newcommand{\wmkpdl}{\qv_{mk'}^{\dl}}

\newcommand{\wmlul}{\qv_{m\ell}^{\ul}}

\newcommand{\gamumlp}{\gamma_{m\ell'}^{\ul}}
\newcommand{\hGmdl}{\hat{\qG}_{\mathcal{S}_m}^{\dl}}

\newcommand{\hGmul}{\hat{\qG}_{m}^{\ul}}
\newcommand{\dm}{\delta_{m}}
\DeclareMathOperator{\Z}{\mathbf{Z}}
\newcommand{\tgmlu}{\tilde{\qg}_{m\ell}^{\ul}}
\newcommand{\SIm}{\sigma^2_{\mathtt{SI},m}}

\title{Ten Years of Research Advances in Full-Duplex Massive MIMO}

\author{Mohammadali Mohammadi,~\IEEEmembership{Senior Member,~IEEE,} Zahra Mobini,~\IEEEmembership{Member,~IEEE,}\\
Hien Quoc Ngo,~\IEEEmembership{Senior Member,~IEEE,} and Michail Matthaiou~\IEEEmembership{Fellow, IEEE}\\
\textit{(Invited Paper)}
\thanks{The authors are with the Centre for Wireless Innovation (CWI), Queen's University Belfast, U.K.
email:\{m.mohammadi, zahra.mobini, hien.ngo, m.matthaiou\}@qub.ac.uk.

This work is a contribution by Project REASON, a UK Government funded project under the Future Open Networks Research Challenge (FONRC) sponsored by the Department of Science Innovation and Technology (DSIT). It was also supported by the U.K. Engineering and Physical Sciences
Research Council (EPSRC) (grant No. EP/X04047X/1). The work of M. Mohamamadi and M. Matthaiou was
supported by the European Research Council
(ERC) under the European Union’s Horizon 2020 research
and innovation programme (grant agreement No. 101001331). The work of Z.~Mobini and  H.~Q.~Ngo was supported by the U.K. Research and Innovation Future Leaders Fellowships under Grant MR/X010635/1, and a research grant from the Department for the Economy Northern Ireland under the US-Ireland R\&D Partnership Programme.
}}

\allowdisplaybreaks
\begin{document}
\bstctlcite{IEEEexample:BSTcontrol}
\maketitle
\thispagestyle{empty}

\begin{abstract}
We present an overview of ongoing research endeavors focused on in-band full-duplex (IBFD) massive multiple-input multiple-output (MIMO) systems and their applications. In response to the unprecedented demands for mobile traffic in concurrent and upcoming wireless networks, a paradigm shift from conventional cellular networks to distributed communication systems becomes imperative. Cell-free massive MIMO (CF-mMIMO) emerges as a practical and scalable implementation of distributed/network MIMO systems, serving as a crucial physical layer technology for the advancement of next-generation wireless networks. This architecture inherits benefits from co-located massive MIMO and distributed systems and provides the flexibility for integration with the IBFD technology. We delineate the evolutionary trajectory of cellular networks, transitioning from conventional half-duplex multi-user MIMO networks to IBFD CF-mMIMO. The discussion extends further to the emerging paradigm of network-assisted IBFD CF-mMIMO (NAFD CF-mMIMO), serving as an energy-efficient prototype for asymmetric uplink and downlink communication services. This novel approach finds applications in dual-functionality scenarios, including simultaneous wireless power and information transmission, wireless surveillance, and integrated sensing and communications. We highlight various current use case applications, discuss open challenges, and outline future research directions aimed at fully realizing the potential of NAFD CF-mMIMO systems to meet the evolving demands of future wireless networks.
\end{abstract}
%***************************************************************************

%%***************************************************************************

\section*{Main Nomenclature}
\addcontentsline{toc}{section}{Nomenclature}
\begin{IEEEdescription}[\IEEEusemathlabelsep\IEEEsetlabelwidth{$V_1,V_2,V_3,V_4$}]
\fontsize{0.33cm}{0.4cm}\selectfont
\item[$3$GPP]     3rd generation partnership project
\item[$5$G]    Fifth generation
\item[$6$G]    Sixth generation 
\item[ADC]     Analog-to-digital converter 
\item[AI]     Artificial intelligence
\item[AP]      Access point
\item[B5G]     Beyond fifth generation
\item[BS]       Base station
\item[CCI]        Co-channel interference
\item[CF-mMIMO]   Cell-free massive MIMO
\item[CLI]      Cross-link interference
\item[CPU]     Central processing unit
\item[CSI]     Channel state information
\item[DL]      Downlink
\item[E-UE]      Energy UE
\item[EE]      Energy efficiency
\item[FZF]     Full zero-forcing
\item[HD]      Half-duplex
\item[IBFD]    In-band full-duplex
\item[IoT]     Internet-of-Things
\item[ISAC]       Integrated sensing and communication 
\item [LTE]   Long-term evolution
\item[MIMO]    Multiple-input multiple-output
\item[ML]      Machine learning
\item[mmWave]  Millimeter wave
\item [MRT]    Maximum ratio transmission
\item [MU-MIMO]    Multiuser MIMO
\item [NAFD]    Network-assisted full-duplex
\item [OFDM]   Orthogonal frequency division modulation 
\item[O-RAN]     Open RAN
\item[PZF]      Partial ZF 
\item[QoS]      Quality-of-service
 \item[RAN]     Radio access network
\item[RF]      Radio-frequency
\item[RIS]     Reconfigurable intelligent surface
\item[SE]      Spectral efficiency
\item[SI]      Self interference
\item[SIC]     Self interference cancellation
\item[SWIPT]   Simultaneous wireless information and power transfer
\item[TDD]     Time-division duplex 
\item[UAV]     Unmanned aerial vehicle
\item[UE]      User equipment 
\item[UL]      Uplink
\item[URLLC]   Ultra-reliable low-latency communication
\item[ZF]      Zero forcing
\end{IEEEdescription}

%%%%%%%%%%%%%%%%%%%%%%%%%%%%%%%%%%%%%%%%%%%%%%
\begin{figure*}[t]
	\centering
  \vspace{0em}
	\includegraphics[width=170mm, height=77mm]{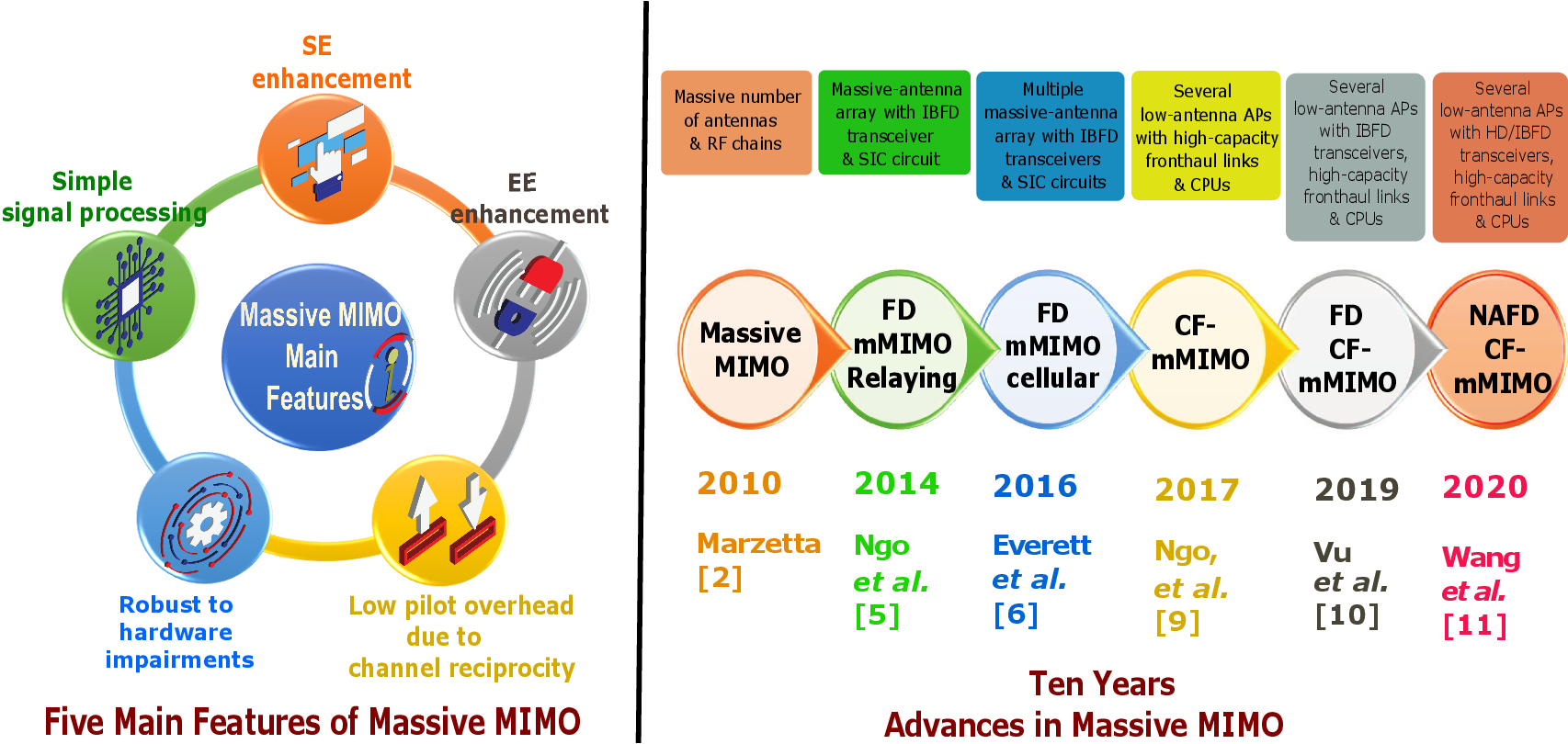}
		\caption{ Illustrative figure showcasing the five unique features of massive MIMO technology and the scientific breakthroughs in the IBFD massive MIMO space.}
 \vspace{-1em}
	\label{fig:Fig1}
\end{figure*}
%%%%%%%%%%%%%%%%%%%%%%%%%%%%%%%%%%%%%%%%%%%%%%

%%%%%%%%%%%%%%%%%%%%%%%%%%%%%%%%%%%%%%%%%%%%%%%%%%
\vspace{-1em}
\section{Introduction}
%%%%%%%%%%%%%%%%%%%%%%%%%%%%%%%%%%%%%%%%%%%%%%%%%%%
The revolution in mobile communication networks has been significantly propelled by advancements in multiple-input multiple-output (MIMO) technology. This transformation is evident as single-input single-output (SISO) transmitters, prevalent in the first generation ($1$G) of mobile networks, have been supplanted by massive MIMO base stations (BSs) in the fifth-generation ($5$G) wireless networks~\cite{Matthaiou:JSAC:2020}. This has been sparked by the promising potential of massive MIMO technology, which includes enhancements in spectral efficiency (SE) and energy efficiency (EE). These enhancements are achieved through massive MIMO's ability to alleviate intra-cell and inter-cell interference via simple linear processing schemes, along with a reduction in the channel estimation overhead thanks to channel reciprocity~\cite{Marzetta:TWC:2010} (cf. left hand side of Fig.~\ref{fig:Fig1}).
Simultaneously, in response to the increasing number of wireless devices and the limited available spectrum resources, orthogonal multiple access techniques have evolved from frequency-division multiple access (FDMA) in $1$G to orthogonal FDMA (OFDMA) in the fourth generation ($4$G). Nevertheless, to address the challenges posed by emerging heterogeneous services and applications in $5$G networks (e.g., virtual reality, augmented reality, and Industry $4.0$), non-orthogonal transmission strategies have been integrated~\cite{Liu:JSAC:022}. In this context, the promise of radical in-band full-duplex (IBFD) operation, demonstrated to double the SE through proof-of-concept results, has gained increasing attention~\cite{kim2024state}.
Beyond enhancing the SE, IBFD technology holds the potential of integration with other communication technologies and catalyze the development of flexible network architectures. Consequently, it can unlock a myriad of new applications within the dynamic realm of wireless networks. Among these, the synergy of IBFD and massive MIMO stands out due to the complementary strengths of both technologies. IBFD communication enhances the SE, while massive MIMO excels in managing multiple data streams simultaneously, thereby optimizing the usage of the scarce spectrum resources. Furthermore, massive MIMO's advanced signal processing capabilities can effectively mitigate the self-interference (SI) challenges inherent in IBFD systems. This combination not only promises substantial improvements in data rates and latency but also supports the growing demand for higher capacity and efficiency in next-generation wireless networks. This synergy was initially introduced for multipair relaying systems in~\cite{Multipair:JSAC:2014}, and subsequently, it was applied to cellular massive MIMO systems in~\cite{Everett:TWC:2016}.

Relying on massive MIMO as core technology, the $5$G landscape has evolved beyond mobile broadband (MBB) in $4$G to encompass enhanced MBB (eMBB) and the Internet of Things (IoT). The IoT comprises massive machine-type communications (mMTC) and ultra-reliable, low-latency communications (uRLLC), catering to a diverse range of connectivity needs. While $5$G networks are being deployed worldwide and are smoothly penetrating our daily lives, both industry and academia have already embarked on the exploration of sixth-generation ($6$G) wireless networks~\cite{Tataria:PROC:2021,Matthaiou:COMMag:2021}. According to the latest insights provided by the International Telecommunication Union (ITU) for $6$G~\cite{Tataria:PROC:2021}, the forthcoming generation is envisioned not only to enhance existing $5$G usage scenarios—such as achieving a $100$-fold increase in peak data rate (from Gbps to Tbps), a $10$-fold reduction in latency with a hyper-reliability requirement of $99.99999\%$, and a $10$-fold enhancement in connection density—but also to usher in a new era in wireless communication. This includes integrated sensing and communication (ISAC) along with the consolidation of artificial intelligence and communication. On the other hand, another key objective of $6$G  is to achieve reliable, high-speed and uniform connectivity \emph{anywhere},  \emph{anytime} to  \emph{anyone} across an extensive coverage area, a demand that $5$G struggles to fully meet  by relying on  co-located topologies of massive MIMO. The limitations arise from the inherent challenges of inter-cell interference within the  cellular network, leading to significant performance degradation for cell-edge users. Therefore, the forthcoming communication networks are poised to undergo paradigm shifts, embracing innovative and disruptive network designs and technologies. 

This renewed interest has given rise to a range of innovative solutions based on massive MIMO. Consequently, a novel network architecture leveraging massive MIMO technology, known as \textit{cell-free massive MIMO (CF-mMIMO)} network, has emerged over the past decade~\cite{Hien:cellfree}. CF-mMIMO, a fusion of massive MIMO and network MIMO, stands out as a promising wireless networking technology to address the cell-edge problem and provide pervasive connectivity in wireless communication networks. In CF-mMIMO, many access points (APs), distributed over a wide area, coherently serve many user equipments (UEs). The total number of service antennas is much larger than the number of UEs. The reduced coverage area of multiple distributed APs within a CF-mMIMO system, resulting in lower transmit power levels at the APs, renders them an ideal infrastructure for deploying IBFD transceivers. This is due to a significant reduction in complexity and power consumption associated with self-interference cancellation (SIC) circuits compared to co-located IBFD massive MIMO BSs. Consequently, IBFD CF-mMIMO networks were introduced in~\cite{tung19ICC}. The lower transmit power at the APs further results in lower amount of inter-AP interference, known as cross-link interference (CLI). 
Therefore, IBFD CF-mMIMO networks are theoretically capable of providing ubiquitous connectivity for all the UEs in any location and at any time.   

Nevertheless, IBFD CF-mMIMO appears to face challenges in becoming a feasible choice for $6$G networks. Specifically, with the proliferation of numerous IBFD transceivers, the APs required for supporting mMTC and uRLLC applications in $6$G, would experience substantial CLI, thereby compromising the overall system performance. To tackle this challenge, a novel concept known as \textit{network-assisted IBFD CF-mMIMO (NAFD CF-mMIMO)} has recently emerged~\cite{Wang:TCOM:2020}. The NAFD concept aims to achieve IBFD functionality by effectively leveraging existing half-duplex (HD) hardware devices in a virtual manner. Therefore, by eliminating the need for SIC at the APs, we achieve a notable reduction in both the computational complexity and power consumption. Furthermore, compared to IBFD CF-mMIMO networks, there is a significant decrease in CLI. The NAFD conceptual framework can be also exploited in other exciting application-oriented opportunities aiming at combining electromagnetic radiation for dual purposes. These include simultaneous wireless information and power transfer (SWIPT), wireless surveillance, and ISAC. The advances in the IBFD massive MIMO space over a period of ten years are summarized in the right-hand side of Fig.~\ref{fig:Fig1}.    

%%%%%%%%%%%%%%%%%%%%%%%%%%%%%%%%%%%%%%%%
\subsection{Existing Surveys and Tutorials}
%%%%%%%%%%%%%%%%%%%%%%%%%%%%%%%%%%%%%%%%
In this subsection, we provide an overview of the existing survey and tutorial papers on IBFD, massive MIMO, and CF-mMIMO. We highlight the main focus of each paper to give a comprehensive understanding of the current research landscape and key areas of interest within these technologies.
%-----------------------------
\subsubsection{IBFD}
%-----------------------------
Several surveys and tutorials on IBFD have been published~\cite{Ashutosh:JSAC:2014, Zhang:PROIEEE:2016, Hong:COMM:2014, Goyal:COMM:2015, Mohammadi:TUT:2024, Smida:JPROC:2024}. These works discuss SIC techniques, key technical challenges, and the integration of IBFD with other wireless technologies. In particular, Ashutosh \textit{et al.}~\cite{Ashutosh:JSAC:2014} explored opportunities for IBFD communication in relay, bidirectional, and cellular topologies, along with their generalizations. They also reviewed various SI reduction methods, including propagation-domain, analog-circuit-domain, and digital-domain SIC. Hong \textit{et al.}~\cite{Hong:COMM:2014} provided an overview of a general SI suppression architecture and discussed emerging applications that may adopt SI suppression with minimal changes to existing standards.  Goyal \textit{et al.}~\cite{Goyal:COMM:2015} reviewed fundamental challenges in incorporating IBFD radios in cellular networks, including new interference scenarios and increased energy consumption due to SI suppression. They proposed scheduling strategies to maximize the joint uplink (UL) and downlink (DL) utility by scheduling either a single DL or UL UE or a pair of DL and UL UEs. Zhang \textit{et al.}~\cite{Zhang:PROIEEE:2016} outlined critical implementation issues, performance enhancements, and design challenges for practical IBFD systems. They analyzed primary impairments such as phase noise, power amplifier nonlinearities, and in-phase and quadrature-phase imbalance. Additionally, this work discussed IBFD-based medium access control-layer protocol designs to address hidden terminal problems, end-to-end delay, and high packet loss due to network congestion, proposing potential solutions to these challenges. Mohammadi \textit{et al.}~\cite{Mohammadi:TUT:2024} offered a comprehensive survey of IBFD technology, highlighting transceiver structures and SIC techniques, and discussing the coexistence of IBFD with existing and emerging wireless technologies. Smida \textit{et al.}~\cite{Smida:JPROC:2024} reviewed key concepts and advancements in the physical-layer design of IBFD communications, followed by an overview of basic concepts and the state of the art in massive MIMO IBFD.

%%%%%%%%%%%%%%%%%%%%%%%%%%%%%%%%%%%%%%%%%%%%%%%%%%%%%%
\begin{table*}[t]
  \centering
  \caption{Summary of the existing survey and tutorial papers.}
  \vspace{-1em}
  \small
  \begin{tabular}{|c|>{\centering\arraybackslash}m{0.7cm}|c|>{\centering\arraybackslash}m{11cm}|}
    \hline
    \textbf{Category} & \textbf{Year} & \textbf{Lit.} & \textbf{\hspace{0em}Main focus} \\
    \hline
    \multirow{6}{*}{IBFD} 
    & 2014 &\cite{Ashutosh:JSAC:2014} 
    & IBFD communication in relay, bidirectional, and cellular topologies
    \\ \cline{2-4}
    & 2014 &\cite{Hong:COMM:2014}     
    & General overview of SI suppression architectures
    \\ \cline{2-4}
    & 2015 &\cite{Goyal:COMM:2015}    
    & Fundamental challenges in incorporating IBFD radios in cellular networks 
    \\ \cline{2-4}
    & 2016 &\cite{Zhang:PROIEEE:2016} 
    &Critical implementation issues and design challenges for practical IBFD systems
    \\ \cline{2-4}
    & 2024 &\cite{Mohammadi:TUT:2024} 
    & Coexistence of IBFD with existing and emerging wireless technologies
    \\ \cline{2-4}
    & 2024 &\cite{Smida:JPROC:2024}   
    & Key concepts and advancements in the physical-layer design of IBFD  
    \\ 
    \hline
    \multirow{6}{*}{\vspace{2em}Massive MIMO} 
    & 2014 & \cite{Larsson:COMMG:2014} 
    & Overview of the advancements and potential of massive MIMO technology 
    \\ \cline{2-4}
    & 2014 &\cite{Lu:JSTSP:2014}       
    & Benefits and challenges associated with massive MIMO technology and key implementation issues 
    \\ \cline{2-4}
    & 2015 &\cite{Zheng:Tut:2015}      
    & Large-scale MIMO channel measurement and modeling, typical application scenarios, and critical techniques at both physical and network layers 
    \\ 
    \cline{2-4}
    & 2019 & \cite{Wang:JSTSP:2019} & Overview of array signal processing
            techniques for massive MIMO systems \\ 
    \hline
    \multirow{6}{*}{\vspace{-6em}CF-mMIMO} 
    & 2022 &\cite{Elhoushy:Tuts:2022} 
    & Precoding and detection techniques for DL and UL transmission under different channel fading models
    \\ \cline{2-4}
    & 2022 &\cite{HeLJCOM:2022}       
    & UE association, pilot assignment, transmitter and receiver design, and power control 
    \\ \cline{2-4}
    & 2022 &\cite{Zhang:Survey:2022}  
    & Signal processing methods to minimize fronthaul load for combined channel estimation and transmit precoding
    \\ \cline{2-4}
    & 2023 &\cite{kassam2023review}   
    & Challenges related to the limited capacity of fronthaul links and the connection between APs and UEs \\ \cline{2-4}
    &\centering 2024 &\cite{Ngo:JPROC:2024}     
    & Ultradense CF-mMIMO networks and important unresolved questions regarding their future deployment \\ \cline{2-4}
    &\centering 2024 &\cite{Zheng:WC:2024}      
    & Practical implementation challenges, including mobility issues, channel calibration, and scalability, along with respective solutions \\ 
    \hline
    \multirow{6}{*}{\vspace{5em}\textbf{This paper}} 
    & 2024 & 
    & Advancements in IBFD massive MIMO systems, ranging from cellular networks to NAFD CF-mMIMO systems
    \\ \hline
  \end{tabular}\label{table:survey}
  \vspace{-0.6em}
\end{table*}

%%%%%%%%%%%%%%%%%%%%%%%%%%%%%%%%%%%%%%%%%%%%%%%%%%%%%%%
%-----------------------------
\subsubsection{Massive MIMO}
%-----------------------------
The seminal survey paper on massive MIMO~\cite{Larsson:COMMG:2014} provided a comprehensive overview of the advancements and potential of massive MIMO technology. The research highlighted significant improvements in EE, spatial multiplexing, and throughput capabilities enabled by large antenna arrays. The authors elaborated on challenges, such as channel estimation and pilot contamination, emphasizing the benefits of time-division duplexing (TDD) and the importance of low-cost, low-power components. Additionally, the paper addressed practical implementation issues, including synchronization, the impact of hardware imperfections, and potential deployment scenarios for future broadband networks.
Lu \textit{et al.}~\cite{Lu:JSTSP:2014} provided a detailed examination of the benefits and challenges associated with massive MIMO technology. By relying on an information-theoretic analysis, they demonstrated the significant improvements in SE and EE achievable through massive MIMO. Key implementation issues such as channel estimation, detection, and precoding were thoroughly discussed, with a particular focus on the impact of pilot contamination. Zheng \textit{et al.}~\cite{Zheng:Tut:2015} covered key aspects, such as large-scale MIMO channel measurement and modeling, typical application scenarios, and critical techniques at both the physical and network layers. The authors also discussed implementation challenges and future research directions, highlighting the importance of EE and SE. Wang \textit{et al.}~\cite{Wang:JSTSP:2019} presented a comprehensive overview of array signal processing in massive MIMO communications. The primary focus of their work is to explore massive MIMO, particularly in sparse scenarios, such as millimeter wave (mmWave) transmission. The authors first examined various techniques to accurately extract physical channel parameters, including the angle of arrival, angle of departure, multi-path delay, and Doppler shift parameters.  Then, they discussed the design of transceiver techniques, such as synchronization, channel estimation, beamforming, precoding, and multi-user access, from the array signal processing viewpoint.

%-----------------------------
\subsubsection{CF-mMIMO}
%-----------------------------
The following survey papers focus on different aspects of CF-mMIMO, each highlighting unique perspectives and challenges within the field. Collectively, they provide a comprehensive understanding of the advancements and future directions of CF-mMIMO technology. Elhoushy \textit{et al.}~\cite{Elhoushy:Tuts:2022} reviewed various aspects of CF-mMIMO systems, focusing on precoding and detection techniques for DL and UL transmission under different channel fading models. They also emphasized the potential integration of CF-mMIMO systems with enabling techniques and technologies for $5$G and beyond $5$G (B5G) networks. He \textit{et al.}~\cite{HeLJCOM:2022} discussed the enabling physical layer technologies for CF-mMIMO, including UE association, pilot assignment, transmitter and receiver design, and power control. Zhang \textit{et al.}~\cite{Zhang:Survey:2022} presented a comprehensive survey and quantified the advantages of CF-mMIMO systems in terms of energy and cost efficiency. They also analyzed signal processing techniques used to reduce the fronthaul burden for joint channel estimation and transmit precoding. Kassam \textit{et al.}~\cite{kassam2023review} provided an overview of the current state-of-the-art in CF-mMIMO systems, focusing on challenges related to the limited capacity of fronthaul links and the connection between APs and UEs. They also discussed prospective cell-free technologies, including reconfigurable intelligent surfaces (RIS), unmanned aerial vehicles (UAVs), and artificial intelligence (AI)-enabled CF-mMIMO systems. Ngo \textit{et al.}~\cite{Ngo:JPROC:2024} provided a contemporary overview of ultradense CF-mMIMO networks and addressed important unresolved questions regarding their future deployment. Zheng \textit{et al.}~\cite{Zheng:WC:2024} outlined an overview of mobile CF-mMIMO communication systems, focusing on the main challenges in practical implementations, including mobility issues, channel calibration, and scalability. They also discussed respective solutions such as channel prediction, hierarchical cancellation and rate-splitting, and dynamic clustering. 

Table~\ref{table:survey} summarizes the key contributions of these surveys, capturing their main findings and insights.

%-----------------------
\begin{table*}[t]
    \centering
    \small
    \caption{Organization of the Paper and Relationships Between Sections.}
    \vspace{-0.5em}
    \begin{tabular}{|c|>{\centering\arraybackslash}m{3.35cm}|>{\centering\arraybackslash}m{6cm}|>{\centering\arraybackslash}m{6cm}|}
        \hline
        \textbf{Section} & \textbf{Title} & \textbf{Description} & \textbf{Relationship to Other Sections} \\
        \hline
        I & Introduction & Introduces the topic and outlines the purpose and structure of the paper. Provides an overview of existing research on IBFD, massive MIMO, and CF-mMIMO technologies. & Sets the stage for the entire paper, providing context for the detailed discussions in subsequent sections and the main motivation of the paper. \\
         \hline
        II & In-Band Full-Duplex and Massive MIMO Techniques: State-of-the-Arts and Review & Describes the fundamental principles, opportunities, and challenges  of IBFD and massive MIMO technology. & Serves as a foundation for understanding the new structures for wireless networks discussed in later sections. \\
        \hline
        III & In-Band Full-Duplex Cellular Networks & Discusses the transition from conventional HD cellular networks to IBFD massive MIMO, highlighting key developments. & Showing the progression and evolution of the cellular technology towards CF-mMIMO. \\
        \hline
         IV &The Path Towards In-Band Full-Duplex Cell-free Massive MIMO  &Explains the concept of CF-mMIMO and the integration of IBFD into it, along with the corresponding opportunities and challenges. &Introduces an innovative network structure in response to the challenges and requirements discussed in Section III. \\
        \hline
        V & Network-Assisted Full-Duplex Cell-Free Massive MIMO & Explores the concept of NAFD CF-mMIMO system and its advantages/challenges. & Explain the ultimate transition in wireless networks' structure discussed in the last two sections. \\
         \hline
        VI & Mathematical Framework and System Design& Includes  system model description and signaling, precoding design, UE association, and SE analysis.  & This model enables analytical and theoretical comparisons between different network structures in Sections III, IV, and V. \\
        \hline
        VII & Interplay Between NAFD CF-mMIMO and Emerging Technologies & Examines current applications of NAFD CF-mMIMO for distributed implementation of SWIPT, ISAC, and wireless surveillance and outlines potential future research areas. & Applies the concepts from Sections IV and V to practical dual-functional scenarios. \\
         \hline
        VIII & Key open Challenges and Future Research Directions & Highlights some open challenges in NAFD CF-mMIMO networks and introduces potential future research directions. & Provides open future research directions based on the discussions in Sections V and VII.  \\
        \hline
        IX & Conclusion & Summarizes the key points of the paper. & Ties together all the sections, reflecting on the discussions and findings presented throughout the paper. \\
        \hline
    \end{tabular}\label{table:organization}
    \vspace{-1.0em}
\end{table*}
%-----------

%%%%%%%%%%%%%%%%%%%%%%%%%%%%%%%%%%%%%%%%
\subsection{Main Contributions and Paper Organization}
%%%%%%%%%%%%%%%%%%%%%%%%%%%%%%%%%%%%%%%%%
Distinct from the aforementioned literature, in this article, we present a comprehensive overview of the decade-long advancements in IBFD massive MIMO systems. Our survey delineates the connectivity characteristics of diverse IBFD massive MIMO systems, ranging from cellular networks to NAFD CF-mMIMO systems. We specifically address the transitions between different architectures, providing clear explanations of the underlying rationale and the benefits achieved with each shift. By evaluating the SE and EE of each scenario, we present a thorough comparison of the different IBFD massive MIMO implementations.

In Section~\ref{sec:intro}, we explore the opportunities and main challenges associated with IBFD and massive MIMO technologies. Section~\ref{sec:IBFDcellular} discusses the transition from HD multiuser MIMO (HD MU-MIMO) to IBFD MU-MIMO and eventually to IBFD massive MIMO. Section~\ref{sec:IBFDcellfree} explains the progression towards IBFD CF-mMIMO systems, which represents a distributed implementation of massive MIMO. Following this, Section~\ref{sec:NAFDcellfree} examines the transition from IBFD CF-mMIMO to NAFD CF-mMIMO. In Section~\ref{sec:mathModel}, we present a mathematical framework for NAFD CF-mMIMO networks.  Section~\ref{sec:NAFDcellfreeApp} includes conceptual discussions  and distributed implementation scenarios. Finally, Section~\ref{sec:openchallenges} addresses the opportunities and challenges that IBFD massive MIMO systems present for future 6G networks, followed by the conclusion remarks in Section~\ref{sec:conc}. Table~\ref{table:organization} describes the organization of this paper and the relationship between different sections.

\textit{Notation:} We use bold upper case letters to denote matrices, and lower case letters to denote vectors. The superscripts $(\cdot)^\dag$ and $(\cdot)^{-1}$ stands for conjugate-transpose (Hermitian) and matrix inverse, respectively;  $\mathbf{I}_N$ denotes the $N\times N$ identity matrix. The circular symmetric complex Gaussian distribution having variance $\sigma^2$ is denoted by $\mathcal{CN}(0,\sigma^2)$.  
%%%%%%%%%%%%%%%%%%%%%%%%%%%%%%%%%%%%%%%%%%%%%%%%%
\section{In-Band Full-Duplex and Massive MIMO Techniques: State-of-the-Arts and Review}
\label{sec:intro}
%%%%%%%%%%%%%%%%%%%%%%%%%%%%%%%%%%%%%%%%%%%%%%%%%
%%%%%%%%%%%%%%%%%%%%%%%%%%%%%%%%%%%%%%%%%%%%%%%%%%%%%%%%%%%%%%%%%%%%
\subsection{In-Band Full-Duplex Communications}
%%%%%%%%%%%%%%%%%%%%%%%%%%%%%%%%%%%%%%%%%%%%%%%%%%%%%%%%%%%%%%%%%%%%
IBFD transmission, with a long history dating back to its applications in continuous wave radar system designs since at least the 1940s~\cite{o1963high}, has experienced a resurgence in interest over the past decade. This renewed attention has sparked efforts to re-architect wireless communication systems, transforming the IBFD concept from a laboratory experimental platform to a significant influence on telecommunications standards and commercial product development~\cite{Jain2011,Katti:Sigcomm:2013}. This heightened attention toward IBFD communication is motivated by the acknowledgment that IBFD transmission holds the potential to double the ergodic capacity of MIMO systems compared to conventional HD communications~\cite{Sabharwal:JSAC:2014,Zhang:CMG:2015,Zhang:PIEE:2016,kim2024state}. HD systems employ either TDD or frequency-division duplexing (FDD) for signal transmission, requiring independent channels for bidirectional end-to-end communication. This configuration results in a reduction in resource utilization. In contrast, IBFD systems overcome this limitation by simultaneously transmitting and receiving data signals within the same frequency band~\cite{Sabharwal:JSAC:2014}. The main advantages offered by the IBFD communication over the conventional HD systems can be summarized as follows:
\begin{itemize}
    \item IBFD communication can theoretically double the SE/ergodic capacity compared to HD transmission.
    \item End-to-end delay in IBFD systems is substantially reduced~\cite{Ju:JSAC:2012}.
    \item IBFD entails shorter latency for feedback signalling reception. 
    \item Simultaneous transmission in bi-directional links or the ability for simultaneous overhearing and jamming transmission makes eavesdropping challenging for malicious nodes, thereby enhancing the performance from a secrecy standpoint~\cite{Mobini:TIFS:2019}. 
    \item IBFD offers spectrum usage flexibility by allowing the use of either a single frequency band (IBFD transmission) or two distinct frequency bands (HD transmission) for UL and DL. Each transceiver has the flexibility to choose between the IBFD and HD transmission modes~\cite{Riihonen:WCOm:2011}.
\end{itemize}

The IBFD transmission capability necessitates each node to be equipped with both transmit and receive radio-frequency (RF) chains. The separation of transmit and receive signals can be achieved through either shared or separated antenna structures~\cite{Katti:Sigcomm:2013}. In shared antenna structures, a circulator acts as a duplexer, effectively isolating the receiver from the transmitter, while allowing them to share common antennas~\cite{Katti:Sigcomm:2013}. Conversely, in scenarios where a node is equipped with more than two antennas, a separated antenna structure (also known as split-array in the literature) can be employed for efficient operation. Therefore, in a multi-antenna node, two groups of antennas and RF chains can be utilized for simultaneous transmission and reception. 

Nevertheless, despite the aforementioned promising futures, IBFD communication in  wireless networks gives rise to the following fundamental challenges:
\begin{enumerate}
\item  \emph{SI}: The highly coupled transmit and receive circuits of an IBFD radio generate a high-power SI signal at the receiver, overwhelming the desired signal from other remote transmitting antennas~\cite{Duarte:TWC:2012}. More specifically, the SI signal can be over $100$ dB stronger than the signal from the intended transmitter. Since practical analog-to-digital converters (ADCs) have finite dynamic range and resolution, canceling this strong SI becomes infeasible. Therefore, the finite dynamic range of the ADC is the main bottleneck in realizing IBFD radios. To make IBFD radios a reality, it is necessary to suppress or cancel the SI to the noise level before the downconverted signal reaches the analog-to-digital stage. The increasing attention garnered by IBFD, both in academia and industry, can be attributed to advancements in SIC techniques, that seek to mitigate this limitation and facilitate the practical integration of IBFD into wireless communication systems. SIC techniques can be generally categorized as propagation-domain, analog- and digital-domain techniques~\cite{Hong2014}. Propagation-domain cancellation can be accomplished passively or actively using techniques, such as antenna separation, coupling networks, and/or surface treatments~\cite{Sabharwal:JSAC:2014}. Analog-domain SIC is implemented within the analog circuits situated between the antennas and digital conversion stages. The prevalent method for analog-domain SIC involves digitally-assisted techniques relying on auxiliary transmit chains~\cite{Sabharwal:JSAC:2014}. The propagation- and/or analog-domain SIC circuits are followed by the digital-domain SIC circuit. Note that digital-domain techniques are ineffective if the receiver is saturated, i.e., the desired signal is overwhelmed by the SI signal. The digital-domain approach is designed based on the premise that the receiver of an IBFD radio has knowledge of its transmitted signal and an estimate of the SI link.  In the digital-domain, SIC occurs post ADC, through techniques like SI subtraction (also known as channel-modeling) and/or spatial suppression (also known as null-space projection). Specifically,  SI subtraction involves deducting the replicated interference signals from the input signal. On the other hand, spatial suppression employs multi-antenna techniques that leverage additional degrees of freedom (DoF), provided by the spatial dimensions, to implement linear receive and transmit filtering. To address the complexity challenges associated with these methods, data-driven machine learning (ML) approaches have been introduced in the literature. ML-based SIC methods encompass neural networks, support vector regressors, and more advanced techniques, including tensor completion, TensorFlow graphs, and random Fourier features. A comprehensive overview of ML-based SIC techniques can be found in~\cite{Elsayed:OPVT:2024}. Leveraging these models and availing of the receiver's knowledge of its transmitted signal, the system can effectively cancel the SI~\cite{Sabharwal:JSAC:2014}. Digital-domain SIC offers relative ease of processing and is less hardware-expensive compared to analog-domain SIC, though it may lead to an increase in the computational complexity of the IBFD transceiver.  A summary of existing SI mitigation circuits and experiments, along with their setups and performance, can be found in~\cite[Table II and Table III]{Mohammadi:TUT:2024}.

\item \emph{CLI}: Another fundamental challenge in IBFD communications is the emergence of new interference terms caused by IBFD transmissions in multiuser scenarios, termed as CLI. For example, in IBFD cellular networks, with only the BS operating in IBFD mode, on top of the inter-cell BS-to-UE and UE-to-BS interference that already exists in HD networks, IBFD networks now experience inter/intra-cell inter-UE (DL-to-UL) interference, inter-cell inter-BS interference, as well as the residual SI after SIC~\cite{Sabharwal:JSAC:2014}. Hence, to reap the anticipated benefits of IBFD operation, it is essential to carefully schedule a suitable combination of DL and UL UEs with corresponding transmission rates and powers. This approach ensures the maximization of the aggregate network utility~\cite{Goyal:COMM:2015}.  

\item \emph{Fairness Control:} Unlike link-level communication, which primarily aims to maximize the data rates, real-world networks require the inclusion of fairness considerations for multiple UEs. Achieving fairness in IBFD networks introduces a more intricate challenge compared to HD networks, as it demands simultaneous attention to both UL and DL communications~\cite{kim2024state}.

\item \emph{EE Challenge}: Though IBFD can notably increase the SE, it can significantly diminish the EE when compared to HD networks. To discuss this  further, fist let us briefly introduce the concept of EE, which is an important metric for designing wireless networks, and it is defined as the aggregated number of bits transmitted in both UL and DL in unit bandwidth per Joule energy consumed~\cite{Feng:TuT:2013}. In particular, the EE can be expressed as
%-----------------------
\begin{align}
   \EE 
   &=  \frac{ \SEtot \times T}{ \Etot}
   =  \frac{\SEtot}{ \Ptot },
\end{align}
%------------------------
where $\SEtot$ is the aggregate SE, $\Etot$ and $\Ptot$ stand for the energy and power consumption, $T$ denotes the transmission time and thus $\Etot=\Ptot\times T$. The overall power consumption primarily comprises two main components: 1) data-dependent transmit power and 2) circuit power consumption. The latter, which is dissipated across various electronic devices responsible for signal processing (including mixers, filters, ADC, digital-to-analog converter, low noise amplifiers, etc.), is a core parameter in the characterization of the EE performance. Notably in short-range communications, like micro- or femto-cells, the circuit power consumption can be comparable to or may even overshadow the actual transmit power during data transmission~\cite{Arnold:2010}.

In IBFD communications, the deployment of advanced signal processing schemes for SIC and user-scheduling  introduces considerable complexity in both hardware and software. Moreover, the CLI management requires increased power consumption to meet the UEs quality-of-service (QoS) requirements. On the other hand, the SIC circuits entail power-intensive hardware in the IBFD devices. To align with green communication principles and leverage the potential advantages of IBFD communications, it is crucial to integrate IBFD with energy-efficient technologies. This integration should be carried out in a manner that inherently alleviates the CLI challenge and improves the SIC effectiveness without requiring additional energy consumption. One of the most viable solutions is the deployment of large-scale multiple antenna arrays in the IBFD devices. 

\item \textit{Hardware Cost and Complexity:} In distributed systems, the BS is replaced by a large number of APs equipped with IBFD transceivers. This imposes significant costs on system developers. While APs are typically designed to perform simple processing tasks, IBFD communication introduces substantial computational complexity required for SIC and interference management. This increased complexity leads to higher energy consumption, reducing the overall EE of the system. Additionally, the simultaneous transmission and reception by all APs increase the amount of CLI in the network, further complicating the system performance management.
\end{enumerate}

%%%%%%%%%%%%%%%%%%%%%%%%%%%%%%%%%%%%%%%%%%%%%%%%%%%
\vspace{-1.1em}
\subsection{ Massive MIMO Communications}
%%%%%%%%%%%%%%%%%%%%%%%%%%%%%%%%%%%%%%%%%%%%%%%%%%%
The concept of massive MIMO has been under development over the years, reshaping the landscape of wireless communication. MIMO technology, which forms the foundation of massive MIMO, entails the use of multiple antennas at both the transmitter  and receiver. Seminal works by Winters, Salz, Foschini, Gans, Cioffi, and Telatar played a pioneering role in the development of MIMO. MIMO is a well-known technology recognized for its effectiveness in  mitigating the effects of fading and enhancing the SE~\cite{PAULRAJ:IPCR:2004,Jiang:JPROC:2007}.  Transmission with MIMO antennas is a well-known diversity technique to enhance the reliability of the communication. Furthermore, with multiple antennas, multiple streams can be sent out and hence, we can obtain a multiplexing gain which significantly improves the communication capacity. To further exploit the multiplexing gain among the single-antenna UEs, multiple UEs must be co-scheduled on the same time-frequency resources. This  paradigm shift from MIMO to MU-MIMO is particularly beneficial as single UE MIMO (SU-MIMO) transmission is often limited by the number of antennas and antenna design constraints at the UE side~\cite{Taesang:JSAC:2006}. Moreover, the UEs cannot support many antennas due to the small physical size and low cost requirements of the terminals, whereas the BS can support many antennas. However, MU-MIMO, with roughly an equal number of service antennas and UEs and FDD operation, is not a scalable technology~\cite{Larsson:COMMG:2014}.

To support a larger number of UEs, the groundbreaking concept of massive MIMO (also referred to as `Large-Scale MIMO' or `Large-Scale Antenna Systems') was introduced by Marzetta in~\cite{Marzetta:TWC:2010}. Massive MIMO systems use a large excess of service antennas over active terminals, and operate in TDD. Deploying an unconventionally large array of antennas at the BS, well beyond the count of UEs, offers a significant increase in DoF. This abundance of DoF allows for precise signal shaping directed towards the intended UE and enables effective interference nullification. The asymptotic analysis in~\cite{Marzetta:TWC:2010} demonstrated that with simple linear signal processing approaches, such as matched-filter precoding/detection, as the number of BS antennas  approaches infinity \textit{``the effects of uncorrelated noise and fast fading vanish, the throughput and the number of terminals are independent of the size of the cells, spectral efficiency is independent of bandwidth, and the required transmitted energy per bit vanishes."}  

Massive MIMO offers a plethora of advantages over conventional MIMO systems: 
\begin{itemize}
   \item Considering realistic propagation conditions, massive MIMO has the potential to enhance capacity. This capacity improvement is attributed to the aggressive spatial multiplexing techniques employed in the context of massive MIMO~\cite{Larsson:COMMG:2014}. 
   \item It can significantly improve the EE compared to conventional MIMO and SISO systems. This improvement arises from its capability to concentrate energy sharply into specific spatial regions. The transmit power of each single-antenna UE in a massive MIMO system can scale down proportionally to the (square root of the) number of antennas at the BS, achieving the same SE performance as the corresponding SISO counterpart with (imperfect) perfect channel state information (CSI)~\cite{Ngo:Massive:2013,Zhang:JSTSP:2014}. 
   \item By exploiting the channel reciprocity, the channel estimation overhead is remarkably reduced as it  
   is independent of the number of BS antennas. The beamforming procedure converts the  massive MIMO effective  channels into almost deterministic scalars. This phenomenon, is known as \textit{``channel hardening"}~\cite{Matthaiou:WCL:2019}.  By virtue of channel hardening, the instantaneous effective channel gain at the UE is close to its average and, hence, UEs can rely on knowledge of that average (via statistical channel information) for signal detection. Therefore, DL pilots are no longer required~\cite{Yang:JSAC:2013,Ngo:TWC:2017}.
   \item Coherent, yet simple, precoders and detectors exhibit near-optimal performance. Linear processing designs, including matched-filter and zero-forcing (ZF), are enough to obtain performance as much as
   the minimum mean-square error (MMSE) filter, even close to the performance of non-linear filters in large-scale MIMO systems~\cite{Rusek:SPM:2013,Ngo:Massive:2013,Yang:JSAC:2013,Zhang:JSTSP:2014}.
   The reason that maximum ratio combining (maximum ratio transmission) works so well for DL (UL) massive MIMO is that the channel responses associated with different terminals tend to be nearly orthogonal when the number of BS antennas is large. This property is referred to as \textit{``favorable propagation"}, wherein the mitigation of inter-user and intra-cell interference becomes more manageable through the application of low-complexity linear signal processing schemes.  
   \item The favorable action of the law of large numbers can  significantly simplify the power control and resource allocation. Equally significant is the ability to analyze large-scale systems using tools of random matrix theory~\cite{Wagner:TIT:2012}.
   \item From the hardware perspective point of view, expensive ultra-linear forty Watt amplifiers are replaced by numerous inexpensive low-power devices, whose combined action, has to meet stipulated tolerances. Moreover, low resolution ADCs and DACs can be used. The findings in~\cite{Matthaiou:TWC:2015} highlight the robustness of massive MIMO systems to hardware imperfections (e.g., implications of the main circuits at the receiver, namely, the ADC, low noise amplifier, and local oscillator). These imperfections arise from the use of low-cost and low-power components in the manufacturing of antenna branches. 
\end{itemize}
%%%%%%%%%%%%%%%%%%%%%%%%%%%%%%%%%%%%%%%%%%%%%%%%%%
\begin{table*}[t]
    \centering
    \caption{Summary of IBFD and massive MIMO technologies.}
    \vspace{-0.5em}
    \small
    \begin{tabular}{| >{\centering\arraybackslash}m{1.5cm} | >{\centering\arraybackslash}m{4.85cm} | >{\centering\arraybackslash}m{4.85cm} | >{\centering\arraybackslash}m{4 cm} |}
        \hline
        \textbf{Technology} &\textbf{Pros} & \textbf{Cons} & \textbf{Current status} \\ \hline
        IBFD
        &
        \begin{itemize}[left=0pt, itemsep=0pt, parsep=0pt, partopsep=0pt, topsep=2pt]
            \item  Improved SE 
            \item  Reduced end-to-end delay and latency
            \item  Enhanced secrecy
            \item  Flexible spectrum usage  
        \end{itemize}        
                &
        \begin{itemize}[left=0pt, itemsep=0pt, parsep=0pt, partopsep=0pt, topsep=6pt]
            \item   SI and its cancellation complexity
            \item   High CLI
            \item   Complicated fairness control
            \item   Increased power consumption and EE reduction  
            \item   Higher hardware cost/complexity 
        \end{itemize}  
        &
        Enhancing duplex capabilities with IBFD are part of the $3$GPP Release $18$, starting from early $2022$
        \\ \hline
        Massive MIMO
        &
        \begin{itemize}[left=0pt, itemsep=0pt, parsep=0pt, partopsep=0pt, topsep=6pt]
            \item  Improved SE 
            \item  Enhanced EE 
            \item  Low channel estimation overhead 
            \item  Simple signal processing
            \item Robust to hardware impairments
        \end{itemize}   
        &
        \begin{itemize}[left=0pt, itemsep=0pt, parsep=0pt, partopsep=0pt, topsep=7pt]
            \item  Pilot contamination
            \item  Synchronization (wideband scenarios)
            \item  Implementation and hardware costs
        \end{itemize}   
        &         
        Telecom operators have been actively rolling out $5$G infrastructure that leverages massive MIMO
        \\ \hline
    \end{tabular}
        \label{table:IBFD_challenges}
                \vspace{-1em}
\end{table*}
%%%%%%%%%%%%%%%%%%%%%%%%%%%%%%%%%%%%%%%%%%%%

Despite the aforementioned advantages, practical implementation of massive MIMO posses several challenges that need to be studied and addressed. Some of these challenges are discussed below:
\begin{enumerate}
    \item \textit{Pilot Contamination:} The CSI is acquired using finite-length pilot sequences amidst intra-cell interference. As shown in~\cite{Marzetta:TWC:2010}, the pilot contamination effect—arising from the reuse of non-orthogonal pilot sequences across neighbouring cells—causes the interference rejection performance to quickly saturate with an increasing number of antennas. This phenomenon undermines the potential benefits of massive MIMO systems in cellular networks. Several techniques have already been proposed to address this challenge~\cite{Elijah:CSTO:2016}. However, the ever-increasing number of users in future wireless networks will impact the performance of these schemes, rendering them insufficient and necessitating the development of new designs.
    \item \textit{Synchronization:} In wideband scenarios, such as mmWave massive MIMO systems using orthogonal frequency division modulation (OFDM) modulation, with perfect time and frequency synchronization, the length of the cyclic prefix  is typically set to be slightly larger than the channel delay spread, and the length of the OFDM symbol is set to be inversely proportional to the Doppler spread. However, in the presence of the beam squint effect, additional cyclic prefix is required to mitigate time delays across the large array aperture, imposing higher overhead on the system. Therefore, synchronization designs should be restructured to balance overhead and accuracy. Additionally, in high-mobility scenarios, these challenges are amplified, necessitating careful attention to synchronization.
    \item \textit{Implementation Costs:} The large number of antennas needed for massive MIMO systems can lead to significant implementation and maintenance costs. This poses a challenge for wireless service providers, who must balance the advantages of massive MIMO with the associated expenses.
\end{enumerate}

%%%%%%%%%%%%%%%%%%%%%%%%%%%%%%%%%%%%%%%%%%%%%
\subsection{Implementation Efforts and Standardization }
%%%%%%%%%%%%%%%%%%%%%%%%%%%%%%%%%%%%%%
\subsubsection{IBFD Radios}
The initial designs and implementations of IBFD radios were documented in the literature in the early 2010s. The design and implementation of MIDU, recognized as the first MIMO IBFD system for wireless networks, were detailed in~\cite{aryafar2012midu}. Notably, this design is effective for small bandwidths, specifically $500$KHz. It relies on achieving $50$dB of cancellation through antenna cancellation, complemented by an additional $30$dB of digital cancellation. However, as we transition to standard bandwidths of $20$ MHz, commonly found in WiFi signals, the antenna cancellation is reduced to a maximum of $40$dB. Consequently, we face limitations reaching a total cancellation of only $70$dB. An implementation of a single-carrier IBFD radio, based on the IEEE 802.11ac standard, supporting a bandwidth of at least $80$ MHz in the $2.4$ GHz range, and featuring an average transmit power of $20$dBm, was reported in~\cite{Katti:Sigcomm:2013}. This design achieves SIC to the receiver noise floor. Specifically, it can attain $50$dB ($60$dB) of analog cancellation, complemented by an additional $60$dB ($50$dB) of digital cancellation when using the Rohde-Schwarz radios (the WARP radios). In~\cite{bharadia2014full}, the implementation of IBFD WiFi-PHY-based MIMO radios, that practically achieve SIC close to the noise floor and the theoretical doubling of throughput, was reported. This architecture relies on analog cancellation circuits with $56$ taps for a 3-antenna IBFD MIMO radio, along with a total of $485$ filter taps in digital cancellation. In 2015, IBFD was first adopted by cable modem industry into DOCSIS $4.0$ which has made next-generation cable modems IBFD-enabled~\cite{DOCSIS}. 

Regarding the latest cellular network standards, discussions are underway from the start of $2022$ onwards, with ongoing updates, around the IBFD network standard for duplex enhancement as part of the 3rd Generation Partnership Project ($3$GPP) Release $18$. Within the $5$G New Radio (NR) standard documents, three IBFD network scenarios are outlined. These options involve selecting a uniform duplex mode, whether it would be IBFD or HD, for all BSs (cells), or integrating both IBFD and HD BSs. 

\subsubsection{Massive MIMO}
The vastly increased coverage, capacity, EE, and per-UE SE that massive MIMO offers have already sparked strong interest in both industry and academia, quickly establishing it as a natural and essential element of wireless network deployments. These initiatives involve the development of prototype systems with substantial antenna arrays, such as the Argos testbed with $96$-antennas~\cite{shepard2012argos}, EURECOM's $64$-antenna long-term evolution (LTE) compatible testbed~\cite{openairinterface_2018}, Facebook’s ARIES project~\cite{facebook_aries_2016}, the 100-antenna LuMaMi testbed from Lund University~\cite{Malkowsky:2017}, SEU’s $128$-antenna testbed~\cite{Yang:2017},  testbeds exploring distributed arrays from KU Leuven~\cite{Chen:GC:2016},  and the $120$-antenna testbed from University of Bristol~\cite{Bristol}. From the perspective of cellular network standardization, a significant milestone has been achieved in the first phase of the $5$G NR standardization. This phase involved the successful incorporation of massive MIMO technology, specifically utilizing a $64$-antenna configuration, as accomplished by the $3$GPP~\cite{Matthaiou:JSAC:2020}.

A summary of the existing pros and cons of IBFD and massive MIMO technologies along with their current status is shown in Table~\ref{table:IBFD_challenges}.

\subsubsection{IBFD Massive MIMO}
An experiment-based prototype of many-antenna beamforming for IBFD systems was reported in~\cite{Everett:TWC:2016}. The platform operates in the $2.4$ GHz ISM band, with a $20$ MHz bandwidth and a $72$-element two-dimensional planar antenna array. This structure, known as \textit{SoftNull}, dedicates a subset of antennas for transmission, while the remaining antennas are used for reception. Additionally, transmit beamforming is employed for SIC, which consequently sacrifices some DoF. With an equal antenna split, i.e., $36$ antennas for transmission and $36$ antennas for reception, and by sacrificing $12$ ($20$) antennas out of the $32$ transmit antennas, it is possible to achieve $50$ dB of SIC for outdoor (indoor) scenarios. Therefore, the IBFD gains over HD are not consistently achievable in all situations, such as low-SNR outdoor environments or high-scattering indoor settings.

\textbf{Takeaway Messages: }
Both IBFD and massive MIMO technologies have the potential to significantly enhance the SE, aligning well with the future demands of wireless networks. The TDD operation in massive MIMO systems makes them well-suited for IBFD communications, as feedback-based closed-loop beamforming can be directly applied for both UL and DL transmissions. However, since IBFD was originally designed for nodes with a few antennas, SIC in massive MIMO setups presents significant practical challenges. To achieve the promised gains of IBFD in massive MIMO systems, SIC designs that balance between complexity and performance loss are required. 

%%%%%%%%%%%%%%%%%%%%%%%%%%%%%%%%%%%%%%%%%%%%%%%%%%%
\section{In-Band Full-Duplex Cellular Networks}\label{sec:IBFDcellular}
%%%%%%%%%%%%%%%%%%%%%%%%%%%%%%%%%%%%%%%%%%%%%%%%%%%
The utilization of massive MIMO technology and IBFD has become widespread across various wireless communication systems. These systems range from point-to-point wireless communication systems~\cite{Koc:JOPCS:2021}, relay setups~\cite{Multipair:JSAC:2014,Jin:TWC:2015,Dai:TWC:2016} to cellular networks~\cite{Everett:TWC:2016,Xia:TVT:2017,Shojaeifard:TCOM:2017, Bai:TWC:2017, Koh:TWC:2018,Kim:TWC:2023}. Notably, cellular networks have experienced substantial improvements in both the SE and EE. This progress is attributed to the gradual adoption of transformative technologies, transitioning from HD MU-MIMO to IBFD MU-MIMO, and ultimately from IBFD MU-MIMO to IBFD massive MIMO.  We elaborate on this step-by-step transformation, aiming to lay a foundation that clearly illustrates the rationale behind each transformative step. Specifically, we highlight the shortcomings of earlier structures and then focus the significant improvements brought by each technological advancement, leading up to the latest, IBFD massive MIMO.

%-------------------------------------
\subsection{Traditional HD-MU-MIMO Cellular Networks}
\begin{figure}[t]
	\centering
  \vspace{0em}
	\includegraphics[width=0.5\textwidth]{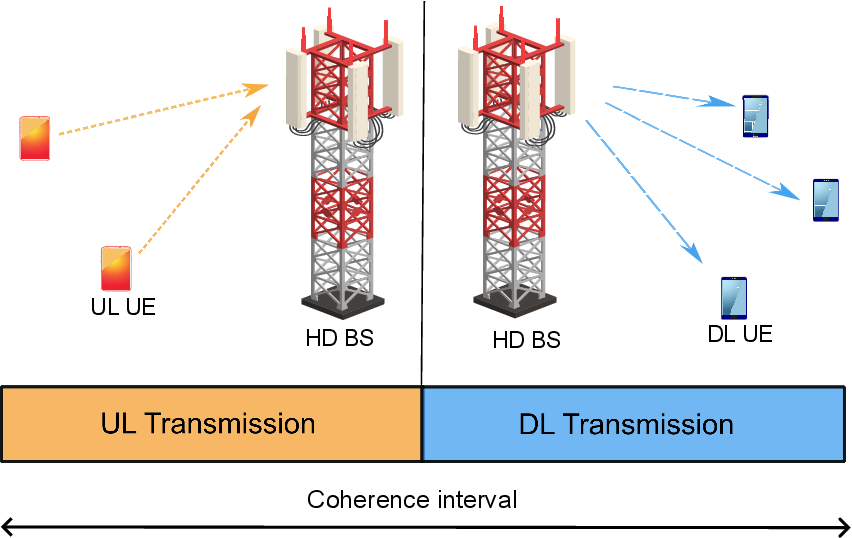}
	\vspace{-1em}
	\caption{Diagram of an HD MU-MIMO cellular system with DL and UL UEs.} 
 \vspace{1em}
	\label{fig:Fig2}
\end{figure}

%--------------------------------------
In conventional cellular networks, the coverage area is divided into cells and each UE is typically associated with one BS within a cell among many cells, specifically the one that provides the strongest signal. These networks have primarily relied on time- or frequency-division-multiple-access  to manage UL and DL transmissions. In this approach, the available time-frequency resource block is exclusively dedicated to either the DL transmission or UL reception (c.f. Fig.~\ref{fig:Fig2}). Furthermore, a MU-MIMO setup may be deployed to exploit the multiplexing gains among the single-antenna UEs. Nevertheless, due to the presence of mutual co-channel interference (CCI) in MU-MIMO setups, it is essential to employ transmission schemes at the DL (reception schemes for UL) that can effectively mitigate the CCI at the UEs (HD BS). The suppression of the CCI can be pursued by space-division multiple access (SDMA). In the context of SDMA and during DL transmission, a BS with $M$ antennas codes each UE's data stream independently. These coded data streams are then multiplied by a precoding weight vector and transmitted through the $M$ antennas. The design of the precoding/decoding weight vectors plays a critical role in minimizing or eliminating the CCI. This is achieved by capitalizing on the spatial separation between the UEs, thus enhancing the overall efficiency and capacity of the network. 

Consider a cellular HD system, as illustrated in Fig.~\ref{fig:Fig2}, where a multi-antenna BS in each cell communicates with $K_{\ul}$ single-antenna UL UEs over half of the coherence interval. The remaining coherence interval is then utilized to serve $K_{\dl}$ single-antenna DL UEs.\footnote{In a more general case and for this setup, the DL transmission with a multi-antenna BS and multiple single-antenna/multi-antenna UEs can be modeled by a multiple-input single-output/multiple-output (MISO/MIMO)
broadcast channel.} We define the sets $\KU\triangleq \{1,\ldots,K_{\ul}\}$ and $\KD\triangleq \{1,\ldots,K_{\dl}\}$ as the collections of indices of the UL and DL UEs within each cell, respectively.  The number of transmit and receive RF chains at the BS are  $\Ntx$ and $\Nrx$, respectively. By using a linear precoding/decoding design at the BS, the SE of the $k$th DL and UL UE ($\forall k\in\KU/\KD$) can be expressed as
%-------------------
\begin{subequations}
  \begin{align}
    \SE^{\HDCEL}_{\dl,k} &\!=\! \frac{1}{2} \log_2\bigg(1+ \frac{\DSHDceldl}{{\CCId} + \NOISE}\bigg),~\label{eq:SE:ULHD1}\\
    \SE^{\HDCEL}_{\ul,k} &\!=\! \frac{1}{2}\log_2\bigg(1+ \frac{\DSHDcelul}{{\CCIu} + \NOISE}\bigg),~\label{eq:SE:ULHD2}
\end{align}  
\end{subequations}
%------------------------------------------
respectively, where the prelog factor $\frac{1}{2}$ is due to the HD constraint at the terminals;\footnote{ The prelog factor of $\frac{1}{2}$ corresponds to the assumption that half of the coherence interval is allocated for  UL transmission, while the other half is designated for DL transmission. In general, this prelog factor may vary depending on the proportion of the coherence interval allocated to UL and DL, as well as other overhead considerations. } $\NOISE$ is the noise power, $\DSHDceldl$ and $\DSHDcelul$ are the desired signals in the DL and UL, respectively; $\CCId$ captures the intra-cell  and  inter-cell DL interference ($\Intrad$ and $\Interd$) and  $\CCIu$ captures the intra-cell  and  inter-cell UL interference ($\Intrau$ and $\Interu$), i.e., 
%--------------
\begin{align}
  \CCId \triangleq \Intrad + \Interd,\\
  \CCIu \triangleq \Intrau + \Interu.
\end{align}
%-----------------
%%%%%%%%%%%%%%%%%%%%%%%%%%%%%%%%%%%%%%%%%%%%%%
%%%%%%%%%%%%%%%%%%%%%%%%%%%%%%%%%%%%%%%%%%%%%%
In the UL (DL), the inter-cell interference at the BS (DL UE) refers to the CCI from the adjacent UL UEs (BSs) transmitting over the same frequency at the same time. The inter-cell interference at the \textit{cell-edge UEs}, i.e., UEs placed in severe inter-cell interference-dominated areas, such as the cell boundary areas, poses a major challenge, causing a significant data rate drop at the UEs and a strong lack of fairness across the cell UEs. 
 
The DL/UL intra-cell and inter-cell CCI are reduced or mitigated by deploying a suitable precoder/decoder at the BS. Achieving optimal precoder and decoder designs typically involves maximizing the SE (or equivalently the signal-to-interference-plus-noise ratio) in the DL and UL, respectively. Nevertheless, due to the coupled nature of these optimization problems, an iterative procedure with high complexity is often required~\cite{Spencer:TSP:2004}. In single-cell systems, linear precoding schemes, such as block diagonalization  and ZF-based precoding have been proposed to perfectly cancel the intra-cell CCI for each UE, such that each UE perceives an interference-free MIMO channel~\cite{Spencer:TSP:2004}.\footnote{In the special case of a MISO broadcast channel, block diagonalization reduces to the ZF precoding.} The maximum number of concurrent interference-free transmissions is referred to as the multiplexing gain of the network, or the number of DoF in the information-theoretic terminology.  However, to ensure adequate DoF for ZF to effectively cancel intra-cell CCI, a system configuration constraint is necessary: \textit{the number of antennas at the BS must be larger than the total number of UEs' antennas.} In multi-cell scenarios, the application of block diagonalization and ZF precoder/decoder to mitigate both intra-cell and inter-cell interference faces similar challenges. Hence, traditional MIMO cellular systems lack scalability and must rely on time-scheduling when the dimensional condition is not satisfied~\cite{Heath:2001}.

The EE of an HD MU-MIMO cellular network can be expressed as
%-----------------------
\begin{align}
   \EE^{\HDCEL} 
   =  \frac{\sum\nolimits_{k\in\KD}\SE^{\HDCEL}_{k,\dl} + \sum\nolimits_{k\in\KU}\SE^{\HDCEL}_{k,\ul}}{ \PtotHDU + \PtotHDD},
\end{align}
%------------------------
where $\PtotHDU$ and $\PtotHDD$ stand for the UL and DL power consumption in the HD mode. The total power consumption of HD cellular systems, i.e.,  $\PtotHD\triangleq\PtotHDU + \PtotHDD$, can be generally written as
%--------------------
\begin{align}
    \PtotHD = &
    \sum\nolimits_{k\in\KU} \bigg(\frac{1}{\epsilon_{\UEk}}\PULUE  + \Pulcir\bigg)
    \nonumber\\
    &+\frac{1}{\epsilon_{\BS}}\PBSk + \PBScir,     
\end{align}
%--------------------
where $\PULUE$ and $\Pulcir$ denote the data-dependent and circuit power consumption of the $k$th UL UE, while $\Pulcir\triangleq (\PULUEDyn + \PULUESta)$, where $\PULUEDyn$ and $\PULUESta$ are the dynamic circuit power consumption, corresponding to the power radiation of all circuit blocks, and  static circuit power spent by the cooling system of the UL UEs; $\PBSk$ and $\PBScir$ denote the data-dependent and circuit power consumption of the BS, where $\PBScir\triangleq(\Ntx\PBSsdyndl + \Nrx\PBSsdynul +\PBSstat)$, while $\PBSsdyndl$ and $\PBSsdynul$ are the dynamic power consumption, associated with the power radiation of all circuit blocks in each active transmit and receive RF chain, respectively; $\PBSstat$ is the static power consumed by the cooling system, power supply, etc. Moreover, $\epsilon_{\UEk}\in(0,1)$ and $\epsilon_{\BS}\in(0,1)$ denote the power amplifier efficiencies of the $k$th UL UE and the BS, respectively. 

%-------------------------
\textbf{Fundamental Challenges for SE and EE Enhancement:}
%-------------------------
The primary bottlenecks hindering the enhancement of the SE in traditional cellular networks include: \textit{i)} prelog factor resulting from the HD constraint of the HD MIMO BS; \textit{ii)} the fairness problem arising from inter-cell CCI; \textit{iii)} the system configuration constraint for intra-cell CCI alleviation. Furthermore, this configuration does not meet the necessary criteria for providing low-latency services. Specifically, an UE with UL (DL) data requirements must wait for a time slot during which the BS is operating in the UL (DL) mode to complete its transmission. However, for widely-used delay-sensitive services like cloud storage, video chat, and innovative IoT applications, where both UL and DL transmissions hold equal priority, simultaneous execution is essential. \emph{Therefore, a natural question arises: How can we improve the  SE of  cellular networks through effective management of UL and DL transmissions, while also meeting the low-latency requirements?}

%-------------------------------------
\subsection{Transition From HD MU-MIMO to IBFD MU-MIMO}
\begin{figure}[t]
	\centering
  \vspace{0em}
	\includegraphics[width=0.5\textwidth]{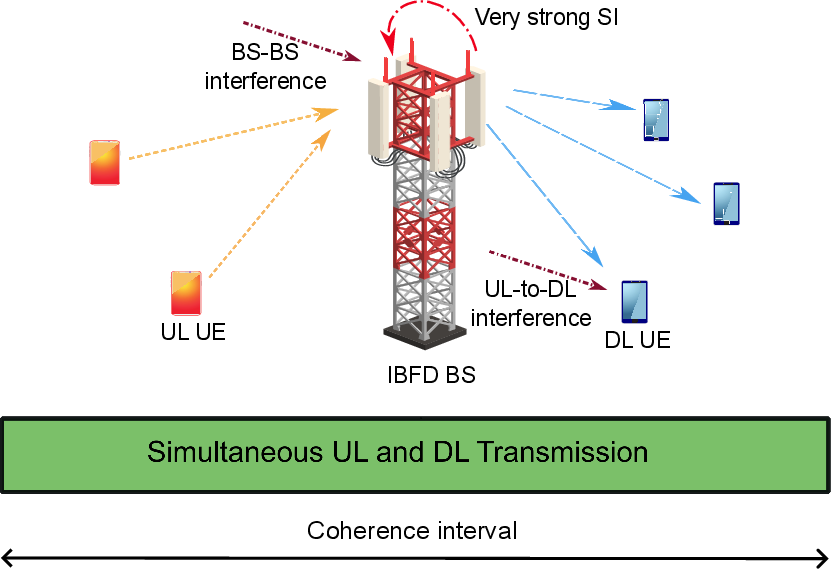}
	\vspace{-1em}
	\caption{Diagram of an IBFD MU-MIMO cellular system with DL and UL UEs.} 
 \vspace{1em}
	\label{fig:Fig3}
\end{figure}

%--------------------------------------
The emergence of IBFD technology has paved the way for concurrent UL and DL communications  over the same frequency within cellular networks, as shown in Fig.~\ref{fig:Fig3}. An IBFD BS is equipped with $\Ntx$ transmit antennas (and transmit RF chains) and $\Nrx$ receive antennas (and receive RF chains) to serve the DL and UL UEs simultaneously, over the same frequency band.\footnote{For a fair comparison between the HD and IBFD operation, we assume that the BS in the HD mode employs the same number of transmit and receive antennas as in the IBFD mode. This assumption corresponds to the ``RF chains conserved" condition, where an equal number of total RF chains is assumed~\cite{aryafar2012midu,Multipair:JSAC:2014}.}
This synergy  offers substantial advantages, including enhanced SE, reduced latency, and a comprehensive increase in the network capacity~\cite{Goyal:COMM:2015,Li:COMMG:2017}. Specifically, the prelog factor $\frac{1}{2}$ in the HD SE expressions is eliminated, as the entire time-frequency resources can now be utilized for concurrent UL and DL transmissions. This leads to the potential doubling of the SE.
Furthermore, the time gap between UL and DL transmissions is eliminated  resulting in reduced latency, providing a more responsive and seamless UE experience. Consequently, the UEs avail of improved SEs and lower latency, leading to a more satisfying and reliable communication experience. This is essential for supporting emerging applications that demand high-performance connectivity.

While the integration of IBFD technology into cellular networks presents numerous advantages, there are inherent challenges to overcome, including interference management, hardware constraints, and compatibility with existing network infrastructure. As shown in Fig.~\ref{fig:Fig3}, in addition to SI at the IBFD BS, new interference terms kick in IBFD cellular networks that are not encountered in conventional HD cellular networks. Specifically, in single-cell scenarios, this type of interference occurs at the received signal of the DL UEs and is caused by the adjacent UL UEs in the cell, which is called intra-cell  UL-to-DL interference. In multicell IBFD cellular networks, inter-UE interference worsens due to the presence of UL UEs in adjacent cells. Additionally, inter-cell interference between BSs occurs at each BS's UL reception due to the adjacent BSs' DL signal, termed as BS-BS interference. 

%%%%%%%%%%%%%%%%%%%%%%%%%%%%%%%%%%%%%%%%%%%%%%
%%%%%%%%%%%%%%%%%%%%%%%%%%%%%%%%%%%%%%%%%%%%%%

The UL SE of the $k$th UL UE, $\forall k\in\KU$, can be expressed as
%-------------------
    \begin{align}~\label{eq:SE:ULFD}
    \SE^{\FDCEL}_{k,\ul} &=  \log_2\bigg(1+ \frac{\DSFDcelul}{ \CCIu +\BBI+ \SIBS +  \NOISE}\bigg), 
\end{align}
%------------------------------------------
where $\CCIu$ denotes the overall inter-cell and intra-cell CCI at the IBFD BS, $\BBI$ is the BS-BS interference, while $\SIBS$ represents the residual SI at the IBFD BS after SIC.

The DL SE of the $k$th DL UE, $\forall k\in\KD$, can be also written as
%-------------------
    \begin{align}~\label{eq:SE:DLFD}
    \SE^{\FDCEL}_{k,\dl} &=  \log_2\bigg(1+ \frac{\DSFDceldl}{\CCId + \UDICel +  \NOISE}\bigg),
\end{align}
%------------------------------------------
where $\CCId$ denotes the overall inter-cell and intra-cell CCI at the UE, while $\UTDI$ represents the overall inter-cell and intra-cell  UL-to-DL interference.

By examining the SE expressions, we observe that the prelog factor is removed in the IBFD scenario, while new interference terms appear. Since these interference terms  affect logarithmically the SE, while the prelog factor has a linear impact, we expect that IBFD will outperform the HD counterpart in most cases. Nevertheless, in ultra-dense networks where the $\BBI$ is strong and the $\SIBS$ is high (due to using low-efficiency SIC), these interference terms might outweigh the prelog factor, making IBFD inferior to HD in the UL. In the DL direction, when an appropriate user-scheduling is not considered, a similar situation can occur. Moreover, in IBFD scenarios, some DoFs at the BS are exploited to manage these new interference terms, potentially lowering the desired signal level compared to HD. This could make the SE in IBFD be lower than in HD. However, by applying user-scheduling, using efficient SIC methods, and resource management, we can expect that IBFD will outperform HD in most cases. Nevertheless, these new processing requirements impose additional levels of energy consumption, which significantly affect the EE of the IBFD network.

From an EE point of view,  implementing IBFD operation at the BSs results in a significant increase in the power consumption, leading to a reduction in the EE. The EE of IBFD cellular system can be obtained as
%-----------------------
\begin{align}
   \EE^{\FDCEL} 
   =  \frac{\sum\nolimits_{k\in\KD}\SE^{\FDCEL}_{k,\dl} + \sum\nolimits_{k\in\KU}\SE^{\FDCEL}_{k,\ul}}{ \PtotFDU + \PtotFDD},
\end{align}
%------------------------
where  $\PtotFDU$ and $\PtotFDD$ stand for the UL and DL power consumption in the IBFD mode.  Noticing that in the IBFD mode the BS and UL UEs are transmitting over the entire transmission time $T$, the total power consumption of the IBFD cellular systems can be generally written as
%--------------------
\begin{align}
    \PtotFD =& 2\PtotHD +  \Nrx\PSICBS,     
\end{align}
%--------------------
where  $\PSICBS$ is the power of the SIC scheme at the receiving part of BS.

%-------------------------
\textbf{Fundamental Challenges for SE and EE Enhancement:}
%-------------------------
Upon comparing the SE of the HD and IBFD cellular systems, the disappearance of the prelog factor suggests a potential doubling of the SE. Nonetheless, the emergence of new interference terms presents a potential source of SE degradation, possibly falling below the HD counterpart in some specific scenarios. Additionally, the substantial increase in the overall power consumption (more than doubling) contributes to a reduction in the EE for the IBFD operation mode in comparison to the HD network.

To address the SE loss due to the appearance of new interference terms,  UE scheduling algorithms and UL power control at the UL UEs have been proposed to alleviate the  UL-to-DL interference, $\UTDI$. In response to the residual SI at the IBFD BS, intra-cell interference between the UEs, and inter-cell BS-BS interference challenges, researchers have explored interference alignment techniques, by capitalizing on the capabilities of MIMO technology~\cite{Nguyen:TSP:2013,Chae:TCOM:2017,Yang:TCOM:2017,Atzeni:TCOM:2017,Luo:TVT:2024}.  Recent advancements have led to the introduction of new interference management approaches in multi-cell IBFD cellular systems. Chen \textit{et al.} \cite{Chen:TWC:2024} proposed deploying RISs at cell boundaries of a multi-cell IBFD cellular system. This approach provides an opportunity to create a favorable wireless environment to manage inter-cell interference and leverage the potential of IBFD transceivers by configuring the phase shift matrices at the RISs and designing precoding at the BS.

However, IBFD MU-MIMO cellular systems with multi-antenna BSs still suffer from \textit{i)} system configuration constraints, i.e., to apply linear precoding/decoding schemes at the BS, the number of transmit and receive antennas at the BS must be greater than the total number of serving DL and UL UEs, respectively. Moreover, when certain DoFs are allocated for SI and CCI (which are now stronger than that of the HD cellular network) cancellation, only a limited number of DoF remains to boost the desired received/transmitted signal, leading to a situation that the strength of the desired signal in the IBFD cellular network becomes weaker than that in the HD network, under fair comparison scenarios.  Finally, deploying RISs at cell edges would increase the channel estimation overhead and system design complexity. This is due to the need for CSI between UEs and RISs for phase shift matrix design (a challenging task given the passive nature of RIS elements) and the necessity for coordination between BSs and RISs for joint precoder and phase shift matrix design.
\textit{ii)} The increased power consumption in the IBFD mode compared to HD mode is unavoidable, given the necessity for consuming some power for running the SIC circuit and the continuous activation of both the BS and UE throughout the entire transmission duration; \textit{iii)} Te fairness problem still holds as cell-edge UEs experience stronger inter/intra interference. \emph{Now this question arises: To what extent can the transition from MU-MIMO to massive MIMO contribute to balancing the trade-off between SE and EE in cellular networks?}

%--------------------------------------------------------
\subsection{Transition From IBFD MU-MIMO to IBFD  Massive MIMO}
%--------------------------------------------------------
As a subsequent development in IBFD MU-MIMO cellular networks, researchers broadened their focus by leveraging the advantages of massive MIMO technology, incorporating large antenna arrays at the IBFD BS~\cite{Everett:TWC:2016,Shojaeifard:TCOM:2017, Bai:TWC:2017, Koh:TWC:2018,Kim:TWC:2023}. The main goal of this expansion is to address the limitations found in IBFD MU-MIMO cellular networks. 

When increasing the number of antennas at the IBFD BS, a substantial surplus of DoF is provided. This surplus enables the establishment of asymptotically orthogonal channels, leading to nearly interference-free signals for each UE. Consequently, the SE can be significantly improved by simultaneously serving more UEs with reduced interference for each UE. This improvement in SE, in turn, offers the potential to further boost the overall EE of the system. At the same time, the extensive array gain,  provided by large-antenna arrays, contributes to transmit power savings.  In the UL, coherent combining can significantly enhance the array gain, allowing for a substantial reduction in the transmit power of each UL UE to achieve the same rate. In the DL, the BS can focus the energy into a specific spatial direction, in which the DL UEs are located. As a result, with massive antenna arrays, the transmit power can be reduced by an order of magnitude, or more, while provisioning the same DL rate and, hence, we can obtain higher EE. 

With the perfect CSI assumption, let us consider that the transmit power of each UL UE is scaled proportionally to $1/\Nrx$ as $\PULUE = \mathtt{E}_{\ul}/\Nrx$, $\forall k\in\KU$. Simultaneously, the transmit power of the BS is scaled proportionally to $1/\Ntx$ as $\PBSk = \mathtt{E}_{\dl}/\Ntx$, where $\mathtt{E}_{\ul}$ and $\mathtt{E}_{\dl}$ are fixed. As $\Ntx\rightarrow \infty$ and $\Nrx \rightarrow \infty$, while $\Ntx/\Nrx$ is kept fixed, the asymptotic UL SE for the $k$th UL UE, $k\in\KU$, can be expressed as~\cite{Bai:TWC:2017}
%-------------------
\begin{align}
    \SE^{\FDCEL}_{k,\ul} &\xrightarrow[]{}  \log_2\bigg(1+ \frac{\DSFDcelul}{  \NOISE}\bigg), ~\label{eq:SE:mmULFD}
\end{align}
%------------------------------------------
where $\DSFDcelul \propto \mathtt{E}_{\ul}$. 

Moreover, the asymptotic DL SE for the $k$th DL UE, $k\in\KD$, can be expressed as~\cite{Bai:TWC:2017}
%-------------------
\begin{align}
    \SE^{\FDCEL}_{k,\dl} &\xrightarrow[]{}
    \log_2\bigg(1+ \frac{\DSFDceldl}{ \NOISE }\bigg),~\label{eq:SE:mmDLFD}
\end{align}
%------------------------------------------
where $\DSFDceldl \propto \mathtt{E}_{\dl}$.

%-------------------------
\textbf{Fundamental Challenges for SE and EE Enhancement:} 
%-------------------------
In existing massive MIMO-enabled IBFD cellular networks, the coverage area is partitioned into disjoint cells, each containing a BS with a large antenna array situated at the center serving UEs located within each respective cell. However, this cellular structure presents inevitable challenges in ensuring high per-UE SE performance across the entire network. In particular: 
\begin{enumerate}
    \item Pilot contamination continues to be a significant factor affecting system performance in multi-cell scenarios. In multi-cell IBFD networks, the probability of pilot contamination increases because a larger number of UEs are served in both the UL and DL directions compared to HD networks, which raises the likelihood of using non-orthogonal pilot sequences.
    \item The promising gains of massive MIMO to cancel the impact of imperfect SIC at IBFD BSs, as well as intra-cell and inter-cell interference in MU-MIMO IBFD networks, \textit{are achievable in the asymptotic antenna region~\cite{Koh:TWC:2018}.} In practical scenarios with a large (yet finite) number of antennas deployed at the IBFD BS, UL-to-DL interference persists in the network, accompanied by residual SI attributable to the SI channel estimation errors~\cite{Kim:TWC:2023}.
    \item  Moreover, practical circumstances present persistent challenges for SIC. The substantial increase in the number of antennas exacerbates the severity of SI, resulting in a highly correlated SI channel. Despite the advanced level of analog SIC, the power from direct paths—such as line-of-sight paths, reflected paths around the antenna, and circuit leakage—remains higher than that from non-line-of-sight paths. Additionally, due to the very small angular spread of the line-of-sight path, the spatial correlation of the SI channel is notably high \cite{Kim:TWC:2023}.
     \item SI subtraction or spatial suppression techniques, which aim to reduce the SI level, require the development of channel estimators. To this end, the SI channel must be probed with DL pilots transmitted from the IBFD BS's transmit antennas. This SI estimation scheme fundamentally requires a DL pilot signal dimension at least equal to the number of transmit antennas at the IBFD BS, i.e., $\Ntx$. However, this necessity imposes a significant overhead on the system, consuming a large percentage of the coherence interval~\cite{Koh:TWC:2018}. 
     \item SI subtraction or spatial suppression not only mandates substantial power consumption in SIC circuits but also highlights the crucial challenge of accurately estimating the SI channel, a matter of utmost significance in SIC. The residual SI due to the estimation error remains in the system and degrades the UL performance. 
    \item  Lastly, employing the digital-domain SIC techniques on the IBFD BS leads to a degradation in both the UL and/or DL transmissions. SI suppression through null-space projection (or ZF design) sacrifices some DL DoF, resulting in a lower DL SE compared to SI subtraction. Nevertheless, the analysis in~\cite{Koh:TWC:2018} demonstrated that the UL SE of spatial suppression improves beyond that of SI subtraction as spatial correlation increases. In~\cite{Koh:TWC:2018}, the authors presented a comprehensive comparison of the UL performance of the system under SI subtraction and spatial SIC, considering specific system constraints, such as the DL/UL traffic ratio, the number of transmit and receive RF chains, total transmit power, and the power scaling law at the BS.
\end{enumerate}

\textbf{Takeaway messages:}
In summary, despite implementing massive MIMO at the IBFD BSs, some level of interference, including intra- and inter-cell interference, as well as SI, persists in the network. Furthermore, pilot contamination remains a significant bottleneck in multi-cell scenarios. Additionally, digital-domain SIC, aimed at mitigating highly correlated SI channels, requires substantial pilot overhead and DoF for spatial SIC, resulting in notable performance degradation and undermining the advantages of IBFD over HD transmissions. Therefore, fundamental changes in the network structure are necessary to mitigate these challenges.

%%%%%%%%%%%%%%%%%%%%%%%%%%%%%%%%%%%%%%%%%%%%%%%%%%
\section{The Path Towards In-Band Full-Duplex Cell-free Massive MIMO }\label{sec:IBFDcellfree}
%%%%%%%%%%%%%%%%%%%%%%%%%%%%%%%%%%%%%%%%%%%%%%%%%%%
%%%%%%%%%%%%%%%%%%%%%%%%%%%%%%%%%%%%%%%%%%%%%%%%%%%
In response to the escalating demands for ubiquitous data traffic and the need to guarantee seamless communication services in $5$G and beyond, cellular networks have undergone a transformative evolution in the past twenty years. The key emphasis has shifted towards increasing network density relying on either \emph{i)} reducing the cell size or  \emph{ii)} increasing the number of antennas per site by using technologies, such as MIMO and massive MIMO~\cite{Andrews:Commag:2016}.

%%%%%%%%%%%%%%%%%%%%%%%%%%%%%%%%%%%%%%%%%%%%%%
\begin{figure}[t]
	\centering
  \vspace{0em}
	\includegraphics[width=0.48\textwidth]{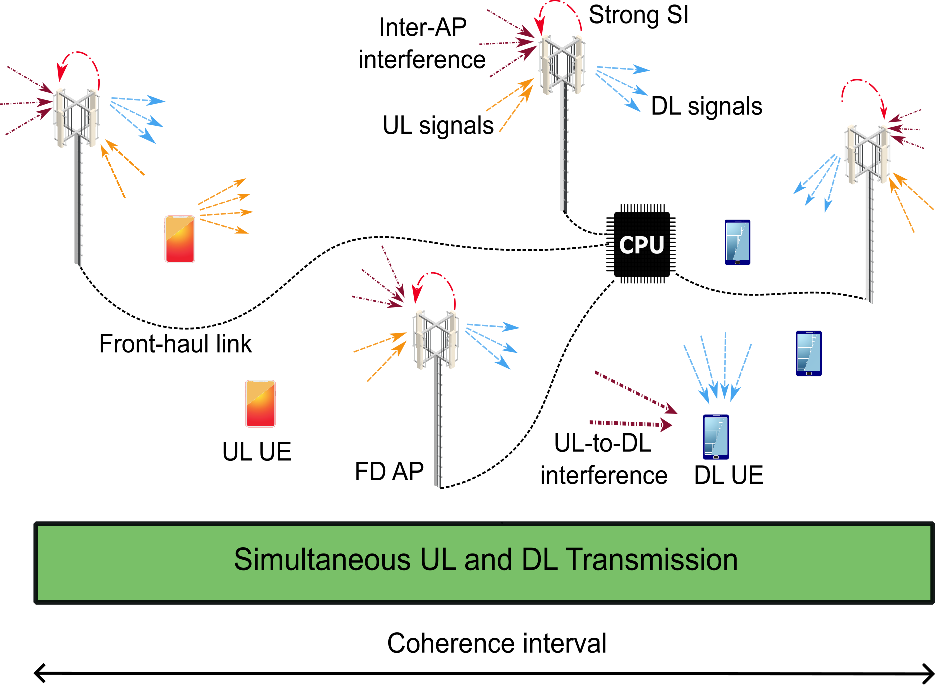}
	\vspace{-0.5em}
	\caption{Diagram of an IBFD CF-mMIMO system with DL and UL UEs. All APs coherently transmit to all DL UEs and simultaneously receive from all UL UEs over the same frequency.}  
 \vspace{1em}
	\label{fig:FDCFmMIMO}
\end{figure}
%%%%%%%%%%%%%%%%%%%%%%%%%%%%%%%%%%%%%%%%%%%%%%

Deployment of smaller and smaller cells (e.g., microcells, picocells, and femtocells) has been acknowledged as an efficient strategy for enhancing the network capacity, in terms of the number of bits per second that can be transmitted in a given area. Theoretically speaking, the network capacity increases in proportion to the number of active serving BSs/APs to the associated served UEs. However, in practice, this tendency diminishes gradually due to the rising interference between neighboring cells~\cite{Shidong:MCOM:2003}. As a matter of fact, beyond a specific threshold, additional network densification  may result in a decrease of the overall network capacity~\cite{Andrews:TCOM:2011}. Further, the UEs located in close proximity to the APs, specifically in the cell center, will encounter a higher SNR compared to those situated near the edge between two cells due to the rapid decay in the received signal power with the propagation distance. Additionally, the cell-edge UEs   contend with inter-cell interference, resulting in a substantially lower signal-to-interference-plus-noise ratio. Consequently, significant variations in data rates exist within each cell in traditional cellular networks. On the other hand,  although by deploying higher number of antennas per site using communication technologies, such as  massive MIMO, the  area associated with  the maximum delivered data rate expands and  the cell-edge UEs experience higher data rates, the rate variations are still a major bottleneck~\cite{demir2021foundations}. 

CF-mMIMO is an innovative technology that merges the strengths of both distributed  networks and  massive MIMO technology, endeavoring to overcome their inherent weaknesses~\cite{Hien:cellfree}. The concept of CF-mMIMO was  initially introduced in 2017 by Ngo \textit{et al.}  in~\cite{Hien:cellfree} with the objective of designing  a new network capable of  improving the user-experienced data rates, instead of the average or peak rates, and hence, providing ubiquitous coverage. This is realized by deploying many geographically distributed APs in the coverage area without dividing the area into disjoint cells to coherently serve the UEs in the same time-frequency
resources. As illustrated in Fig.~\ref{fig:FDCFmMIMO}, in a CF-mMIMO network, the APs are connected through fronthaul links to central processing units (CPUs), while the CPUs are interconnected by back-haul links. Moreover, CF-mMIMO employs the TDD operation, and hence,  by leveraging the channel reciprocity, channel estimation and precoding can be performed locally at each AP. In particular, for the UL, data detection can take place locally at individual APs, centrally at the CPU, or undergo a two-step process where it begins at each AP and is then completed at the CPU. In the DL, the UEs can coherently receive service from either all APs or a subset of APs while experiencing no cell boundaries. 

Early versions of CF-mMIMO were previously referred to as  distributed wireless communication systems~\cite{Choi:TWC:2007}, network MIMO~\cite{Venkatesan:ACSSC:2007}, and coordinated multi-point ~\cite{Irmer:COMG:2011}. The main distinctions between these initial technologies and CF-mMIMO lie in the operational setup, characterized by a larger number of APs than active UEs, and the physical-layer processes relying on recent advancements in massive MIMO technology~\cite{interdonato2019ubiquitous}. The three distinct characteristics of CF-mMIMO can be summarized as:
%===========
\begin{itemize}
\item  The APs cooperate to jointly serve the UEs in a user-centric fashion, as shown in Fig.~\ref{fig:FDCFmMIMO}, and hence, CF-mMIMO can  address the interference challenges that constraint traditional  dense cellular networks.
\item  By employing numerous distributed APs with fewer number of antennas, as opposed to a limited number of APs with large antenna arrays, CF-mMIMO has significantly greater DoF as well as macro-diversity gain to eliminate the large rate variations  that hinder the effectiveness of conventional cellular massive MIMO. Indeed, CF-mMIMO is capable of addressing the ubiquitous connectivity requirements in next generation wireless networks.
\item Thanks to the TDD operation in CF-mMIMO, channel information can be solely estimated during the UL training phase, making it applicable to both UL and DL data transmission phases. This, in turn, avoids costly estimation overhead and outdated CSI challenges.
\end{itemize}

Due to these characteristics, CF-mMIMO has become a popular subject of interest for various applications, including satellite-terrestrial scenarios~\cite{Chien:IoT:2024}, ISAC systems~\cite{Mao:TWC:2024}, and high mobility scenarios~\cite{Mohammad:OTFS}.
%%%%%%%%%%%%%%%%%%%%%%%%%%%%%%%%%%%%%%%%%%%%%%%%
\subsection{In-Band Full-Duplex Cell-Free Massive MIMO }
%%%%%%%%%%%%%%%%%%%%%%%%%%%%%%%%%%%%%%%%%%%%
Owing to the all aforementioned advantages of CF-mMIMO and IBFD technology, it is totally reasonable to envision a wireless system integrating these technologies, referred to as IBFD CF-mMIMO, contributing to the improvement of SE in future wireless networks~\cite{tung19ICC,hieu20JSAC}. Recently, some efforts are put upon  exploring more fundamental interests in IBFD CF-mMIMO~\cite{tung19ICC,hieu20JSAC,Anokye:TVT:2021,Anokye:TWC:2023}. In particular, by deriving closed-form SE expressions for the UL and DL UEs, the authors in~\cite{tung19ICC} confirmed that the IBFD CF-mMIMO provides significant gains in terms of SE  compared to the HD CF-mMIMO and IBFD massive MIMO systems. Nevertheless, the results in~\cite{tung19ICC} show that the residual SI and inter-AP interference  remain key challenges for practical CF-mMIMO systems with a finite number of APs.  

Consider an IBFD CF-mMIMO system as shown in Fig.~\ref{fig:FDCFmMIMO}, comprising $M$ multi-antenna IBFD APs, each one equipped with $\NrxAP$ and $\NtxAP$ receive and transmit antennas, respectively.  Denote the set of APs with $\MM\triangleq\{1,\ldots,M\}$, while the two sets $\KU$ and $\KD$ denote the collections of indices of the UL and DL single-antenna UEs, respectively. By using a linear decoding design at each AP, the SE of the $k$th UL UE, $\forall k\in\KU$,  can be expressed as
%-------------------
\begin{subequations}
    \begin{align}~\label{eq:SE:FDUP}
    \SE^{\FDCF}_{k,\ul} &\!\!= \! \log_2\bigg(1\!+ \!\frac{\DSFDCFul}{\CCIuCF \!+\IAPICF \!+\! \SIAP \!+ \! \NOISE}\bigg),
\end{align}
\end{subequations}
%------------------------------------------
where $\CCIuCF$ denotes the overall CCI at the IBFD AP from other UL UEs, $\IAPICF$ is the inter-AP interference (the interference between  APs), while $\SIAP$ represents the residual SI at the serving IBFD APs after SIC.

The DL SE of the $k$th DL UE, $\forall k\in\KD$, can be also written as
%-------------------
\begin{subequations}
    \begin{align}~\label{eq:SE:FDDL}
    \SE^{\FDCF}_{k,\dl} &=  \log_2\bigg(1+ \frac{\DSFDCFdl}{\CCIdCF + \UDIdCF +  \NOISE}\bigg), 
\end{align}
\end{subequations}
%------------------------------------------
where $\UDIdCF$ denotes the overall CCI at the $k$th DL UE, from other DL UEs;  $\UTDI$ is the UL-to-DL interference, i.e., the interference from the UL UEs to the $k$th DL UE.  

A comparison between the SE of the IBFD CF-mMIMO system and IBFD massive MIMO systems reveals that IBFD operations are more practical for CF-mMIMO deployment than conventional massive MIMO, owing to various factors:
\begin{itemize}
\item \textit{Manageable SI:} The smaller coverage area of numerous distributed APs in  IBFD CF-mMIMO system, makes APs an ideal host to deploy energy- and cost-effective IBFD transceivers. In fact, APs are low-power compared to the BSs, thus SI becomes more manageable.
\item \textit{Manageable CCI:} The typical range of the CCI and intra-AP interference is significantly lower than the corresponding interference terms in co-located IBFD massive MIMO systems due to the distributed network topology of CF-mMIMO. In fact, the  geographical distances among the distributed APs result in a lower CCI level due to the effect of path loss.
\item \textit{Fairness Among UEs:}  The CF-mMIMO is capable of providing uniform connectivity to all the UEs in any location and, hence, IBFD techniques exhibit enhanced efficacy within the context of CF-mMIMO, as contrasted with co-located massive MIMO configurations. In fact, in IBFD co-located massive MIMO systems, the performance of cell-center UEs is  jeopardized by the  cell-edge users, thereby introducing deleterious effects on the whole system performance. 
\end{itemize}

In CF-mMIMO networks, data exchange between the APs and CPU over the fronthaul link introduces a new source of power consumption compared to traditional cellular networks. This new power consumption component primarily comprises two main components: 1) a fixed power consumption term, $\Pbfixdm$, which is traffic-independent and may depend on the distances between the AP $m$ and the CPU; 2) A traffic-dependent term, which is proportional to the fronthaul rate between the AP $m$ and CPU.  Therefore, the power consumption of the fronthaul signal load to each AP $m$ can be expressed as~\cite{ngo18TGN}:
%--------------------------
\begin{align}
    \PAPbhm = \Pbfixdm +R^{\FDCF}\Pbtm,
\end{align}
%---------------------------
where $\Pbtm$ is the traffic-dependent fronthaul power (in Watt/bit/s) and $R^{\FDCF} = B\Big( \sum\nolimits_{k\in\KD}\SE^{\FDCF}_{k,\dl} + \sum\nolimits_{k\in\KU}\SE^{\FDCF}_{k,\ul} \Big)$ is the fronthaul rate between the AP $m$ and CPU,  while $B$ is the system bandwidth. 

The EE of IBFD CF-mMIMO can be expressed as
%------------------------------------
\begin{align}
    \EE^{\FDCF} =\frac{\sum\nolimits_{k\in\KD}\SE^{\FDCF}_{k,\dl} + \sum\nolimits_{k\in\KU}\SE^{\FDCF}_{k,\ul}}{\PtotCFFC},
\end{align}
%-----------------------------------
where the total power consumption is given by 
%--------------------------
\begin{align}
  \PtotCFFC &= 
  \sum\nolimits_{k\in\KU} \bigg(\frac{1}{\epsilon_{\UEk}}\PULUE  + \Pulcir\bigg)
    \nonumber\\
    &\hspace{-1em}+\!\sum\nolimits_{m\in\MM} \!\!\bigg(\frac{1}{\epsilon_{\AP}}\PAPm \!+\!  \PAPcir \!+\! 
     \NrxAP\PSICAP+\PAPbhm\!\bigg), 
\end{align}
%-------------------------
where $\PAPm$ and $\PAPcir$ denote the data-dependent and circuit power consumption of the $m$th AP, where $\PAPcir\triangleq(\NtxAP\PAPdyndl + \NrxAP\PAPdynul +\PAPstat)$, while $\PAPdyndl$ and $\PAPdynul$ are the dynamic power consumptions, associated with the power radiation of all circuit blocks in each active transmit and receive RF chain, respectively; $\PAPstat$ is the static power consumed by the cooling system, power supply, etc. Moreover, $\epsilon_{\AP}\in(0,1)$ is the power amplifier efficiency of each AP.

An IBFD CF-mMIMO system results in an increased power consumption within the network when compared to conventional cellular networks. This is attributed to the deployment of additional APs and, consequently, a greater number of RF chains. Moreover, the power consumption at the fronthaul link, which is directly proportional to the cumulative SE, increases the overall power consumption. Therefore, to characterize the intricate relationship between the EE and SE, we must consider the power consumption at the fronthaul link. In addressing the inherent challenges of CF-mMIMO, whether it is in pursuit of maximizing SE, EE, or striking a balance between the two, three robust techniques—namely, power control, AP-user association, and AP selection—are widely employed~\cite{hieu20JSAC}. Another promising solution to improve the EE is to reduce the power consumption associated with both RF chains and the fronthaul operations of the APs by employing low-resolution ADCs. The power dissipation of ADCs increases linearly with the sampling frequency and exponentially with the ADC resolution. Utilizing low-resolution ADCs helps to mitigate the power consumption. Moreover, transferring the quantized version of the received data from the APs to the CPU reduces the power consumption at the fronthaul link. Note that using the low-resolution ADCs introduces quantization noise, which affects the SE performance of the system. The authors in~\cite{Anokye:TVT:2021,Anokye:TWC:2023} characterized the UL/DL SE of the IBFD CF-mMIMO systems with low-resolution ADCs at the APs and DL UEs. Their findings reveal that the effects of the residual SI, inter-AP interference, and UL-to-DL interference are amplified by the low-resolution ADCs.

Nevertheless, in spite of the various efforts made in CF-mMIMO networks, the expansion of the number of APs in CF-mMIMO systems still introduces four significant bottlenecks for IBFD communications:
%===============================
\begin{enumerate}
\item  \emph{Power Consumption:} IBFD CF-mMIMO leads to an increase in the total network power consumption and, in turn, lower EE because of the additional number of  APs and, hence, more RF chains.
\item \emph{CLI:}  By deploying  IBFD transceivers at the APs, IBFD CF-mMIMO systems are exposed to an additional source of interference compared to the HD variant. These new interference sources share the same similarities as CLI in IBFD cellular networks. Specifically, CLI in IBFD CF-MIMO is the interference either received by the receiving antennas of one AP
originated from the transmitting antennas of another AP ($\IAPICF$) or
the interference received by a DL UE from other UL UEs ($\UDIdCF$).
\item \emph{Residual SI:}  Residual SI after using the SIC at the APs remains a challenging issue in the design of IBFD CF-mMIMO. This SI can significantly degrade the UL performance of the network, while its cancellation requires power-hungry circuits, thereby, degrading the EE.  
\item \emph{High-Cost IBFD Transceiver Deployment:} In general, the implementation of IBFD CF-mMIMO  requires complicated signal processing tasks at all APs, accompanied by substantial IBFD transceivers' deployment costs, especially when the number of APs grows large.
\end{enumerate}
%===================================

\textbf{Takeaway Messages: } Reducing the number of antennas and lowering transmit power at IBFD APs, compared to IBFD massive MIMO BSs, improves the feasibility of implementing SIC at the APs. However, increasing the number of APs introduces new challenges, significantly raising the implementation costs and power consumption in the network. These challenges are driving the development of the next generation of CF-mMIMO networks, which aim to suppress the power consumption over the fronthaul link, reduce the AP power while maintaining target rates, and effectively manage interference within the network.

%%%%%%%%%%%%%%%%%%%%%%%%%%%%%%%%%%%%%%%%%%%%%%%%%%
\section{Network-Assisted Full-Duplex Cell-Free Massive MIMO}\label{sec:NAFDcellfree}
\begin{figure}[t]
	\centering
  \vspace{0em}
	\includegraphics[width=0.5\textwidth]{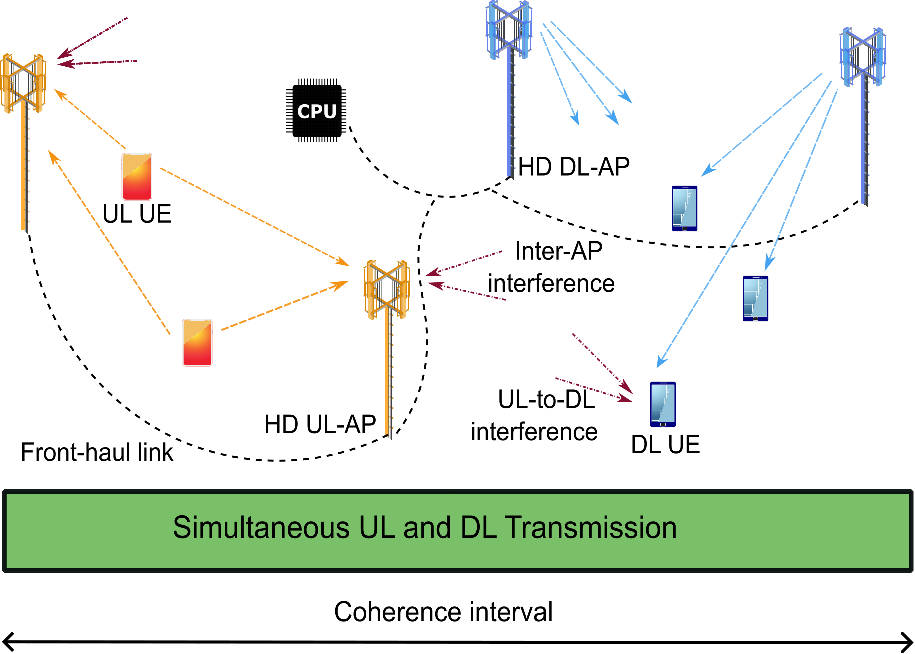}
	\vspace{-1.5em}
	\caption{Diagram of an NAFD CF-mMIMO system with HD UL-/DL-APs and DL and UL UEs. The HD UL-APs serve UL UEs, while the  HD DL-APs coherently transmit to the DL UEs.} 
 \vspace{0.5em}
	\label{fig:NAFD}
\end{figure}

%%%%%%%%%%%%%%%%%%%%%%%%%%%%%%%%%%%%%%%%%%%%%%%%%%%
The prevalence of contemporary massive machine-type communications, where large numbers of devices are communicating with each other or with a central network,  raises the likelihood of asymmetric demands between DL and UL. To provide a clearer depiction of this, consider a crowded public event, where attendees are using their smartphones to share photos and videos.  In this context, two distinct groups of UEs coexist simultaneously: \textbf{1) DL operation UEs:} People attending the event may be predominantly consuming data by downloading event-related content, such as live-streaming video feeds, event updates, or multimedia content shared by organizers. These UEs  are in DL operation, as they receive data from the network. \textbf{2) UL operation UEs:} Meanwhile, another group of attendees might be actively uploading their own photos and videos to social media platforms, contributing to the event's coverage. These UEs are in UL operation, as they are sending data to the network. This creates an asymmetric and dynamic demand, as the network needs to manage the simultaneous requirements of both types of UEs in a real-time manner. Efficiently handling this imbalance is crucial for providing a satisfactory UE experience and optimizing the network resources. While an IBFD CF-mMIMO network appears to be an ideal solution for handling these asymmetric demands in the network, the above-mentioned challenges have prompted researchers to investigate new architectures for CF-mMIMO networks. This has led to a recent paradigm shift known as NAFD CF-mMIMO. The concept of NAFD aims to achieve IBFD functionality using current HD hardware devices virtually, as shown in Fig.~\ref{fig:NAFD}. The NAFD concept offers a significant advantage by eliminating SI within the APs, thereby simplifying the complexity of transceivers.

%%%%%%%%%%%%%%%%%%%%%%%%%%%%%%%%%%%%%%%%%%%%%%
%%%%%%%%%%%%%%%%%%%%%%%%%%%%%%%%%%%%%%%%%%%%%%

%%%%%%%%%%%%%%%%%%%%%%%%%%%%
\subsection{Historic Background}
%%%%%%%%%%%%%%%%%%%%%%%%%%%%
The concept of NAFD exhibits similarities with established terminologies in the literature. These include dynamic-TDD~\cite{Haas:JSAC:2001,Liu:TCOm:2017}, dynamic UL-DL configuration in time-division LTE~\cite{Shen:MCOM:2012}, coordinated multipoint for IBFD~\cite{Thomsen:WCL:2016}, and bidirectional dynamic networks~\cite{Xin:ACCESS:2017}. These techniques were introduced to address the challenges posed by asymmetric UL-DL data traffic in wireless (cellular) networks. The overarching theme among these concepts is the adaptation of network configurations to efficiently manage the imbalances in UL-DL data flows, providing valuable insights into addressing the complexities associated with asymmetric communication patterns.

In dynamic-TDD, based on the UL-DL traffic demands within each cell, a specific time slot can be dynamically allocated for either reception or transmission. Nevertheless, this scheme does not fully cater to heterogeneous data demands within the cells, as UL (DL) UEs within a cell assigned a DL (UL) time slot must wait for the UL (DL) time slot to transmit their data. Moreover, the effectiveness of the dynamic-TDD is constrained by inter-BS interference and the need for strict synchronization among cells, which necessitates cell-cooperation. This results in significant overhead and complexity. The rationale behind the dynamic UL-DL configuration in time-division LTE is to establish opposite transmission directions across various small cells in heterogeneous cellular networks~\cite{Shen:MCOM:2012}. Coordinated multipoint for IBFD, which is motivated by the coordinated multipoint concept in cellular networks, involves emulating an IBFD BS and using two spatially separated and coordinated HD BSs~\cite{Thomsen:WCL:2016}. In bidirectional dynamic networks, the number of UL and DL remote radio heads is adjusted flexibly to facilitate simultaneous UL and DL communications~\cite{Xin:ACCESS:2017}. 

%%%%%%%%%%%%%%%%%%%%%%%%%%%%
\vspace{-1em}
\subsection{Fundamentals of NAFD CF-mMIMO}
%%%%%%%%%%%%%%%%%%%%%%%%%%%%
NAFD unifies all duplex modes in the CF-mMIMO space, i.e.,  the APs can operate in hybrid-duplex (both IBFD and HD APs exist in the network) or flexible-duplex (APs operate in HD mode)~\cite{Wang:TCOM:2020,Jiamin:TWC:2021}. The transmission mode of each AP (UL reception and/or DL transmission) can be dynamically designed based on the network conditions and requirements. Accordingly, both UL and DL services could be supported simultaneously and flexibly.  When NAFD is forced to be established via pure flexible-duplex, the SI and related challenges are completely removed, as the APs are working in the UL (termed as UL-AP) or in the DL (termed as DL-AP). Moreover, the amount of CLI becomes fairly less than that in the IBFD counterpart. This is due to the fact that only a part of  the APs contribute to DL transmissions, whilst the remaining APs are dedicated for UL reception. This can help to reduce the amount of intra-AP interference in NAFD compared to IBFD networks. Furthermore, AP mode assignment offers a flexible DoF to better manage the CCI compared to IBFD.  For instance, the APs located near the DL UEs can transition to DL mode, while those in proximity to the UL UEs can switch to UL mode. Hence, the distance between the transmitting and receiving nodes becomes quite small. Accordingly, by properly selecting the power control coefficients at the DL-APs and UL UEs, and optimizing the weights of large-scale fading, we can effectively manage the CLI.

Consider an NAFD CF-mMIMO system shown in Fig.~\ref{fig:NAFD}, where the HD APs are partitioned into UL-APs and DL-APs, supporting the UL and DL UEs, respectively. We denote the set of UL-APs and DL-APs as $\MAPDL$ and $\MAPUL$. Therefore, the SE of the $k$th UL UE, $k\in\KU$, can be expressed as
%-------------------
    \begin{align}~\label{eq:SE:ULCF-VFD}
    \SE^{\NAFD}_{k,\ul} &=  \log_2\bigg(1+ \frac{\DSNAFDul}{\CCIuNA +\IAPINAFD+  \NOISE}\bigg), 
\end{align}
%------------------------------------------
where $\CCIuNA$ denotes the CCI from the other UL UEs, while $\IAPINAFD$ denotes the inter-AP interference. In comparison to the IBFD CF-mMIMO system, it can be reasonably inferred that both $\CCIuNA$ and $\IAPINAFD$ are significantly lower than their counterparts, $\CCIuCF$ and $\IAPICF$, respectively. Furthermore, since the UL reception is not affected by SI, it is anticipated that the NAFD structure will provide higher UL SE compared to the IBFD configuration.

The DL SE of the $k$th DL UE, $\forall k\in\KD$, can be also written as
%-------------------
    \begin{align}~\label{eq:SE:DLCF-VFD}
    \SE^{\NAFD}_{k,\dl} &=  \log_2\bigg(1+ \frac{\DSNAFDdl}{\CCIdNA + \UDIdNAFD +  \NOISE}\bigg), 
\end{align}
%------------------------------------------
where $\CCIdNA$ denotes the CCI from the other DL UEs and $\UDIdNAFD$  is the UL-to-DL interference. Due to the proximity of UL UEs to UL-APs and the lower inter-AP interference at the UL-APs compared to the inter-AP interference in the IBFD APs of the IBFD CF-mMIMO counterpart, UL UEs can transmit with lower power in NAFD compared to IBFD CF-mMIMO. Consequently, we can infer that $\UDIdNAFD$ is less than $\UDIdCF$, suggesting that NAFD has the potential to enhance the DL SE as well.

The EE of the NAFD CF-mMIMO can be expressed as
%------------------------------------
\begin{align}
    \EE^{\NAFD} =\frac{\sum\nolimits_{k\in\KD}\SE^{\NAFD}_{k,\dl} + \sum\nolimits_{k\in\KU}\SE^{\NAFD}_{k,\ul}}{\PtotNAFD},
\end{align}
%-----------------------------------
where the total power consumption is given by 
%--------------------------
\begin{align}
 \PtotNAFD &= 
  \sum\nolimits_{k\in\KU} \bigg(\frac{1}{\epsilon_{\UEk}}\PULUE  + \Pulcir\bigg)
    \nonumber\\
    &+\!\sum\nolimits_{m\in\MAPDL} \!\!\bigg(\frac{1}{\epsilon_{\AP}}\PAPm \!+\!  \PAPcir  
     \!\bigg) + \PAPbhNAFD, 
\end{align}
%-------------------------
where $\PAPbhNAFD$ is the fronthaul power consumption, which is proportional to the overall network rate. When comparing $ \PtotNAFD$ and $\PtotCFFC$, it becomes evident that the power consumption needed for SIC is vanished thanks to the NAFD structure. Additionally, both the data-dependent transmit power and circuit power consumption of the APs are limited to a subset of APs, specifically the DL-APs. Therefore, we can conjecture that NAFD CF-mMIMO can in principle offer better EE than IBFD CF-mMIMO. 

%%%%%%%%%%%%%%%%%%%%%%%%%%%%
%\vspace{0.5em}
\subsection{State-of-the-Art}
%%%%%%%%%%%%%%%%%%%%%%%%%%%%
Considering the notable improvements in the SE and EE offered by NAFD CF-mMIMO compared to IBFD CF-mMIMO, there has been a significant focus on exploring and understanding various facets of this emerging paradigm. This research pursues a comprehensive analysis of the NAFD CF-mMIMO network's performance, including aspects, such as AP mode assignment and resource allocation. Additionally, efforts have been directed towards integrating, investigating, and optimizing the performance of other technologies within NAFD CF-mMIMO systems.

% %%%%%%%%%%%%%%%%%%%%%%%%%%%%%%%%%%%%%
 \subsubsection{Interference Mitigation and user-scheduling}
% %%%%%%%%%%%%%%%%%%%%%%%%%%%%%%%%%%%%%%
The preliminary investigations have focused on assessing the UL and DL SE of NAFD CF-mMIMO systems. These studies were conducted under various conditions, including scenarios with correlated channels and the presence of CSI~\cite{Wang:TCOM:2020,Jiamin:TWC:2021}. It is worth mentioning that these investigations were conducted within a consistent framework, maintaining a fixed setup throughout. Specifically, this type of analysis was carried out under the assumption that the modes of the APs, i.e., UL reception or DL transmission, had been pre-designed and remained fixed throughout the network. These studies shed light on the challenges posed by inter-AP interference on the UL reception at the UL-APs, as well as the impact of UL-to-DL interference on the performance of the DL UEs. 

In response to the challenge of inter-AP interference, an interference cancellation scheme was introduced in~\cite{Jiamin:TWC:2021}. The core concept behind this scheme involves regenerating interference signals and subsequently subtracting them from the received data signals. To achieve this, an estimate of the CSI between the DL-APs and UL-APs at the CPU and the CSI between the DL-APs and DL UEs at the DL UEs is required. A beamforming training approach was adopted for this purpose, aiming to estimate the effective CSI, defined by the inner products of beamforming and channel vectors. Notably, the advantage of this beamforming training lies in its overhead, which is determined solely by the number of DL UEs, rather than the number of antennas deployed across all DL-APs. It is important to note, however, that this method operates in a centralized manner, necessitating the estimation of inter-AP channels at the CPU to execute the interference cancellation process.

In order to minimize the UL-to-DL interference in an NAFD CF-mMIMO network, a user-scheduling method based genetic algorithm was proposed in~\cite{Wang:TCOM:2020}. However, this algorithm does not take into account the priority of UL and DL UEs. In other words, during a given coherence interval, it is possible for a high-priority UL or DL UE to be queued.

% %%%%%%%%%%%%%%%%%%%%%%%%%%%%%%%%%%%%%%%%%%%%%%%%%%%%%%%%%%%%%%%%
\subsubsection{AP Mode Selection and Resource Allocation} 
% %%%%%%%%%%%%%%%%%%%%%%%%%%%%%%%%%%%%%%%%%%%%%%%%%%%%%%%%%%%%%%%
To efficiently exploit the available UL/DL resources, the operation mode of the APs can be flexibly determined. In pursuit of this objective, the most recent research has shifted its focus to explore the challenges associated with operational mode selection in NAFD CF-mMIMO systems~\cite{Zhu:COML:2021,Xia:China:2021,Xia:SYSJ:2022,Song:SYST:2023,Mohammad:JSAC:2023,Chowdhury:TCOM:2024,Sun:TCOM:2024}. This investigation takes into account various criteria, such as the sum SE maximization and EE enhancement. The duplex mode assignment problem is inherently challenging. By considering a more general class of optimization problems, which includes joint duplex mode selection, UL/DL power control design, and precoding/decoding design, the resulting problems fall into the category of mixed-integer non-convex optimization problems. In these problems, both types of binary and continuous variables are interconnected within the UL/DL UE's SE. The allocation of operating modes for each antenna of the AP, encompassing UL reception mode, DL transmission mode, and sleep mode, was addressed in~\cite{Xia:China:2021}. The objective was to maximize the aggregate UL and DL SE while accommodating QoS constraints and power budget constraints. Furthermore, in~\cite{Mohammad:JSAC:2023}, a different approach was taken by considering channel estimation errors and fronthaul energy consumption. This involved a joint design of the AP mode assignment, UL/DL power control coefficients, and the design of large-scale fading decoding weights for the UL APs to maximize the SE and EE, while ensuring that the QoS requirements of UL/DL UEs are met. In the presence of pilot contamination, a comparative study reported in ~\cite{Chowdhury:TCOM:2024} demonstrated that NAFD CF-mMIMO outperforms IBFD CF-mMIMO under similar antenna density, leveraging pilot assignment and power control. In~\cite{Sun:TCOM:2024}, a dual time-scale resource scheduling scheme for NAFD CF-mMIMO networks was proposed, which consists of long-term AP mode optimization and short-term power allocation optimization to enable massive URLLC.

 Table~\ref{table:IBFD} summarizes the key papers on IBFD massive MIMO in cellular and cell-free networks.
%%%%%%%%%%%%%%%%%%%%%%%%%%%%%%%%%%%%%%%%%%%%%%%%%%%%%%

\begin{table*}[t]
  \centering
 \caption{Summary of the existing IBFD massive MIMO papers.}
 \vspace{-0.5em}
  \small
  \begin{tabular}{|c|>{\centering\arraybackslash}m{0.7cm}|c|>{\centering\arraybackslash}m{11cm}|}
    \hline
    \textbf{Category} & \textbf{Year} & \textbf{Lit.} & \textbf{\hspace{0em}Main focus} \\
    \hline
    \multirow{6}{*}{\vspace{-4em}Cellular} 
    & 2016 &\cite{Everett:TWC:2016} 
    & First experiment-based prototype, known as SoftNull, analyzing the impact of precoding on the SIC capability for indoor and outdoor applications
    \\ \cline{2-4}
    &2017  &\cite{Xia:TVT:2017}     
    &Propose beam-domain IBFD concept and investigate key components for practical implementation: UEs grouping and effective beam-domain channel estimation
    \\ \cline{2-4}
    &2017  &\cite{Shojaeifard:TCOM:2017}     
    & Provided a theoretical framework using tools from stochastic geometry and point processes for the study of massive MIMO-enabled IBFD cellular networks
    \\ \cline{2-4}
    &2017  &\cite{Bai:TWC:2017}    
    & Asymptotic performance analysis for multi-Cell MU-MIMO network considering imperfect SIC, channel estimation error, training overhead and pilot contamination
    \\ \cline{2-4}
    & 2018 &\cite{Koh:TWC:2018} 
    & Pilot overhead problem induced by SI channel estimation, and propose an effective pilot transmission scheme
    \\ \cline{2-4}
    & 2023  &\cite{Kim:TWC:2023} 
    & Performance analysis of SIC methods, SI subtraction and spatial suppression, by considering the SI channel’s estimation error
    and spatial correlation
    \\\hline
    \multirow{6}{*}{\vspace{-1em}CF-mMIMO} 
    &2019  & \cite{tung19ICC} 
    & Introduce the application of IBFD into CF-mMIMO with UL and DL (asymptotic) performance analysis 
    \\ \cline{2-4}
    &2020  &\cite{hieu20JSAC}       
    & SE and EE maximization by jointly optimizing the power control, AP-UE association and AP selection
    \\ \cline{2-4}
    &2021  &\cite{Anokye:TVT:2021}      
    & Characterizing the impact of low-resolution ADCs at the APs and the DL UEs on the sum-SE and EE
    \\ 
    \cline{2-4}
    &2023  & \cite{Anokye:TWC:2023} 
    & Analyzing the combined effects of the Ricean 
$K$-factor, residual SI, intra-UE/AP interference, and quantization noise of low-resolution ADCs on the UL/DL SEs
    \\ 
    \hline
    \multirow{6}{*}{\vspace{-10em}NAFD CF-mMIMO} 
    &2020  &\cite{Wang:TCOM:2020}       
    & Introduction of the NAFD concept for CF-mMIMO networks and derivation of UL and DL SE expressions, considering imperfect CSI and spatial correlations
    \\ 
    \cline{2-4}
    &2021  &\cite{Jiamin:TWC:2021}  
    & Proposal to estimate effective CSI (inner products of beamforming and channel vectors) and implement power allocation based on slowly varying large-scale fading
    \\ 
    \cline{2-4}
    &2021  &\cite{Zhu:COML:2021} 
    & Introduce duplex mode selection (UL or DL) at APs  
    \\ \cline{2-4}
    &2021  &\cite{Xia:China:2021}   
    & Duplex mode selection (UL and/or DL) at APs and transceiver design
    \\ 
    \cline{2-4}
    &2022 &\cite{Xia:SYSJ:2022}     
    & Duplex mode selection (UL and/or DL) at APs for secure transmission in both noncolluding and colluding eavesdropper scenarios
    \\ \cline{2-4}
     &2023  &\cite{Song:SYST:2023}   
    & Analysis of the impact of low-resolution ADCs and design of an efficient bit allocation algorithm for low-resolution ADCs
    \\ \cline{2-4}
     &2023  &\cite{Mohammad:JSAC:2023}   
    & SE and EE optimization through duplex mode selection at APs, DL (UL) power control design at APs (UEs), and large-scale fading decoding design at the CPU
    \\ \cline{2-4}
    &2024  &\cite{Chowdhury:TCOM:2024}   
    & Comparative study of HD APs with dynamic TDD and IBFD APs, focusing on joint UL-DL power allocation and UL/DL mode scheduling of APs to maximize the sum SE
    \\ \cline{2-4}
    &2024  &\cite{Sun:TCOM:2024}      
    & NAFD multi-level CF-mMIMO architecture and dual time-scale resource scheduling scheme to meet  massive URLLC requirements in industrial massive IoT scenarios
    \\ 
    \hline
  \end{tabular}\label{table:IBFD}
    \vspace{-0em}
\end{table*}

%%%%%%%%%%%%%%%%%%%%%%%%%%%%%%%%%%%%%%%%%%%%%%%%%%%%%%

%%%%%%%%%%%%%%%%%%%%%%%%%%%%%%%%%%%
\section{Mathematical Framework and System Design}~\label{sec:mathModel}
%%%%%%%%%%%%%%%%%%%%%%%%%%%%%%%%%%%%%%%
In this section, we provide a general mathematical framework for CF-mMIMO networks to enable analytical and theoretical comparisons between different network structures. Additionally, we present the required steps to achieve results  for IBFD/HD small-cell networks, highlighting their competitiveness with conventional cellular massive MIMO systems.

\vspace{-1em}
\subsection{System Model and Assumptions}
Consider a NAFD CF-mMIMO system, where $M$ APs serve $K_u$ UL UEs and $K_d$ DL UEs, as shown in Fig.~\ref{fig:NAFD}. For notational simplicity, we define the sets $\mathcal{M}\triangleq\{1,\ldots,M\}$, $\K_d\triangleq \{1,\dots,K_d\}$ and  $\K_u\triangleq\{1,\ldots,K_u\}$ as collections of indices of the APs, DL UEs, and UL UEs, respectively. Each AP is connected to the CPU via a high-capacity fronthaul link. DL and UL transmissions are performed simultaneously and in the same frequency band via HD (and IBFD) APs.\footnote{This design is a general model encompassing all possible architectures for NAFD CF-mMIMO. The APs can operate in: 1) IBFD mode, where all APs perform DL and UL simultaneously over the same frequency (IBFD CF-mMIMO); 2) hybrid-duplex mode, where both IBFD and HD APs exist in the network; or 3) flexible-duplex mode, where the APs operate in HD mode. Thus, our model unifies all duplex modes in the network~\cite{Wang:TCOM:2020,Jiamin:TWC:2021}.} Each UE is equipped with one single antenna, while each AP is equipped with  $N$ transmit, and $N$ receive RF chains. 

The channel coefficient vector between the $k$-th DL UE ($\ell$-th UL UE) and the $m$-th AP, $\gmkd\in\mathbb{C}^{N \times 1}$ ($\gmlu\in\mathbb{C}^{N \times 1}$), $\forall k \!\in \!\K_d$, $\ell \!\in\! \K_u$, $ m \!\in\! \MM $, is modeled as $   \gmkd=\sqrt{\betamkd}\tgmkd,~(\gmlu=\sqrt{\betamlu}\tgmlu) $, where $\betamkd$ ($\betamlu$) is the large-scale fading coefficient and $\tgmkd\in\mathbb{C}^{N \times 1}$ ($\tgmlu\in\mathbb{C}^{N \times 1}$) is the small-scale fading vector whose elements are independent and identically distributed (i.i.d.) $\mathcal{CN} (0, 1)$ random variables (RVs). Moreover, the channel gain between the UL UE $\ell$ to the DL UE $k$ is denoted by $h_{k\ell}$. It can be modelled as $h_{k\ell}=(\betakldu)^{1/2}\tilde{h}_{k\ell}$, where $\betakldu$ is the large-scale fading coefficient and $\tilde{h}_{k\ell}$ is a $\mathcal{CN}(0,1)$ RV. Finally, the interference links among the APs are modeled as Rayleigh fading channels. Let $\qZ_{mi}\in \mathbb{C}^{N\times N}$, $i\neq m$, be the channel matrix from AP $m$ to AP $i$, $\forall m,i\in\MM$, whose elements are  i.i.d. $\mathcal{CN}(0,\beta_{mi})$ RVs. Here, $\Z_{mm}, \forall m$ denotes the SI channel at the IBFD APs, whose elements are i.i.d. $\mathcal{CN}(0,\beta_{mm}\triangleq \SIm)$ RVs.

The binary variables to indicate the mode assignment for each AP $m$ are defined as
%-----------------------------------------
\vspace{-0.0em}
\begin{align}
\label{a}
a_{m} \triangleq
\begin{cases}
  1, & \text{if AP $m$ operates in the DL mode,}\\
  0, & \mbox{otherwise}.
\end{cases}, \forall m
\\
\label{b}
b_{m} \triangleq
\begin{cases}
  1, & \text{if AP $m$ operates in the UL mode,}\\
  0, & \mbox{otherwise}.
\end{cases}, \forall m
\end{align}
%----------------------------------------------
According to these assignment variables and to have a  NAFD CF-mMIMO with HD APs (flexible design), we have $a_m + b_m = 1$  $\forall m \in \MM$. When $a_m b_m = 1$, $\forall m \in \MM$, this model captures the IBFD CF-mMIMO.

%%%%%%%%%%%%%%%%%%%
\vspace{-.8em}
\subsection {DL and UL Signaling}

\subsubsection{DL Payload Data Transmission}
The received signal at DL UE $k$ is written as
%----------------------
\begin{align}~\label{eq:ykdl}
y_k^{\dl}
&=
\sqrt{\rho_d}
\sum\nolimits_{m \in \mathcal{M}} a_m\theta_{mk}
\left(\gmkd\right)^\dag\wmkdl
s_{k}^{\dl}
\nonumber\\
&+
\sqrt{\rho_d}
\sum\nolimits_{m \in \mathcal{M}}
\sum\nolimits_{k'\in\mathcal{K}_d \setminus k} 
a_m\theta_{mk'}
\left(\gmkd\right)^\dag \wmkpdl
s_{k'}^{\dl}
\nonumber\\
&+
\sum\nolimits_{\ell\in \mathcal{K}_{u}}h_{k\ell}x_{\ell}^{\ul}+w_{k}^{\dl},
\end{align}
%----------------------
where $\rho_d$ denotes the normalized DL SNR, $\theta_{mk}$ denotes the power control coefficient at AP $m$ corresponding to UE $k$, $\wmkdl\in\mathbb{C}^{N\times 1}$ is the precoding vector at AP $m$ for DL UE $k$,  $s_k^\dl\sim\mathcal{CN}(0,1)$ denotes the intended symbol for DL UE $k$, $w_{k}^{\dl}\sim\mathcal{CN}(0,1)$ is the additive white Gaussian noise (AWGN) at DL UE $k$. We notice that the third term in~\eqref{eq:ykdl}, is the CLI caused by the UL UEs due to concurrent transmissions of DL and UL UEs over the same frequency band and $x_\ell^\ul$ denotes the transmit signal from the $\ell$-th UL UE. 

\subsubsection{UL Payload Data Transmission}
The transmit signal  from  UL UE $\ell$ is represented by $x_{\ell}^\ul  = \sqrt{\rho_u {\varsigma}_\ell} s_{\ell}^{\ul}$, where $s_{\ell}$ (with $\Ex\left\{|s_{\ell}^\ul|^2\right\}=1$) and $\rho_u$ denote,  respectively, the transmitted symbol by the $\ell$-th UL UE and  the normalized transmit power at each UL UE, while ${\varsigma}_{\ell}$ is the transmit power control coefficient at UL UE $\ell$ with $ 0\leq {\varsigma}_{\ell} \leq 1, \forall \ell$. The UL APs with $b_m=1, \forall m$, receive transmit signal from all UL UEs. The received signal $\qy_{m}^{\ul}\in\mathbb{C}^{N \times 1}$ at AP $m$ in the UL mode can be written as
%----------------------
\begin{align}\label{eq:ymul}
\qy_{m}^{\ul}
&=
\sqrt{\rho_u}\sum\nolimits_{\ell\in \mathcal{K}_{u}}\sqrt{b_m \tilde{\varsigma}_{\ell}}\qg_{m\ell}^{\ul} s_{\ell}^{\ul}
\nonumber\\
&
+
\underbrace{\sqrt{\rho_d}\sum\nolimits_{i\in\mathcal{M}\setminus m}\sum\nolimits_{k\in \mathcal{K}_d}
\sqrt{b_m a_i } \theta_{ik}
\qZ_{mi}
\wikdl s_k^\dl}_{\text{Intra-AP interference}}\nonumber\\
&
\hspace{0em}
+
\underbrace{\sqrt{\rho_d}\sum\nolimits_{k\in \mathcal{K}_d}
\sqrt{b_m a_m } \theta_{mk}
\qZ_{mm}
\wmkdl s_k^\dl
}_{\text{Inter-AP (SI) interference}}+\sqrt{b_m}\qw_{m}^{\ul},
\end{align}
%----------------------------------
where $\qw_{m}^{\ul}$ is the AWGN vector with $\mathcal{CN}(0,1)$ distributed elements. In~\eqref{eq:ymul}, the second term captures the interference from APs transmitting towards DL UEs and the third term represents the SI term in case of AP $m$ is IBFD-enabled (i.e., $b_m a_m=1$). 

Then, AP $m$ performs linear processing to the received signal in~\eqref{eq:ymul}. Let $\wmlul\in\mathbb{C}^{N\times 1}$ denote the linear precoding vector at AP $m$. The resulting $(\wmlul)^\dag\qy_{m}^{\ul}$ is then forwarded to the CPU for signal detection. 
In order to improve the achievable UL SE, the forwarded signal is further multiplied by large-scale fading decoding weight  $\alphml, \forall m,\ell,$ at the CPU. The aggregated received signal for UL UE $\ell, \forall \ell$ with $|\alphml|^2\leq 1, \forall \ell, m$ at the CPU can be written as~\cite{Hien:Asilomar:2018,Bashar:TWC:2019}
%=========================
\begin{align}\label{eq:rul}	\qr_{\ell}^{\ul}=\sum\nolimits_{m\in\MM}\alphml(\wmlul)^\dag\qy_{m}^{\ul}. 
\end{align}
%=========================
Finally, $s_{\ell}^{\ul}$ is detected from $\qr_{\ell}^{\ul}$. 

%%%%%%%%%%%%%%%%%%%%%%%%%%%%%%%%%%%%%%%%%%
\subsection{Precoding Design } 
%%%%%%%%%%%%%%%%%%%%%%%%%%%%%%%%%%%%%%%%%%
For both DL and UL transmissions, we consider a local partial ZF (PZF) scheme, that effectively mitigates interference in a distributed and scalable manner, providing a flexible balance between interference suppression and large array gain~\cite{Interdonato:TWC:2020}. The principle of this scheme is that each AP suppresses interference only for the strongest UEs—those with the highest channel gain and presumably the most interference. Conversely, interference to the weakest UEs is tolerated. In its special case, this model encompasses both fully ZF (FZF) and maximum ratio transmission (MRT)/ maximum ratio combining designs.

To serve the DL UEs, each AP $m$, operating in DL mode (IBFD AP or HD AP in DL mode) divides the DL UEs into two groups: $\Sm \subset \{1, \ldots, K_d\}$, which includes the index of strong DL UEs, and $\Wm \subset \{1, \ldots, K_d\}$, which includes the index of weak DL UEs, respectively. Then, AP $m$ employs the ZF precoding for DL UEs in $\Sm$  and MRT precoding for DL UEs in $\Wm$. The local PZF transmit precoding at AP $m$, is given by  $\wmkdlzf = \gamdmk \hGmdl\big( \big(\hGmdl\big)^\dag \hGmdl\big)^{-1} \qe_k$, where $\hGmdl$ is an $N \times |\Sm|$ collective channel estimation matrix from all the UEs in $\Sm$ to AP $m$, given by $\hGmdl=[\hat{\qg}_{mk}^\dl: k \in \Sm]$. Moreover, $\qe_ k$ is the $k$-th column of $\qI_{K_d}$. Therefore, for any pair of DL UE $k$ and $k'\in\Sm $, we have  
%===
\begin{align}\label{eq:PZF_prec2}
(\hgmkpd)^\dag \wmkdlzf = 
\begin{cases} \displaystyle \gamdmk & \text {if } k=k',\\ \displaystyle 0 & \text {otherwise}.
\end{cases}
\end{align}
%====
Moreover,  the MRT precoding vector constructed locally by AP $m$ for DL UE $k \in \Wm$ is given in by  $\wmkdlmr = \hgmkd$.

To support the UL transmissions, each IBFD/UL HD AP $m$, divides the UL UEs into two groups: $\Smu \subset \{1, \ldots, K_u\}$, which includes the index of strong UL UEs, and $\Wmu \subset \{1, \ldots, K_u\}$, which includes the index of weak UL UEs, respectively. 

Accordingly, PZF is performed at the IBFD/UL HD AP for UL data reception.  The local PZF combining vector at AP $m$, is given by  $\wmlulzf = \gamuml\hGmul\big( \big(\hGmul\big)^\dag \hGmul\big)^{-1}\qe_{\ell} $, where $\qe_{\ell}$ is the $\ell$-th column of $\qI_{K_u}$ and $\hGmul = [\hgmk: k\in \Smu]$. Moreover,  for the maximum ratio combining vector, we set $\wmlulmr=\hgmlu$.

\subsection{UE Grouping and DL/UL SE} The  UE grouping can be based on different criteria. Inspired by~\cite{Interdonato:TWC:2020}, the UEs grouping strategy in PZF relies on the following rule,
%------------------
\begin{align}~\label{eq:groupin:criterion}
   \sum\nolimits_{k=1}^{\vert\hat{\mathcal{S}}_m \vert} \frac{\bar{\beta}_{m,k}}{\sum\nolimits_{t=1}^{K_d} \beta_{m,t}} \geq \upsilon\%,
\end{align}
%------------------
where $\hat{\mathcal{S}}_m$ denotes the set constructed by AP $m$ by selecting the UEs that contribute at least $\upsilon\%$ of the overall channel
gain. In~\eqref{eq:groupin:criterion}, $\{\bar{\beta}_{m,1},\ldots,\bar{\beta}_{m,K_d}\}$ indicates the set of the large-scale fading coefficients sorted in descending order.

For the simplicity of notation, we introduce a pair of binary variables to indicate the group assignment for each DL UE $k$ and DL AP $m$ in the local PZF precoding scheme as
%==========
\begin{align*}
\dm^{\Zk} = \begin{cases} \displaystyle 1 & \text {if } m \in \Zk,\\ \displaystyle 0 & \text {otherwise},
\end{cases}
\qquad
\dm^{\Tk} = \begin{cases} \displaystyle 1 & \text {if } m \in \Tk,\\ \displaystyle 0 & \text {otherwise},
\end{cases}
\end{align*}
%==========
where $\Zk$ and $\Tk$ denote the set of indices of APs that assign the $k$-th DL UE into $\Sm$ for  ZF precoding and the set of indices of APs that assign $k$-th DL UE into $\Wm$ for MRT precoding, respectively, defined as:
%=====
\begin{align}
\Zk &\triangleq \{m: k \in \Sm, m=1, \ldots, M\}\nonumber\\
\Tk 
&\triangleq \{m: k \in \Wm, m=1, \ldots, M\},
\end{align}
%=====
with $\Zk\cap \Tk =\emptyset$ and $\Zk\cup \Tk = \mathcal{M}^{\dl}\subset \mathcal{M}$, where $\mathcal{M}^{\dl}$ includes all FD APs and HD APs operating in DL.

Moreover, we introduce a pair of binary variables to indicate the group assignment for each UL UE $\ell$ and UL AP $m$ in our PZF combining scheme as
%==========
\begin{align*}
\dm^{\Zlup} = \begin{cases} \displaystyle 1 & \text {if } m \in \Zlup,\\ \displaystyle 0 & \text {otherwise},
\end{cases}
\qquad
\dm^{\Tlup} = \begin{cases} \displaystyle 1 & \text {if } m \in \Tlup,\\ \displaystyle 0 & \text {otherwise},
\end{cases}
\end{align*}
%==========
where $\Zlup$ and $\Tlup$ denote the set of indices of APs that assign $\ell$-th UL UE into $\Smu$ for ZF precoding and the set of indices of APs that assign $\ell$-th UL UE into $\Wmu$ for maximum ratio precoding, respectively, defined as:
%=====
\begin{align}
\Zlup &\triangleq \{m: \ell \in \Smu, m=1, \ldots, M\}\nonumber\\
\Tlup &\triangleq \{m: \ell \in \Wmu, m=1, \ldots, M\},
\end{align}
%=====
with $\Zlup\cap \Tlup =\emptyset$ and $\Zlup\cup \Tlup = \mathcal{M}^{\ul}\subset \mathcal{M}$, where $\mathcal{M}^{\ul}$ consists of all IBFD APs and HD APs operating in UL.
%----------------------

%%%%%%%%%%%%%%%%%%%%%%%%%%%%%%%%%%%%%
\begin{table*}[t]
\centering
\caption{SE parameters for different CF-mMIMO structures.}
\small
\begin{tabular}{|>{\centering\arraybackslash}m{1.4cm}|c|>{\centering\arraybackslash}m{14cm}|}
\hline
Structure &Direction & Expressions \\ \hline
%---------------------
\multirow{5}{*}{\vspace{-8em} NAFD } & \multirow{2}{*}{\vspace{-2em}DL} 
&$\Xi_k^\NAFD (\boldsymbol \theta, \qa) \triangleq 
    \sqrt{\rho_d}\sum\nolimits_{m \in \MM}\Big( \dm^{\Zk} a_m\theta_{mk} \gamdmk
+
\dm^{\Tk}N  a_m\theta_{mk} \gamdmk\Big)$
\\ \cline{3-3}
&                        
& $\Omega_k^\NAFD (\boldsymbol \theta, {\boldsymbol{\varsigma}}, \qa) \triangleq
    {\rho_d}
    \sum\nolimits_{m \in \MM} 
    \sum\nolimits_{k'\in\mathcal{K}_d}    
    \Big(\dm^{\Zk} \!\!a_m(\betamkd-\gamdmk)\frac{\theta_{mk'}^2 \gamdmkp}{N - \vert \Sm\vert} 
\!+\!
 \dm^{\Tk} N  a_m\theta_{mk'}^2
\gamdmkp\betamkd \Big) 
+\rho_u\sum\nolimits_{\ell\in\K_u}{ {\varsigma}_\ell}\betakldu$ 
\\ \cline{2-3}
                         & \multirow{3}{*}{\vspace{-6em}UL} 
&\hspace{2em}  $\Psi^\NAFD_\ell(\boldsymbol{\varsigma}, \qb, \boldsymbol{\alpha})\triangleq\sqrt{\rho_{u}} \Big(
 \sum\nolimits_{\substack{m\in\MM}} 
 \dm^{\Zlup}\alphml \sqrt{b_m  \varsigma_{\ell}} \gamuml +N
 \dm^{\Tlup}\alphml \sqrt{b_m \varsigma_{\ell}} {\gamuml}\Big)$ \\ \cline{3-3}
                         &                         
& $\Lambda_{\ell}^\NAFD(\boldsymbol{\varsigma}, \qb, \boldsymbol{\alpha})\triangleq
  \sum\nolimits_{m\in\MM}
  \bigg(\rho_{u}
		 \sum\nolimits_{\ell'\in\K_u}\!
   \Big(
   \dm^{\Zlup}
				\alphml^2 b_m \varsigma_{\ell'}
	\frac{\gamuml(\beta_{m\ell'}^{\ul}-
		\gamumlp)}{N-|\Smu|}
 +
	\dm^{\Tlup}
     N
     \alphml^2 b_m \varsigma_{\ell'}
	{\gamuml\beta_{m\ell'}^{\ul}}\Big)
  +
		\dm^{\Zlup}
  \alphml^2b_m\frac{\gamuml}{N-|\Smu|}+
  \dm^{\Tlup}N
		\alphml^2b_m\gamuml\bigg)$ \\ \cline{3-3}
                         &                         
& $\Phi_{\ell}^\NAFD(\boldsymbol{\varsigma}, \qa, \qb, \boldsymbol \theta, \boldsymbol{\alpha}) \triangleq 
 \rho_d 
		\sum\nolimits_{\substack{i\in\MM}}
		\sum\nolimits_{q\in\K_d}
		 a_{i}\theta_{iq}^2 \gamma_{iq}^{\dl}\Big(\sum\nolimits_{m\in\MM}
 \dm^{\Zlup}
\alphml^2 b_m 
\frac{\gamuml}{N - \vert \Smu\vert}
     \Big(\delta_i^{\Zqdl}  \frac{\beta_{mi}} {N - \vert \Sm\vert }   +  \delta_i^{\Tqdl}N\beta_{mi}  \Big)
+\dm^{\Tlup} N
 \alphml^2 b_m
 {\gamuml}
     \Big(\delta_i^{\Zqdl} \frac{\beta_{mi} } {N - \vert \Sm\vert }   +  \delta_i^{\Tqdl}N \beta_{mi} \Big)\Big)$ 
     \\ \hline
\multirow{5}{*}{\vspace{-9em}FD} 
%------------------------
& \multirow{2}{*}{\vspace{-2em}DL} 
& $\Xi_k^\FD (\boldsymbol \theta) \triangleq 
    \sqrt{\rho_d}\sum\nolimits_{m \in \MM}\Big( \dm^{\Zk} \theta_{mk} \gamdmk
+
\dm^{\Tk}N  \theta_{mk} \gamdmk\Big)$ \\ \cline{3-3}
                         &                 
& $\Omega_k^\FD (\boldsymbol \theta, {\boldsymbol{\varsigma}}) \triangleq
    {\rho_d}
    \sum\nolimits_{m \in \MM} 
    \sum\nolimits_{k'\in\mathcal{K}_d}    
    \Big(\dm^{\Zk} (\betamkd-\gamdmk)\frac{\theta_{mk'}^2 \gamdmkp}{N - \vert \Sm\vert} 
\!+\!
 \dm^{\Tk} N  \theta_{mk'}^2
\gamdmkp\betamkd \Big) 
+\rho_u\sum\nolimits_{\ell\in\K_u}{ {\varsigma}_\ell}\betakldu$  \\ 
\cline{2-3}
                         & \multirow{3}{*}{\vspace{-6em}UL} 
& $\Psi^\FD_\ell(\boldsymbol{\varsigma}, \boldsymbol{\alpha})\triangleq\sqrt{\rho_{u}} \Big(
 \sum\nolimits_{\substack{m\in\MM}} 
 \dm^{\Zlup}\alphml \sqrt{  \varsigma_{\ell}} \gamuml +N
 \dm^{\Tlup}\alphml \sqrt{ \varsigma_{\ell}} {\gamuml}\Big)$ \\ \cline{3-3}
                         &                 
                         &
$\Lambda_{\ell}^\FD(\boldsymbol{\varsigma},  \boldsymbol{\alpha})\triangleq
  \sum\nolimits_{m\in\MM}
  \bigg(\rho_{u}
		 \sum\nolimits_{\ell'\in\K_u}\!
   \Big(
   \dm^{\Zlup}
				\alphml^2  \varsigma_{\ell'}
	\frac{\gamuml(\beta_{m\ell'}^{\ul}-
		\gamumlp)}{N-|\Smu|}
 +
	\dm^{\Tlup}
     N
     \alphml^2  \varsigma_{\ell'}
	{\gamuml\beta_{m\ell'}^{\ul}}\Big)
  +
		\dm^{\Zlup}
  \alphml^2\frac{\gamuml}{N-|\Smu|}+
  \dm^{\Tlup}N
		\alphml^2\gamuml\bigg)$ \\ \cline{3-3}
                         &                         & 
$\Phi_{\ell}^\FD(\boldsymbol{\varsigma}, \boldsymbol \theta, \boldsymbol{\alpha}) \triangleq 
 \rho_d 
		\sum\nolimits_{\substack{i\in\MM}}
		\sum\nolimits_{q\in\K_d}
		 \theta_{iq}^2 \gamma_{iq}^{\dl}\Big(\sum\nolimits_{m\in\MM}
 \dm^{\Zlup}
\alphml^2 b_m 
\frac{\gamuml}{N - \vert \Smu\vert}
     \Big(\delta_i^{\Zqdl}  \frac{\beta_{mi}} {N - \vert \Sm\vert }   +  \delta_i^{\Tqdl}N\beta_{mi}  \Big)
+\dm^{\Tlup} N
 \alphml^2 
 {\gamuml}
     \Big(\delta_i^{\Zqdl} \frac{\beta_{mi} } {N - \vert \Sm\vert }   +  \delta_i^{\Tqdl}N \beta_{mi} \Big)\Big)$ 
     \\ \hline
%---------------------
\multirow{5}{*}{\vspace{-5em}HD} & \multirow{2}{*}{\vspace{-1em}DL} & 
 $\Xi_k^\HD (\boldsymbol \theta) \triangleq 
    \sqrt{\rho_d}\sum\nolimits_{m \in \MM}\Big( \dm^{\Zk} \theta_{mk} \gamdmk
+
\dm^{\Tk}N  \theta_{mk} \gamdmk\Big)$ \\ \cline{3-3}
                         &                         
                         & 
$\Omega_k^\HD (\boldsymbol \theta, {\boldsymbol{\varsigma}}) \triangleq
    {\rho_d}
    \sum\nolimits_{m \in \MM} 
    \sum\nolimits_{k'\in\mathcal{K}_d}    
    \Big(\dm^{\Zk} (\betamkd-\gamdmk)\frac{\theta_{mk'}^2 \gamdmkp}{N - \vert \Sm\vert} 
\!+\!
 \dm^{\Tk} N  \theta_{mk'}^2
\gamdmkp\betamkd \Big) $
\\ \cline{2-3}
                         & \multirow{3}{*}{\vspace{-4em}UL} 
& $\Psi^\HD_\ell(\boldsymbol{\varsigma}, \boldsymbol{\alpha})\triangleq\sqrt{\rho_{u}} \Big(
 \sum\nolimits_{\substack{m\in\MM}} 
 \dm^{\Zlup}\alphml \sqrt{  \varsigma_{\ell}} \gamuml +N
 \dm^{\Tlup}\alphml \sqrt{ \varsigma_{\ell}} {\gamuml}\Big)$ \\ \cline{3-3}
                         &                         
& $
   \Lambda_{\ell}^\HD(\boldsymbol{\varsigma},  \boldsymbol{\alpha})\triangleq
  \sum\nolimits_{m\in\MM}
  \bigg(\rho_{u}
		 \sum\nolimits_{\ell'\in\K_u}\!
   \Big(
   \dm^{\Zlup}
				\alphml^2  \varsigma_{\ell'}
	\frac{\gamuml(\beta_{m\ell'}^{\ul}-
		\gamumlp)}{N-|\Smu|}
 +
	\dm^{\Tlup}
     N
     \alphml^2  \varsigma_{\ell'}
	{\gamuml\beta_{m\ell'}^{\ul}}\Big)
  +
		\dm^{\Zlup}
  \alphml^2\frac{\gamuml}{N-|\Smu|}+
  \dm^{\Tlup}N
		\alphml^2\gamuml\bigg), 
$ \\ \cline{3-3}
                         &                        
    & $\Phi_{\ell}^\HD = 0$ \\ \hline
\end{tabular} \label{tab:NAFDFDHD}
\end{table*}
%%%%%%%%%%%%%%%%%%%%

\begin{proposition}~\label{Prop:SE:DLPZF}
With PZF precoding, the achievable DL SE at the $k$-th DL UE is given by
%-----------------------------------
\begin{align}~\label{eq:DL:SE}
&\mathcal{S}_{\dl,k}^{i} (\qa, \boldsymbol \theta, {\boldsymbol{\varsigma}}) =  \mu^i\left(1-\frac{\tau_t}{\tau_c}\right)
\log_2 \left(1  
    + \frac{(\Xi_k^i(\boldsymbol \theta, \qa))^2}{\Omega_k^i(\boldsymbol \theta, {\boldsymbol{\varsigma}}, \qa) + 1}
	\right),
\end{align}  
%---------------------- 
where $\qa \triangleq \{a_m\}$, $\boldsymbol{\theta}\triangleq \{\theta_{mk}\}$, and ${\boldsymbol{\varsigma}} \triangleq \{{\varsigma}_{\ell}\}, \forall m,k,\ell$. Moreover, $i\in\{\NAFD, \FD,\HD\}$ denotes different CF-mMIMO structures, while $\mu^\NAFD=\mu^\FD=1$ and $0\leq \mu^\HD\leq 1$. The corresponding expressions of $\Xi_k^i$ and $\Omega_k^i$ are provided in Table~\ref{tab:NAFDFDHD}.    
\end{proposition}

\begin{proof}
    The proof follows similar steps as in~\cite{Mohammad:JSAC:2023} and is thus omitted here.
\end{proof}

\begin{proposition}\label{Prop:SE:ULZFZF}
The achievable UL SE for $\ell$-th UL UE at the CPU with PZF design is given by
%=====================
\begin{align}\label{eq:UL:SE}
&\mathcal{S}_{\ul,\ell}^i (\qa,\qb, \boldsymbol{\varsigma}, \boldsymbol{\theta}, \boldsymbol{\alpha} )
	=\bar{\mu}^i\left(1-\frac{\tau_t}{\tau_c}\right)\nonumber\\
 &\hspace{2em}\times\log_2
	\Bigg(1+ 
 \frac{ \big(\Psi^i_\ell(\boldsymbol{\varsigma}, \qb , \boldsymbol{\alpha})\big)^2
		}
	{\Lambda_{\ell}^i(\boldsymbol{\varsigma}, \qb, \boldsymbol{\alpha})+
		\Phi_{\ell}^i(\boldsymbol{\varsigma}, \qa, \qb, \boldsymbol \theta, \boldsymbol{\alpha})
}\Bigg),
\end{align}  
%=====================
where $\qb \triangleq \{b_m\}$, $\bar{\mu}^\NAFD=\bar{\mu}^\FD=1$ and $\bar{\mu}^\HD = 1-\mu^\HD$. Moreover, $\Psi_{\ell}^i$, $\Lambda_{\ell}^i$, and $\Phi_{\ell}^i$, for $i\in\{\NAFD,\FD,\HD\}$ are provided in Table~\ref{tab:NAFDFDHD}.
\end{proposition}

\begin{proof}
    The proof follows similar steps as in~\cite{Mohammad:JSAC:2023} and is thus omitted here.
\end{proof}

The analytical results in  Proposition~\ref{Prop:SE:DLPZF} and~\ref{Prop:SE:ULZFZF}  indicate that the interference terms in IBFD and NAFD CF-mMIMO networks are higher compared to their HD counterparts. However, NAFD CF-mMIMO (unlike IBFD networks) offers the advantage of managing these interference terms by selectively activating APs for DL and UL operations, thus providing better interference control opportunities.

\begin{Remark}
\textbf{Conventional precoding/decoding designs:} 
The expressions in Proposition~\ref{Prop:SE:DLPZF} and Proposition~\ref{Prop:SE:ULZFZF} are comprehensive, encompassing different combinations of precoding designs at UL and DL. For instance, by setting $\dm^{\Zlup}=\dm^{\Zk}=0$ and $\dm^{\Tlup}=\dm^{\Tk}=1$, $\forall m\in\MM$, $\mathcal{S}_{\dl,k}^{\NAFD}$ and $\mathcal{S}_{\ul,\ell}^\NAFD$ reduce to the results in~\cite{Mohammad:JSAC:2023}, while $\mathcal{S}_{\dl,k}^{\FD}$ and  $\mathcal{S}_{\ul,\ell}^\FD$ reduce to the results in~\cite{tung19ICC}. Moreover, when $\dm^{\Zlup} = \dm^{\Zk} = 1$ and $\dm^{\Tlup} = \dm^{\Tk} = 0$ for all $m \in \MM$, and $\vert \Sm \vert = K_d$ with $N > K_d$, then $\mathcal{S}_{\dl,k}^{\FD}$ and $\mathcal{S}_{\ul,\ell}^{\FD}$ are the same as the results in~\cite{Interdonato:TWC:2020} and~\cite{Zhang:TCOM:2021}, respectively, referring to the FZF precoding and combining designs. 
\end{Remark}

\textbf{{Small-cell networks:}} For IBFD small-cell systems, where both UL and DL UEs are served via IBFD APs, each UL (DL) UE is served by only one AP. AP selection is performed based on a specific criterion (normally strength of the received signal) in UL and DL. In the DL direction, AP selection is performed by the UEs, while in the UL direction, a single AP locally decodes the received signal from a specific UE. This decentralized approach ensures that the network operates in a fully distributed manner. The macro diversity achieved by selecting the best AP from multiple options has the potential to make IBFD small-cell systems competitive with conventional cellular massive MIMO systems that utilize larger cells~\cite{emil20TWC}. 

Now, we can invoke a similar approach as in~\cite{ngo16}, to obtain UL and DL capacity bounds for IBFD small-cell networks. To this end, the following modifications in the system model are required: 1) Denote the home cell index by $m$ and assume that different cells are using mutual orthogonal pilots;\footnote{For the case of pilot contamination, we refer readers to \cite[Chapter 4]{ngo16}.} 2) Replace $\K_d$ and $\K_u$ with $\mathcal{K}_d^m$ and $\mathcal{K}_u^m$, respectively, which denote the sets of DL and UL UEs served by AP $m$.  Now, following these assumptions, the capacity bounds for UE $k$ in DL and UE $\ell$ in UL at the home cell $m$ can be expressed by~\eqref{eq:DL:SE} and~\eqref{eq:UL:SE}, where the desired DL and UL signals are reduced to $\Xi_{m,k}^\FD (\boldsymbol \theta)={\rho_d}\big( \dm^{\Zk} \theta_{mk} \gamdmk + \dm^{\Tk}N  \theta_{mk} \gamdmk\big)$ and $\Psi^\FD_{m,\ell}(\boldsymbol{\eta})={\rho_{u}} \big(\dm^{\Zlup}  \eta_{m\ell} \gamuml +N  \dm^{\Tlup}  \eta_{m\ell} {\gamuml}\big)$, respectively, where $\eta_{m\ell}$ denotes the UL power control coefficient for UL UE $\ell$ at $\mathcal{K}_u^m$. Moreover, the variance of the effective noise, which includes independent terms of inter-/intra-cell interference and SI for the UL, as well as inter-/intra-cell interference and inter-UE interference for the DL, can be easily obtained. These variances follow the same expressions as those in the denominators of the SINRs in~\eqref{eq:DL:SE} and~\eqref{eq:UL:SE}, which are omitted here for the sake of brevity.

Finally, capacity bounds for HD small-cell networks, can be found in~\cite[Table 4.1]{ngo16}. From the analytical results for small-cell networks, we observe that the lack of coherent transmission in the DL and coherent decoding in the UL significantly reduces the effective SINR compared to CF-mMIMO counterparts, resulting in smaller SE values. 

%%%%%%%%%%%%%%%%%%
\vspace{-0.5em}
\subsection{Case Study and Simulation Results}
%%%%%%%%%%%%%%%%%%%%%%%%%%%%%%%%
To compare the EE of the different CF-mMIMO systems, we consider a CF-mMIMO network, where the APs and UEs are randomly distributed in a square of $0.5 \times 0.5$ km${}^2$, whose edges are wrapped around to avoid the boundary effects. The distances between adjacent APs are at least $50$ m\cite{emil20TWC}. We model the large-scale fading coefficients $\beta_{mk}$ as~\cite{emil20TWC}
%-----------------------------------------------------
\begin{align}\label{fading:large}
\beta_{mk} = 10^{\frac{\text{PL}_{mk}^d}{10}}10^{\frac{F_{mk}}{10}},
\end{align}
%-----------------------------------------------------
where $10^{\frac{\text{PL}_{mk}^d}{10}}$ represents the path loss, and $10^{\frac{F_{mk}}{10}}$ represents the shadowing effect with $F_{mk}\in\mathcal{N}(0,4^2)$ (in dB).  Here, $\text{PL}_{mk}^d$ (in dB) is given by  \cite{emil20TWC}
%-----------------------------------------------------
\begin{align}\label{PL:model}
\text{PL}_{mk}^d = -30.5-36.7\log_{10}\left(\frac{d_{mk}}{1\,\text{m}}\right),
\end{align}
%-----------------------------------------------------
and the correlation among the shadowing terms from the AP $m, \forall m\in\mathcal{M}$ to different UEs $k$ ($\forall k\in\KU/\KD$) is expressed as:
%-----------------------------------------------------
\begin{align}\label{corr:shadowing}
\mathbb{E}\{F_{mk}F_{jk'}\} \triangleq
\begin{cases}
 4^22^{-\delta_{kk'}/9\,\text{m}},& \text{if $j=m$}\\
 0, & \mbox{otherwise},
\end{cases}, \forall j\in\mathcal{M},
\end{align}
%-----------------------------------------------------
where $\delta_{kk'}$ is the physical distance between UEs $k$ and $k'$. Regarding the power consumption parameters,  we use the parameters of the power consumption in \cite{Emil:TWC:2015:EE,Hien:TGCN:2018,bashar19TGCN}, which are shown in Table~\ref{tab:PowerconsumptionParameter}.

%======================================
\begin{table*}[t]
\vspace{0.5em}
	\caption{CF-mMIMO simulation setup parameters.} 
	\vspace{-0.5em}
	\centering 
	\begin{tabular}{|c | c |}
		\hline
		\textbf{Parameter} & \textbf{Value}  \\ [0.5ex]
		\hline
Traffic-dependent backhaul power  ($\Pbtm, \forall m$)~\cite{Emil:TWC:2015:EE,Hien:TGCN:2018} & $0.25$ W/(Gbits/s) \\
		\hline
Power amplifier efficiency at the APs  ($\epsilon_{\AP}$)~\cite{Hien:TGCN:2018} & $0.4$   \\
		\hline
Power amplifier efficiency at the UL UEs  ($\epsilon_{\UEk}, \forall k\in\KU$)~\cite{bashar19TGCN} & $0.3$   \\
		\hline
Fixed power consumption UL and DL UE  ($\PULUE, \PAPm$)~\cite{bashar19TGCN} & $0.1$ W \\
\hline	
	\end{tabular}
	\label{tab:PowerconsumptionParameter}
	\vspace{0.0em}
\end{table*}
%=============================================

\begin{figure}[t]
	\centering
	\vspace{0em}
	\includegraphics[width=0.45\textwidth]{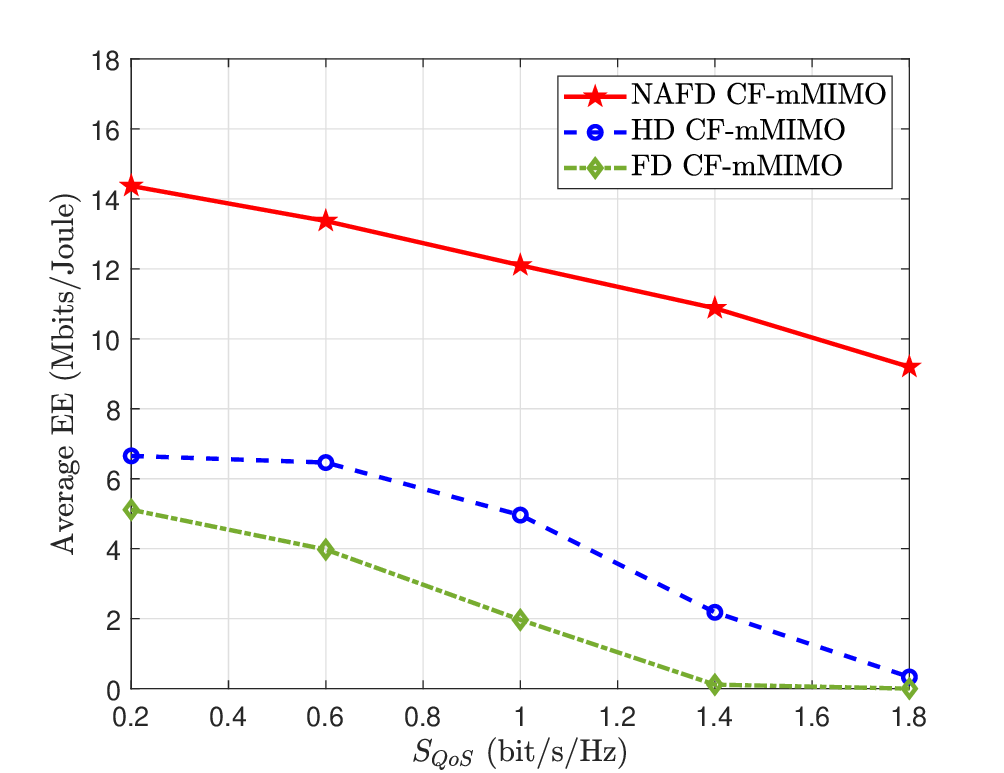}
	\vspace{-0.3em}
	\caption{Average EE versus SE requirements of the UL and UL UEs in a CF-mMIMO network with different structures ($M=40, K_{\dl}=K_{\ul}=4$, $\sigma_{\SI}^2/\sigma_n^2 = 50$ dB). Here, $\sigma^2_{\SI}$ reflects the strength of the residual SI after employing SIC techniques and $\sigma^2_n$ is the noise power.}
	\vspace{0.5em}
	\label{fig:FigEE}
\end{figure}

Figure~\ref{fig:FigEE} shows the EE of CF-mMIMO systems with NAFD, IBFD, and HD structures versus the individual SE requirement $\mathcal{S}_{QoS}$ for DL and UL UEs. Results for NAFD were obtained through the joint optimization of the operational mode of the APs and power control design in both the UL and DL, as well as the optimization of large-scale fading coefficients in the UL. However, in the IBFD and HD scenarios, the power control design and optimization of large-scale fading coefficients were carried out.  We can observe that both  HD and IBFD schemes fail to satisfy the SE requirement for $\mathcal{S}_{QoS} \geq 1.8$ bit/s/Hz,  while the NAFD scheme still meets the individual SE requirements of the system. This result shows that the NAFD scheme can be much more energy-efficient than the HD and IBFD schemes while achieving a high SE target.

\textbf{Takeaway Messages: } NAFD CF-mMIMO represents the most comprehensive form of CF-mMIMO networks, encompassing all operating modes at the APs. In its general form, a mixture of IBFD and HD APs (operating either on UL or DL direction) can simultaneously accommodate asymmetric UL and DL rate requirements across the network. This innovative structure effectively manages interference and transmissions over fronthaul links, resulting in a more energy-efficient design compared to traditional IBFD and HD CF-mMIMO networks.

%%====================================================
%%================================================================

%%%%%%%%%%%%%%%%%%%%%%%%%%%%%%%%%%%%%%%%%%%%%%%%%%
\section{Interplay Between NAFD CF-mMIMO and Emerging Technologies}\label{sec:NAFDcellfreeApp}
%%%%%%%%%%%%%%%%%%%%%%%%%%%%%%%%%%%%%%%%%%%%%%%%%%%
From another point of view, the NAFD conceptual framework can be expanded to facilitate the implementation of dual-functional applications, exemplified by SWIPT, wireless surveillance, and ISAC. To elaborate further within these systems, a distributed approach can be employed wherein two distinct tasks are assigned to two different groups of APs within the CF-mMIMO networks.  For dual-functional applications, it is crucial to strategically allocate resources between the two tasks to enhance and meet the performance requirements of both. This exercise is challenging due to the presence of inter-system interference. The NAFD CF-mMIMO concept offers the opportunity to replace the traditional serving infrastructures in the coverage/service area with low-power APs. This replacement results in reduced transmit power levels and, consequently, lower interference. Additionally, by appropriately selecting the AP operation mode as an additional DoF in system design, inter-system interference can be managed more efficiently. Furthermore, coherent transmission from multiple APs, each responsible for a specific task, can further enhance the performance of the corresponding tasks. This approach enables a more efficient and distributed execution of multiple tasks in the applications mentioned below, outperforming their co-located counterparts.

%%%%%%%%%%%%%%%%%%%%%%%%%%%%%%%%%%%%%%%%%%%%%%%%%%
\subsection{NAFD CF-mMIMO and SWIPT}
%%%%%%%%%%%%%%%%%%%%%%%%%%%%%%%%%%%%%%%%%%%%%%%%%%%
The application of wireless power transfer in scenarios involving energy-constrained devices, like IoT devices, overlaid on communication networks such as cellular systems, emerges as a promising scenario for SWIPT. RF energy harvesting offers significant advantages, including wireless functionality, accessibility through transmitted energy sources (such as TV/radio broadcasters, mobile BSs/APs, and handheld radios), cost-effectiveness, environmental friendliness, and the potential for compact implementation~\cite{Ponnimbaduge:tut:2018}. In an overlaid network, IoT devices exploit all the communication RF signals for energy harvesting. However, due to the substantial signal attenuation in wireless communication channels, the RF energy harvested by the devices is typically constrained~\cite{Clerckx:JSAC:2019}. As a result, the implementation of the network becomes crucial in order to ensure the efficient management of available resources and address the energy requirements of IoT devices for long-term operation. This is essential for achieving sustainable and scalable IoT networks. 

Relying on the high degrees of macro diversity, CF-mMIMO enables the minimization of propagation path loss. This is achieved with a high probability, as there are numerous APs in close proximity to the IoT devices, termed as energy UEs (E-EUs). The synergistic deployment of CF-mMIMO and SWIPT has been recently advocated by various authors, as discussed in~\cite{Wang:JIOT:2020,Zhang:IoT:2022,mohammadi2023cell}. Under different network conditions, the authors proposed power control strategies to  optimize the performance of CF-mMIMO IoT networks, like provisioning fairness among the IoT devices or communication UEs (C-UEs).  However, even  with an optimal power control design, all these designs would still suffer from the fundamental limitation of how to simultaneously increase both the SE and HE for separate C-UEs and E-UEs. This is due to the inefficient use of available resources, as DL wireless power transfer towards the E-UEs and  DL (UL) wireless information transfer towards the C-UEs (APs) occur over orthogonal time slots. A straightforward approach to enhance both the SE and HE would be to  deploy a large number of APs, but this is not energy efficient due to the large fronthaul burden and transmit power requirements~\cite{ngo18TGN}. \textit{So, the question arises: Is there a practical architecture that supports SWIPT in CF-mMIMO IoT networks?}

\begin{figure}[t]
	\centering
	\includegraphics[width=0.5\textwidth]{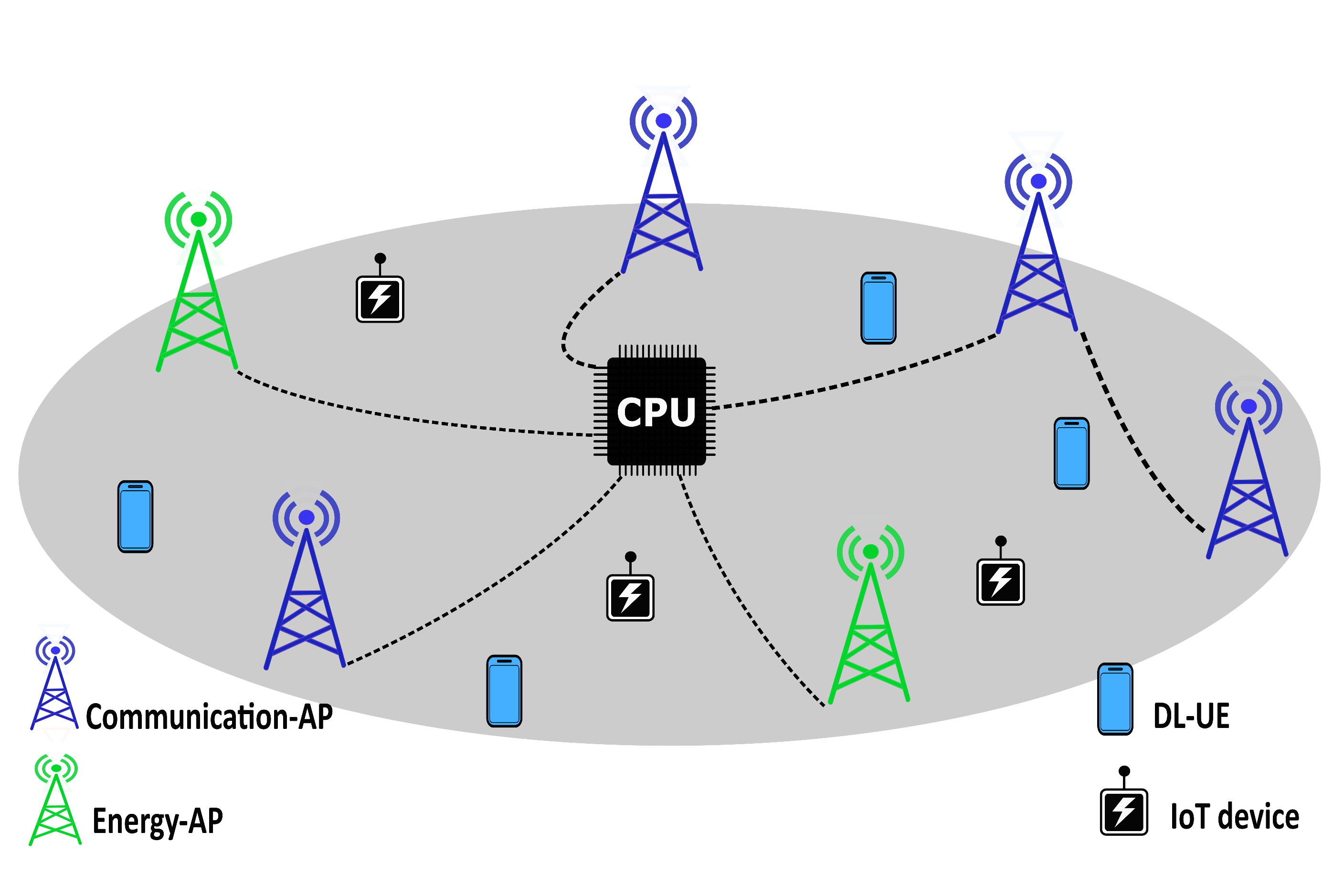}
	\caption{Diagram of an NAFD CF-mMIMO SWIPT system with $K_E$ E-UEs in the charging zone and $K_d$ C-UEs. The HD E-APs serve the E-UEs, while the HD DL-APs coherently transmit to the C-UEs.} 
 \vspace{1em}
	\label{fig:SWIPT}
\end{figure}

Borrowing the conceptual design of NAFD in CF-mMIMO networks, we can cluster the APs and implement linear processing schemes to simultaneously support C-EUs and E-EUs without causing interference between them as shown in Fig.~\ref{fig:SWIPT}. In our recent work~\cite{mohammadi2023cell}, we proposed a novel network architecture that jointly designs the AP operation mode selection and power control strategy to maximize the harvested energy at the E-EUs under the constraints on per C-UE SE (denoted by $\SEQoS$ as the minimum SE required by the $k$th C-UE)  and per E-UE harvested energy (denoted by $\Gamma_{\ell}$ as a minimum power requirement at the $\ell$th EUs). Specifically, relying on the long-term CSI, the APs are divided into communication APs (termed as C-APs) and energy transmission APs (termed as E-APs), which  simultaneously serve C-UEs and E-UEs over the whole time slot period. While this new architecture provides E-UEs with an opportunity to harvest energy from all APs, it also creates increased interference at the C-UEs due to  concurrent transmissions of E-APs. To deal with this problem, local partial zero-forcing  precoding and protective maximum ratio transmission can be deployed at the C-APs and E-APs, respectively, to guarantee full protection for the C-UEs against  energy signals intended for E-EUs. 

%%%%%%%%%%%%%%%%%%%%%%%%%%%%%%%%%%%%%%%%%%%%%%%%%%
%%%%%%%%%%%%%%%%%%%%%%%%%%%%%%%%%%%%%%%%%%%%%%%%%%%

\begin{figure}[t]
	\centering
        \includegraphics[width=0.45\textwidth]{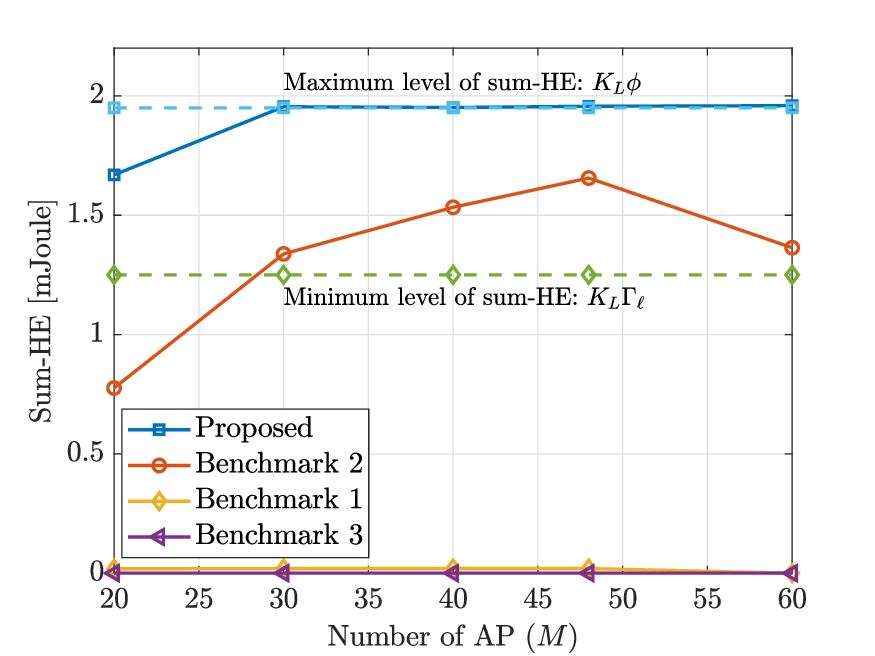}
	\vspace{-0.3em}
	\caption{Impact of the number of APs on the average of sum harvested power for $K_d=3$ C-UEs and $K_E=5$ E-UEs ($\Gamma_{\ell}=250$ $\mu$W,  and $\SEQoS = 18$ bit/s/Hz).}
	\vspace{0.4em}
	\label{fig:FigSWIPT}
\end{figure}

To demonstrate the effectiveness of the proposed architecture for SWIPT NAFD CF-mMIMO IoT networks in~\cite{mohammadi2023cell}, we consider three benchmarks as: \textit{i) Benchmark 1},  random AP operation mode selection without power control; \textit{ii) Benchmark 2}, random AP operation mode selection with power control; and \textit{iii) Benchmark 3}, SWIPT with orthogonal transmission as in~\cite{Wang:JIOT:2020,Zhang:IoT:2022}. Moreover, we  consider a non-linear energy harvesting model with the maximum output DC power $\phi=0.39$ mW. Figure~\ref{fig:FigSWIPT} shows the average sum harvested energy (HE) achieved by the proposed scheme in~\cite{mohammadi2023cell} and the three benchmark schemes as a function of the number of APs, for given fixed number of antenna services, i.e., $MN=480$. We observe that our proposed scheme yields substantial performance gains in terms of energy harvesting efficiency over the benchmarks, especially when the number of APs is small. This highlights the importance of distributed implementation as well as joint AP operation mode selection and power control design in the proposed architecture, as Benchmarks 1 and 3 cannot meet the network requirements.

%%%%%%%%%%%%%%%%%%%%%%%%%%%%%%%%%%%%%%%%%%%%%%%%%%%
%%%%%%%%%%%%%%%%%%%%%%%%%%%%%%%%%%%%%%%%%%%%%%%%%%%

%%%%%%%%%%%%%%%%%%%%%%%%%%%%%%%%%%%%%%%%%%%%%%%%%%
\subsection{NAFD CF-mMIMO and Surveillance}
%%%%%%%%%%%%%%%%%%%%%%%%%%%%%%%%%%%%%%%%%%%%%%%%%%%
It is widely understood that recent advances in wireless communication systems over the past decades have enabled the establishment of high-speed connections virtually anywhere and at any time between any pair of devices. These technologies not only enhance the convenience of our daily lives but also create opportunities for applications and innovations in diverse areas, spanning from education, health, agriculture to the industry. Nonetheless, the accessibility of these wireless systems also exposes them to potential misuse by malicious entities, enabling activities such as illegal actions, (cyber)-crime, and posing risks to public safety. To this end, there is a growing need for legitimate parties, such as governmental institutes, to monitor and even disable  the  malicious communication links. The \textit{proactive monitoring technique}, also known as proactive eavesdropping, has garnered significant attention in recent years~\cite{Zhang:TWC:2017}. In this approach, the authorized monitor operates in an IBFD mode, wherein it observes suspicious links and concurrently transmits a jamming signal. The deliberate interference with the suspicious receiver results in a degradation of the data transmission rate of the untrusted link. In practice, traditional physical-layer proactive monitoring of suspicious communication links relying on an IBFD monitor is challenging, mainly due to 1) the distributed deployment of the multiple  suspicious users over a geographically wide area and 2) acquisition of the CSI of all untrusted links at the monitor  for system level design. More specifically, it is not feasible to address the direct monitoring of multiple  untrusted pairs by using  only one single monitor. On the other hand, by relying on the instantaneous CSI knowledge, the system strategies need to be re-executed  in order to adapt to rapidly changing  small-scale fading in both time and frequency. \textit{So, the key question is: What is a practical architecture to monitor and/or invert multiple  distributed untrusted communication links?}

\begin{figure}[t]
	\centering
  %\vspace{-8em}
	\includegraphics[width=0.51\textwidth]{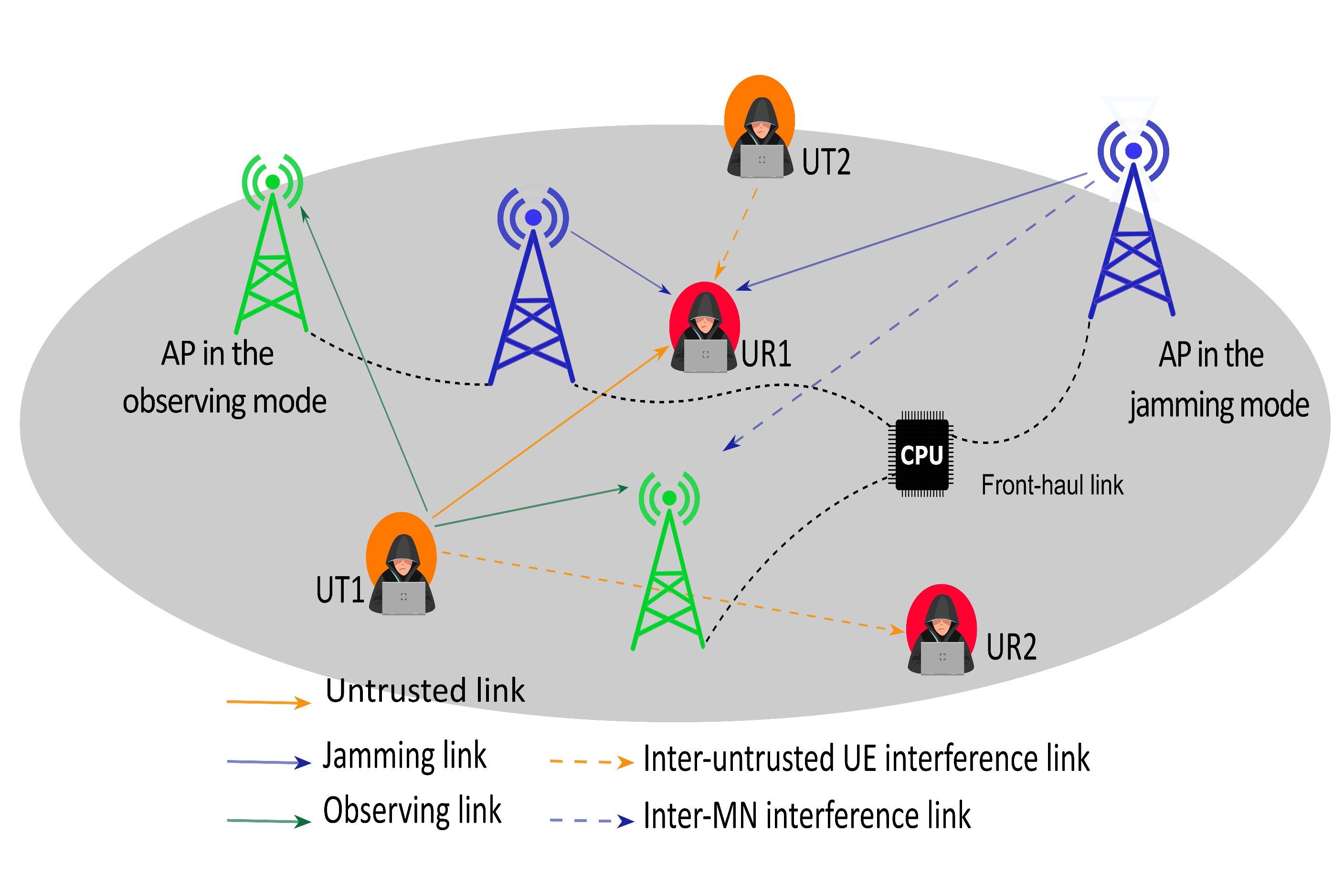}
	\vspace{-1em}
	\caption{Diagram of an NAFD CF-mMIMO surveillance system model with the assigned MNs in observing mode and jamming mode  along with the received desired and interference signals at a typical untrusted pair (UT $1$ and UR $1$).} 
 \vspace{1em}
	\label{fig:Surv}
\end{figure}

Recently, the innovative concept of NAFD CF-mMIMO has been effectively harnessed to introduce a  promising method for proactive monitoring~\cite{mobini2023cell,Mobini:IoT:2024}.  The proposed NAFD CF-mMIMO surveillance system consists of numerous legitimate APs acting as monitoring nodes (MNs)  dispersed across an area. These MNs collaborate in a coordinated manner to perform surveillance over multiple distributed untrusted/malicious pairs. Therefore, almost any untrusted pair  is expected to be within the coverage area of several MNs. This configuration avails of high macro-diversity gain and low path loss, leading to improved observation channel rates and the degradation of the untrusted links. In this system, as shown in Fig.~\ref{fig:Surv}, it is feasible to realize virtual IBFD proactive monitoring by only utilizing HD MNs. In particular,  two distinct types of MNs are taken into account: 1) A subset of MNs is dedicated solely to observing the untrusted transmitter; 2) The remaining MNs collaborate to jam the untrusted receiver. Accordingly, CF-mMIMO surveillance  can reap all benefits of NAFD (cost-effective HD MNs and less sensitive to residual SI) and provide successful surveillance and intervention across all suspicious pairs. Importantly, thanks to the channel hardening property of  CF-mMIMO systems, the observing vs. jamming mode assignments for the MNs and other system tasks, such as transmit power control, can be dynamically adjusted for maximizing the overall monitoring performance based on only long-term CSI. 

%%%%%%%%%%%%%%%%%%%%%%%%%%%%%%%%%%%%%%%%%%%%%%%%%%
%%%%%%%%%%%%%%%%%%%%%%%%%%%%%%%%%%%%%%%%%%%%%%%%%%%

\begin{figure}[t]
	\centering
	\includegraphics[width=0.45\textwidth]{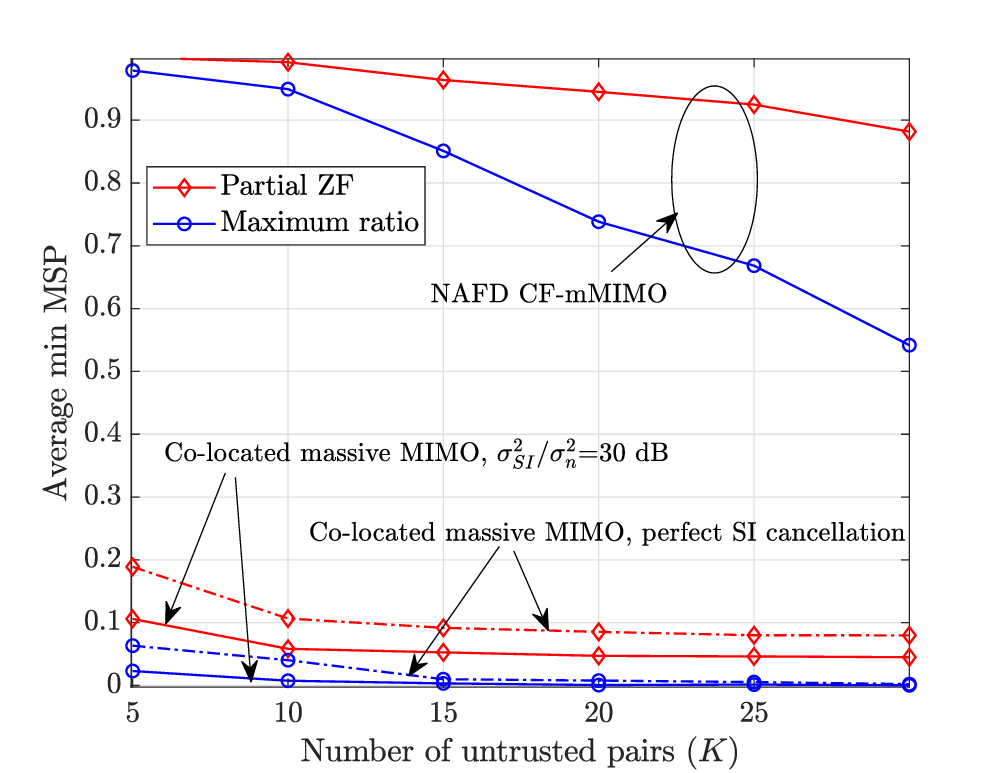}
 \vspace{0em}
	\caption{Average minimum monitoring successful probability (MSP) versus the number of untrusted links, $K$, where $M=40$ and $\Ntx=6$. Here, $\sigma^2_{\SI}$ reflects the strength of the residual SI after employing SIC techniques and $\sigma^2_n$ is the noise power. }
 \vspace{0.7em}
	\label{fig:Fig5}
\end{figure}

To compare the monitoring performance of the NAFD CF-mMIMO with conventional IBFD co-located massive MIMO surveillance system, we carry out an example.   Figure~\ref{fig:Fig5} shows the minimum monitoring successful probability   of the NAFD CF-mMIMO  and co-located surveillance systems relying on both maximum ratio and partial zero-forcing combining designs versus the number of untrusted pairs, $K$. We observe that an NAFD CF-mMIMO surveillance system yields huge performance gains over a conventional co-located surveillance system, even under the idealized assumption of having perfect SIC for the co-located surveillance system. Hence, it is noteworthy that  \emph{surveillance systems can potentially benefit much more from the inherently higher macro-diversity gain in an NAFD CF-mMIMO network, rather than from the higher array gain in co-located massive MIMO networks.}

%%%%%%%%%%%%%%%%%%%%%%%%%%%%%%%%%%%%%%%%%%%%%%%%%%
\subsection{NAFD CF-mMIMO and ISAC}
%%%%%%%%%%%%%%%%%%%%%%%%%%%%%%%%%%%%%%%%%%%%%%%%%%%
Future wireless networks are expected to deliver site-specific services to surrounding UEs in addition to conventional communication-only functionality. This evolution, spurred by recent advancements in signal processing and communications, advocates for the seamless integration of radio sensing functionalities into B5G/6G wireless networks with cost-effectiveness and expeditious implementation. As a result, prospective wireless networks are poised to capture and assess the surrounding environment, paving the way for advanced location-aware services including autonomous vehicles, environmental monitoring, wearable IoT devices, as well as indoor services, such as human activity recognition. This research paradigm is commonly referred to as ISAC in the literature~\cite{Liu:JSAC:2022}. ISAC possesses two primary advantages compared to dedicated sensing and communication functionalities: \textit{i)} Integration gain enables the efficient utilization of congested resources; \textit{ii)}  Coordination gain allows for the balance of dual-functional performance and mutual assistance, adding an intriguing dimension to its capabilities.

%%%%%%%%%%%%%%%%%%%%%%%%%%%%%%%%%%%%%%%%%%%%%%%%%%%
%%%%%%%%%%%%%%%%%%%%%%%%%%%%%%%%%%%%%%%%%%%%%%%%%%%

Within the existing body of literature, two distinct ISAC architectures have been identified. The first, known as the coexisting radar and communication architecture, involves the separate positioning of the radar and communication BS. In this setup, both the sensing beam and communication beam are generated independently~\cite{Liu:JSAC:2022}. In contrast, the second architecture is termed as dual functional radar-communication architecture. In this configuration, the radar and communication BS are colocated, forming an integrated ISAC BS with shared hardware components~\cite{Liu:JSAC:2022}. This integration enables the joint generation of both the sensing beam and the communication beam. Both architectures share a common aspect: radar functionality is based on IBFD operation for the transmission and reception of sensing signals.

Recently, the concept of a network ISAC has emerged, where multiple ISAC transceivers, such as distributed BSs in the cloud RAN, can underpin distributed radar sensing and coordinated wireless communication~\cite{Zhang:MVT:2021,Huang:TVT:2022,Behdad:GC:2022}. The network ISAC offers various benefits. From the sensing perspective, it can expand the surveillance area, improve sensing coverage, and capture richer sensing information. From a communication perspective, advanced coordinated multi-point transmission/reception techniques can be implemented to mitigate or even utilize CCI among different C-UEs and properly control interference between sensing and communication signals. \textit{Notably, a network ISAC presents a viable solution to tackle the IBFD issue in single-cell ISAC by enabling some BSs to function as dedicated sensing receivers. In light of this, the NAFD CF-mMIMO emerges as an ideal host for network ISAC.} 

\begin{figure}[t]
	\centering
  \vspace{0em}
	\includegraphics[width=0.5\textwidth]{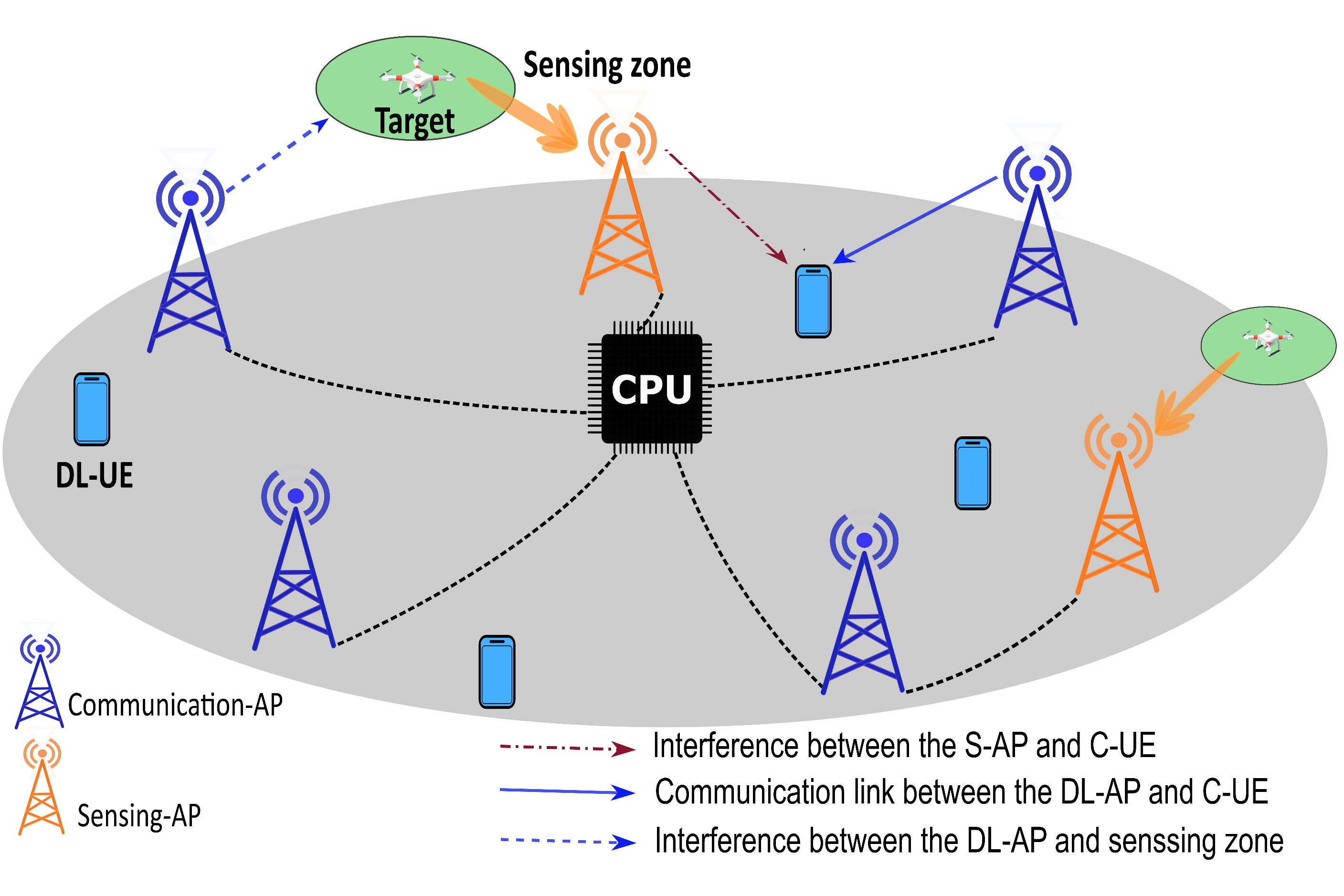}
	\vspace{-1em}
	\caption{Diagram of an NAFD CF-mMIMO ISAC system with multiple sensing zones and C-UEs. The S-APs transmit sensing signals towards the sensing zones, while the HD DL-APs coherently transmit to the C-UEs.} 
 \vspace{0.5em}
	\label{fig:ISAC}
\end{figure}

Expanding on the NAFD concept, an ISAC system, shown in Fig.~\ref{fig:ISAC}, with dynamic AP operation mode selection was proposed in~\cite{mohamed2023cell}. The APs' operation mode is designed to maximize the minimum SE of the DL C-UEs, while satisfying the sensing requirement to detect the presence of a single target in a certain location.  Relying on the long-term CSI, the APs are divided into DL-APs and sensing APs (S-APs) to  support DL communication and sensing operations simultaneously. With the aim of providing fairness among the C-UEs, while adhering to a prescribed mainlobe-to-average-sidelobe ratio level for target detection, the AP operation modes are determined according to a greedy algorithm followed by a power control design at the DL-AP and S-AP. As an intriguing avenue for future exploration, one may consider the joint optimization of AP mode selection and power control design. In a broader context, this model can be extended to accommodate both UL and DL UEs simultaneously and over the same frequency. Concurrently, a group of APs can be strategically assigned to receive and process sensing observations in a distributed manner. 

%%%%%%%%%%%%%%%%%%%%%%%%%%%%%%%%%%%%%%%%%%%%%%
%%%%%%%%%%%%%%%%%%%%%%%%%%%%%%%%%%%%%%%%%%%%%%

%%%%%%%%%%%%%%%%%%%%%%
\subsection{Integration of Miscellaneous Technologies into NAFD CF-mMIMO}
%%%%%%%%%%%%%%%%%%%%%%
Advanced new technologies can be integrated into NAFD CF-mMIMO networks to further enhance the system performance. Moreover, for dual-functional applications, discussed in previous subsections, these technologies can facilitate the implementation of different tasks. RISs, beyond diagonal RISs (BD-RISs), and stacked RISs are potential technologies that can be incorporated into NAFD CF-mMIMO systems to direct signals to locations subject to blockage or to provide specific UEs with additional copies of the signals. In ISAC applications, these additional copies can enhance the sensing and/or communication capability, while in SWIPT systems, they can boost the level of harvested energy for energy UEs. In~\cite{Hua:WCNC:2024}, the authors investigated the performance of a CF-mMIMO SWIPT system with a fixed-phase designed BD-RIS. They examined the improvements in harvested energy achieved by BD-RIS compared to the system without BD-RIS. However, this line of study is still in its infancy, with several interesting directions yet to be explored.

Air-ground communications, wherein UAVs share the same spectrum resources with conventional ground UEs, can also benefit from the NAFD CF-mMIMO structure. The coexistence of aerial and ground UEs will be a common feature of future air-ground communication systems. In such an environment, supporting the simultaneous large UL traffic of UAVs and large DL traffic of ground UEs poses significant challenges in system-level design. The line-of-sight links between UAVs and ground UEs would inevitably cause severe aerial-ground interference to the concurrent transmissions of ground UEs over the same frequency band, deteriorating the system performance. NAFD CF-mMIMO provides the opportunity to benefit from a high degree of macro-diversity for UAVs, while managing air-to-ground interference and addressing UL and DL traffic asymmetry. Wan~\ettall~\cite{Wan:TWC:2024} considered a cell-free radio access network with NAFD to support air-ground communications. They studied the problem of AP clustering and AP mode selection to maximize both the UL and DL sum-rates. Nevertheless, there are still several open problems and challenges in this area, making it an interesting direction for future research.    

New antenna technologies like holographic and fluid antennas~\cite{Huang:WC:2020,Shojaeifard:IPC:2022} represent promising advancements in NAFD CF-mMIMO systems. These architectures, already deployed in cellular and satellite communications, offer significant potential to reduce the complexity and  deployment costs while maintaining high performance standards. However, their integration into CF-mMIMO systems remains largely unexplored. Specifically, their ability to manage intra-user/AP interference in NAFD CF-mMIMO networks, as well as their potential in dual-function systems like SWIPT and ISAC-enabled CF-mMIMO, could greatly enhance the network design capabilities.
%%%%%%%%%%%%%%%%%%%%%%%%%%%%%%%%%%%%%%%%%%%%%%%%%%
\section{Key Open Challenges and Future Research Directions}\label{sec:openchallenges}
%%%%%%%%%%%%%%%%%%%%%%%%%%%%%%%%%%%%%%%%%%%%%%%%%%%
In this section, we highlight some open challenges in NAFD CF-mMIMO networks that need to be taken into consideration. We describe these challenges and introduce potential future research directions in this regard.\\
\textbf{Scalability:}
The demand for data throughput and the quantity of UL and DL UEs (network densification) are expected to persistently increase. In theory, addressing these substantial demands may involve the continuous addition of more APs~\cite{Andrews:TCOM:2011}. However, critical questions arise: What is the fundamental limit as the network density approaches infinity? Can we simply keep increasing the number of APs and UEs, while maintaining a relatively constant SE for each UE? These questions are crucial because, as the UE density increases up to a limit, resulting in small distances between the APs and UEs, the path loss exponent can be smaller than $2$, leading to an increase in the interference power. We anticipate that the available high array gain  will help NAFD CF-mMIMO to maintain a nearly constant SE per DL/UL UE up to a certain level of network densification. However, beyond this threshold, new challenges emerge, particularly concerning the escalating computational complexity, inter-AP interference, and fronthaul capacity, which grow rapidly with the number of UEs, rendering NAFD CF-mMIMO unscalable. A crucial question to delve into is: \emph{What is the limiting network density beyond which the advantages of NAFD CF-mMIMO start to diminish, and how to address this issue?} 

\textbf{Inter-AP Interference:} The UL performance of an NAFD CF-mMIMO is predominantly hindered by inter-AP interference. Although the approach of inter-AP channel estimation and subsequent interference subtraction was explored in~\cite{Jiamin:TWC:2021}, this method introduces a substantial overhead. Furthermore, it adopts a centralized scheme at the CPU that demands significant fronthaul capacity. More importantly, the scheme proposed in~\cite{Jiamin:TWC:2021} relies on pre-assigned AP transmission modes. However, when contemplating AP mode selection, the CPU is required to obtain channel estimates between all pairs of APs, resulting in a significant rise in complexity. Additionally, if orthogonal pilot sequences are employed by the APs, the DL pilot dimension would consume the entire coherence block. On the contrary, the presence of co-pilot APs in the network, leading to pilot contamination, is inevitable unless careful pilot assignment between the APs is undertaken. Therefore, strategic pilot assignment among the APs holds crucial importance.

\textbf{Accuracy of Hardening Bounding Technique:} The hardening bounding technique, extensively used in CF-mMIMO literature, is intimately effective under specific assumptions regarding the propagation environments, such as independent Rayleigh fading channels. However, in practical scenarios, especially within ultra-dense networks, the independence of channels from a given UE to different APs may not be guaranteed. Consequently, the channel hardening approach faces challenges. In NAFD CF-mMIMO networks, where only a subset of APs serves a given UE and channels from different APs possess distinct large-scale fading coefficients, the extent of channel hardening may not be as pronounced~\cite{Chen:TCOM:2018}. Furthermore, the hardening bound necessitates the codeword to span numerous small-scale fading realizations. In real-world systems, particularly those with bursty traffic and dynamic UE buffers, UE scheduling occurs at every transmission time interval and over narrow frequency bands. This dynamic nature leads to rapid changes in the set of active UEs over time and frequency. Consequently, the hardening-based rate might underestimate the practical rate~\cite{Ngo:JPROC:2024}. To address this challenge in NAFD CF-mMIMO networks, it becomes imperative to explore more suitable bounding techniques, including side-information rate techniques. Additionally, due consideration must be given to the associated channel estimation schemes and their corresponding overhead.

\textbf{Synchronization:}  When coherent joint transmission is considered for DL transmission in NAFD CF-mMIMO networks, the data intended for a specific UE must be available at all the DL-APs, while must be also strictly synchronized. Moreover, the implementation of joint reciprocity-based beamforming in the DL, require that the DL-APs be jointly reciprocity-calibrated (or phase synchronized)~\cite{Ngo:JPROC:2024}. The practical non-reciprocity of the UL and DL channels is attributed to the inherent frequency response mismatches of the involved transmitters and receivers. Over-the-air reciprocity calibration plays a pivotal role in advancing future 6G-oriented CF-mMIMO systems~\cite{Larsson:CLET:2023}.  In practice, over-the-air reciprocity calibration can be achieved by transmitting known reference signals between the APs. Then, coherent joint transmission becomes feasible by utilizing the UL CSI and the calibration coefficients obtained from the collected calibration signals. Nevertheless, to realize the NAFD CF-mMIMO systems, it is crucial to realize the ultrafast calibration methods with low complexity and low overhead. This is particularly challenging due to the constant switching of APs between transmit and receive modes during operation. Moreover, analyzing the performance of NAFD CF-mMIMO networks in the presence of imperfect calibration is an interesting future research direction.     

\textbf{Can We Depend on O-RAN as a Future Solution?}
In recent developments,  the O-RAN alliance has sharpened its focus on standardizing RAN interfaces in conjunction with existing $3$GPP 4G   and $5$G RANs (LTE and NR). The aim is to introduce a virtualized centralized RAN  with an open interface, fostering collaboration among diverse vendor products. This initiative is set to facilitate resource pooling, streamline network management, develop advanced clock and time synchronization mechanism for multiple radio resources, and enhance the coordination of radio resources,  strengthening the overall efficiency of the network. The core of the O-RAN architecture lies on four main elements: \textit{i)} disaggregation, \textit{ii)} cloudification, \textit{iii)} intelligence, and \textit{iv)} an open internal RAN interface. These features would provide flexibility and intelligent control opportunities for next-generation wireless systems, including NAFD CF-mMIMO. More specifically, as discussed above, to overcome the scalability, inter-AP interference, and synchronization issues, the wireless transmission technologies of NAFD  CF-mMIMO architectures need to be enhanced. In particular, an NAFD  CF-mMIMO should rely on \textit{1)} different processing options at either the APs or CPU, \textit{2)}  inter-CPU coordination, \textit{3)} flexible fronthaul and cluster processors and \textit{4)} time-frequency synchronization. In pursuit of this goal, O-RAN architectures harmonize effectively with NAFD CF-mMIMO, offering support in addressing these potential challenges and fostering the feasible implementation of future NAFD CF-mMIMO systems. However, the endorsement of NAFD CF-mMIMO within the O-RAN framework, presents substantial open issues and research questions (such as defining new signaling mechanisms,  domain-specific optimization, and deployment of efficient and scalable fronthaul interference), needed to be tackled.

%%%%%%%%%%%%%%%%%%%%%%%%%%%%%%%%%%%%%%%%%%%%%%%%%%
\section{Conclusion}\label{sec:conc}
%%%%%%%%%%%%%%%%%%%%%%%%%%%%%%%%%%%%%%%%%%%%%%%%%%%
This paper provided a comprehensive overview of the state-of-the-art of IBFD massive MIMO communication. We thoroughly explored the evolution of wireless cellular network design, starting from conventional MU-MIMO cellular networks, transitioning to IBFD massive MIMO cellular systems, then progressing to IBFD CF-mMIMO, and culminating in NAFD CF-mMIMO. Then, we introduced some emerging use cases of NAFD CF-mMIMO networks, including SWIPT, wireless surveillance, and ISAC applications. Our work articulated that NAFD CF-mMIMO is a promising solution for providing energy-efficient services in communication systems with asymmetric UL and DL data traffic demands. Moreover, this structure is well-suited for dual-functionality systems, efficiently exploiting time-frequency resources alongside the spatial dimension. Finally, we discussed key open questions pertaining to the practical implementation of NAFD CF-mMIMO, particularly in ultra-dense scenarios and realistic constraints.
%==============================================================================
\bibliographystyle{IEEEtran}
\bibliography{IEEEabrv,references}
%==============================================================================

\begin{IEEEbiography}[{\includegraphics[width=1in,height=1.25in,clip,keepaspectratio]
{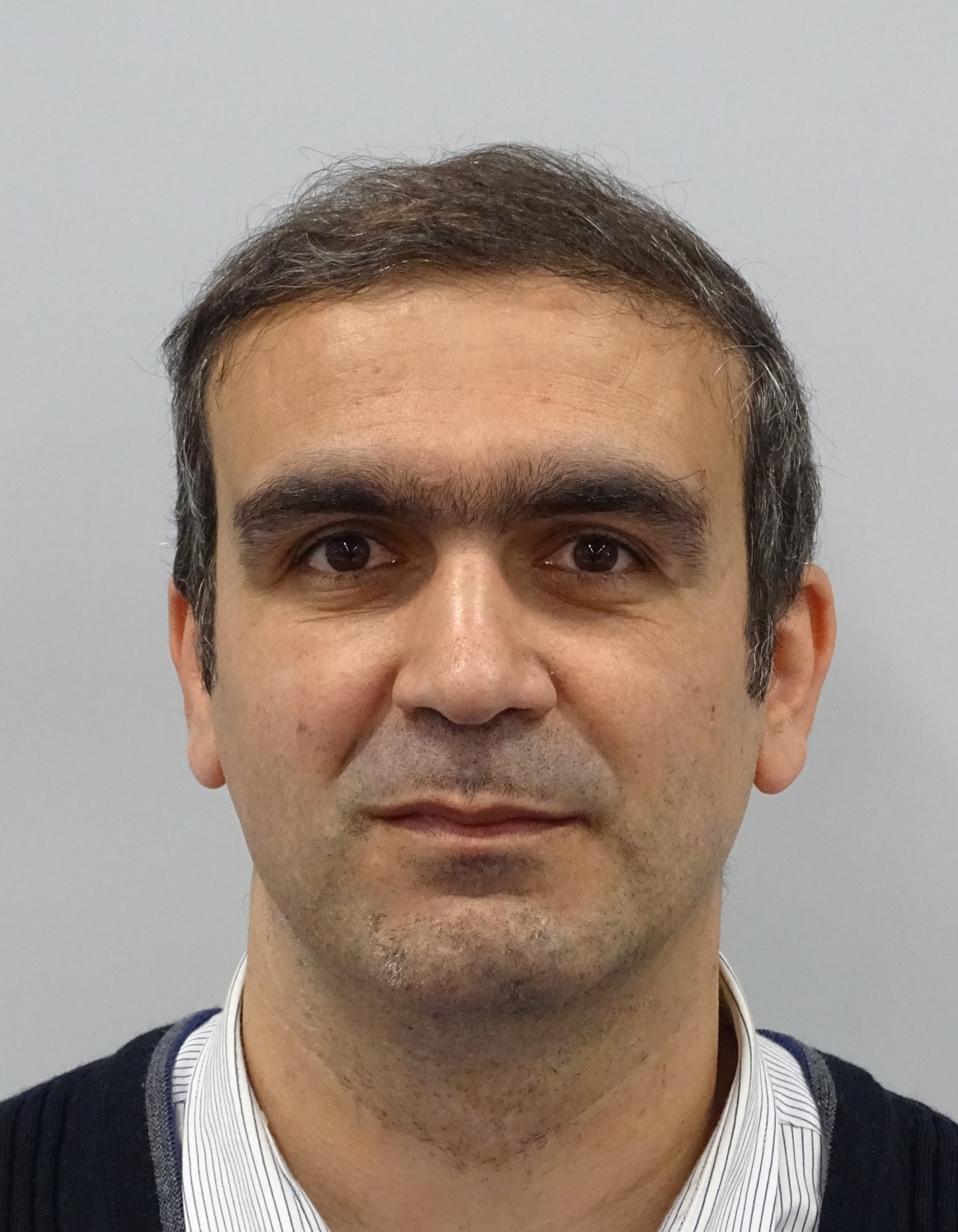}}]
{Mohammadali Mohammadi} (Senior Member, IEEE) is currently a Lecturer at the Centre for Wireless Innovation (CWI), Queen’s University Belfast, U.K. He previously held the position of Research Fellow at CWI from 2021 to 2024. His research interests include signal processing for wireless communications, cell-free massive MIMO, wireless power transfer, OTFS modulation, reconfigurable intelligent surfaces, and full-duplex communication. He has published more than 70 research papers in accredited international peer reviewed journals and conferences in the area of wireless communication. He has co-authored two book chapters, ``Full-Duplex Non-orthogonal Multiple Access Systems," invited chapter in Full-Duplex Communication for Future Networks (Springer-Verlag, 2020) and ``Full-Duplex wireless-powered communications", invited chapter in Wireless Information and Power Transfer: A New Green Communications Paradigm (Springer-Verlag, 2017). He was a recipient of the Exemplary Reviewer Award for IEEE Transactions on Communications in 2020 and 2022, and IEEE Communications Letters in 2023. He has been a member of Technical Program Committees for many IEEE conferences, such as ICC, GLOBECOM, and VTC.
\end{IEEEbiography}

\begin{IEEEbiography}[{\includegraphics[width=1in,height=1.25in,clip,keepaspectratio]{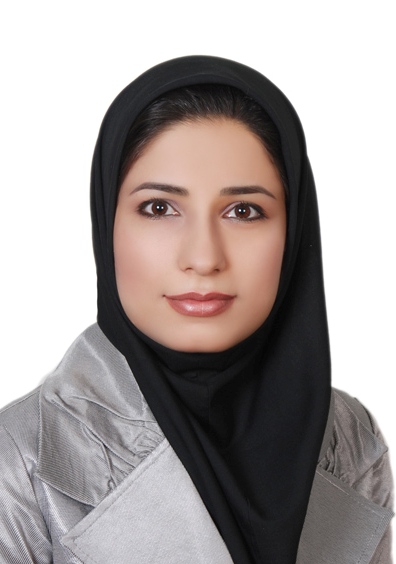}}]{Zahra Mobini}
 received the B.S. degree in electrical engineering from Isfahan University of Technology, Isfahan, Iran, in 2006, and the M.S and Ph.D.
degrees, both in electrical engineering, from the M. A. University of Technology and K. N. Toosi University of Technology, Tehran, Iran, respectively. From November 2010 to November 2011, she was a Visiting Researcher at the Research School of Engineering, Australian National University, Canberra, ACT, Australia. She is currently  a Post-Doctoral Research Fellow at the Centre for Wireless Innovation (CWI), Queen's University Belfast (QUB). Before joining QUB,  she was an Assistant and then Associate Professor with the Faculty of Engineering, Shahrekord University, Shahrekord, Iran (2015-2021). 
Her research interests include physical-layer security, massive  MIMO, cell-free massive  MIMO, full-duplex communications, and resource management and optimization. She has co-authored many research papers in wireless communications. She has actively served as the reviewer for a variety of IEEE journals,  such as TWC, TCOM, and TVT.
\end{IEEEbiography}

\begin{IEEEbiography}[{\includegraphics[width=1in,height=1.25in,clip,keepaspectratio]{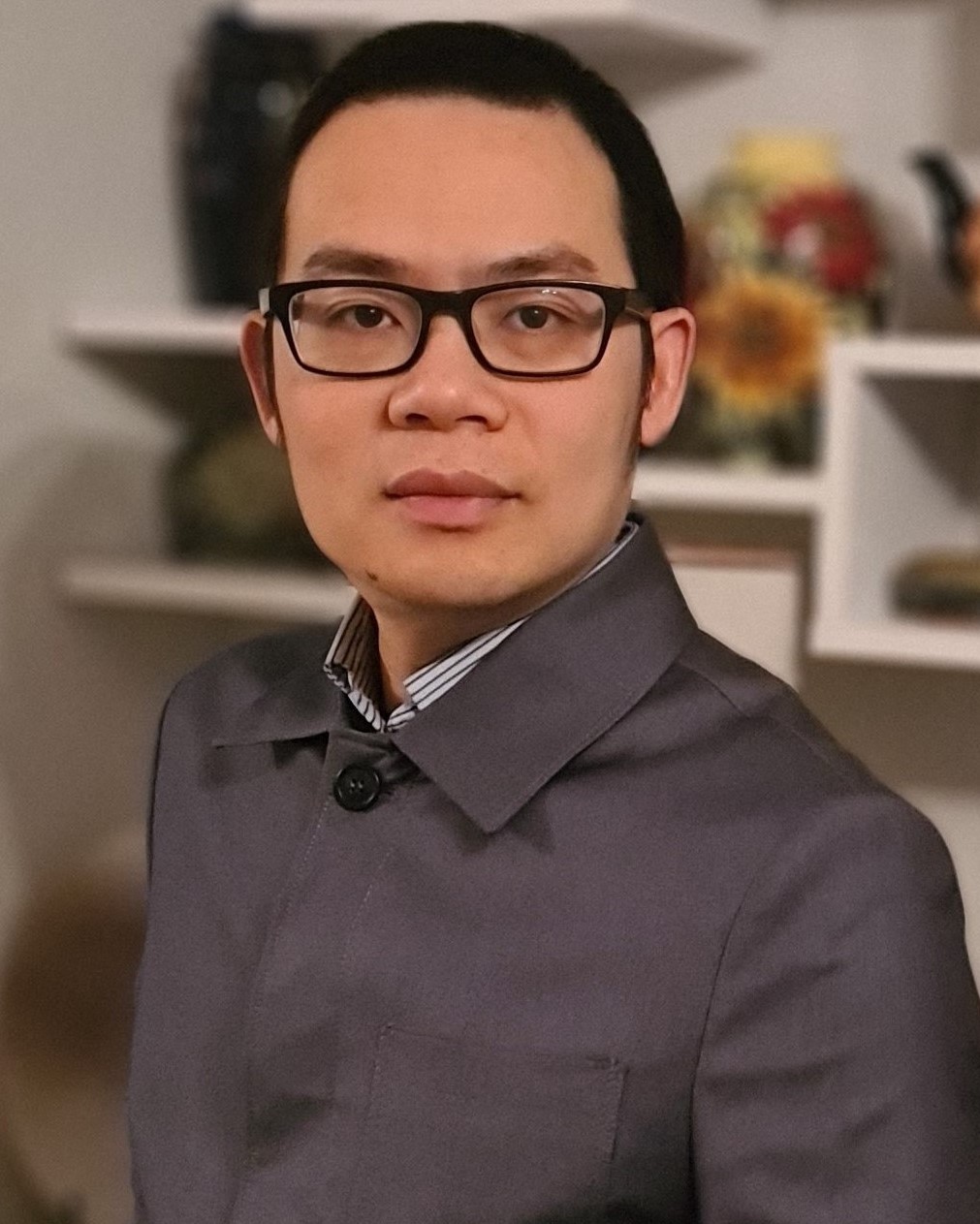}}]
{Hien Quoc Ngo}  received the B.S. degree in electrical engineering from the Ho Chi Minh City University of Technology, Vietnam, in 2007, the M.S. degree in electronics and radio engineering from Kyung Hee University, South Korea, in 2010, and the Ph.D. degree in communication systems from Link\"oping University (LiU), Sweden, in 2015. In 2014, he visited the Nokia Bell Labs, Murray Hill, New Jersey, USA.

Hien Quoc Ngo is currently a Reader (Associate Professor) at Queen's University Belfast, UK. His main research interests include cellular/cell-free massive MIMO systems, integrated sensing and communications, reconfigurable intelligent surfaces, and physical layer security. He has co-authored many research papers in wireless communications and co-authored the Cambridge University Press textbook \emph{Fundamentals of Massive MIMO} (2016).

Dr. Hien Quoc Ngo received the IEEE ComSoc Stephen O. Rice Prize in 2015, the IEEE ComSoc Leonard G. Abraham Prize in 2017,  the Best Ph.D. Award from EURASIP in 2018, and the IEEE CTTC Early Achievement Award in 2023. He also received the IEEE Sweden VT-COM-IT Joint Chapter Best Student Journal Paper Award in 2015.  He was awarded the UKRI Future Leaders Fellowship in 2019.
Dr. Hien Quoc Ngo currently serves as an Editor for the IEEE Transactions on Communications, the IEEE Transactions on Wireless Communications, the Digital Signal Processing, the Elsevier Physical Communication (PHYCOM). He was an Editor of the IEEE Wireless Communications Letters, a Guest Editor of IET Communications, and a Guest Editor of IEEE ACCESS in 2017. 
\end{IEEEbiography}

\begin{IEEEbiography}[{\includegraphics[width=1in,height=1.35in,clip,keepaspectratio]{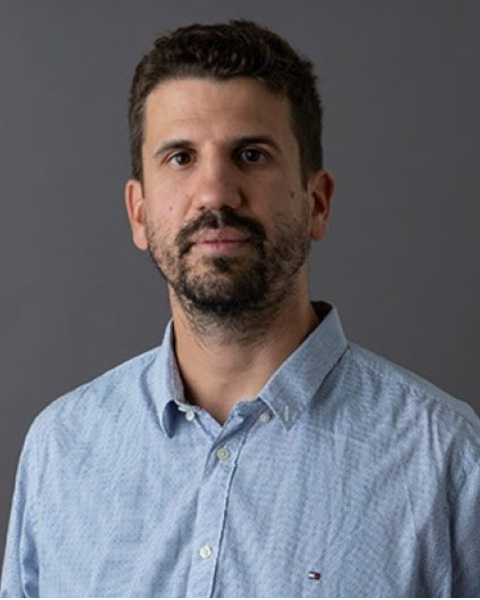}}]
{Michail Matthaiou}(Fellow, IEEE) obtained his Ph.D. degree from the University of Edinburgh, U.K. in 2008. 
He is currently a Professor of Communications Engineering and Signal Processing and Deputy Director of the Centre for Wireless Innovation (CWI) at Queen’s University Belfast, U.K. He has also held research/faculty positions at Munich University of Technology (TUM), Germany and Chalmers University of Technology, Sweden.
His research interests span signal processing for wireless communications, beyond massive MIMO, reflecting intelligent surfaces, mm-wave/THz systems and AI-empowered communications.

Dr. Matthaiou and his coauthors received the IEEE Communications Society (ComSoc) Leonard G. Abraham Prize in 2017. He currently holds the ERC Consolidator Grant BEATRICE (2021-2026) focused on the interface between information and electromagnetic theories. To date, he has received the prestigious 2023 Argo Network Innovation Award, the 2019 EURASIP Early Career Award and the 2018/2019 Royal Academy of Engineering/The Leverhulme Trust Senior Research Fellowship. His team was also the Grand Winner of the 2019 Mobile World Congress Challenge. He was the recipient of the 2011 IEEE ComSoc Best Young Researcher Award for the Europe, Middle East and Africa Region and a co-recipient of the 2006 IEEE Communications Chapter Project Prize for the best M.Sc. dissertation in the area of communications. He has co-authored papers that received best paper awards at the 2018 IEEE WCSP and 2014 IEEE ICC. In 2014, he received the Research Fund for International Young Scientists from the National Natural Science Foundation of China. He is currently the Editor-in-Chief of Elsevier Physical Communication, a Senior Editor for \textsc{IEEE Wireless Communications Letters} and \textsc{IEEE Signal Processing Magazine}, and an Area Editor for \textsc{IEEE Transactions on Communications}. He is an IEEE and AAIA Fellow.
\end{IEEEbiography}
\end{document}